\documentstyle[eqsecnum,prb,aps]{revtex}
\input epsf
\epsfverbosetrue

\newcommand{\resetcounters}{
	\setcounter{equation}{0}
        \setcounter{figure}{0}
        \setcounter{table}{0} }

\newcommand{\up}{\uparrow}
\newcommand{\down}{\downarrow}
\newcommand{\sig}{\sigma}

\begin{document}

\draft

\author{$\mbox{M. Dub\'e}^{1,2} \, \mbox{and} \,
\mbox{P. C. E. Stamp}^{3,4}$}

\title{Dynamics of a Pair of Interacting Spins Coupled to an 
       Environmental Sea}

\address{$^{1}$Helsinki Institute of Physics, P.O. Box 9
(Siltavuorenpenger 20
C),
FIN-00014, \\ University of Helsinki, Finland \\
$^{2}$Laboratory of Physics,  P.O. Box 1000, FIN-02150 HUT,  \\
Helsinki University of Technology, Espoo, Finland \\  
$^{3}$Physics and Astronomy Department, and
Canadian Institute for Advanced Research, \\
University of British Columbia, 6224 Agricultural
Rd., Vancouver, BC, Canada V6T 1Z1 \\
$^{4}$ Max Planck Institute-L.C.M.I., Ave. des Martyrs, Grenoble
38042, France}

\maketitle

\vspace{1cm}

\begin{abstract}
We solve for the dynamics of a pair of spins, coupled to each other 
and also to an environmental sea of oscillators. 
The environment mediates an indirect interaction between the spins, 
causing both mutual coherence effects and dissipation. This model describes a
wide variety of physical systems, ranging from 2 coupled microscopic systems
(eg., magnetic impurities, bromophores, etc), to 2 coupled macroscopic
quantum systems. 

We obtain analytic results for 3 regimes, viz., 
(i) The ``locked'' regime, where the 2 spins lock together; 
(ii) The ``correlated relaxation'' regime (mutually correlated incoherent
relaxation); and
(iii) The ``mutual coherence'' regime, with correlated damped oscillations. 

These results cover most of the parameter space of the system. 
\end{abstract}

\section{Introduction}
\resetcounters

In this paper we wish to establish the dynamics of a model which
describes a pair of spin-1/2 systems (ie., 2-level systems), which are
coupled both to each other and to an environmental sea of oscillators.
Our Hamiltonian thus has the form
\begin{equation}
H = H ( \mbox{\boldmath $\tau$}_{1}, \mbox{\boldmath $\tau$}_{2} ) +
H_{osc} (\{ {\bf x_{k}} \} ) + H_{int} (\mbox{\boldmath $\tau$}_{1},
\mbox{\boldmath $\tau$}_{2} ; \{ {\bf x_{k}} \} )
\label{htotal}
\end{equation}
where $\mbox{\boldmath $\tau$}_{1}$ and $\mbox{\boldmath $\tau$}_{2}$
are 2 Pauli spins and the ${\bf x_{k}}$, with ${\bf k} = 1,2,...N$, 
are the oscillator coordinates. This model is interesting for many 
reasons, and we begin by listing some of them: 

(i) We shall see that the dynamics of $\mbox{\boldmath $\tau$}_{1}$ and
$\mbox{\boldmath $\tau$}_{2}$ can actually be solved (once we have
integrated out the environment). Such solvable models of quantum
dissipative systems are rare. Previous examples include the
``spin-boson'' model \cite{leggettB} 
where a single 2-level system couples to an
oscillator bath, the ``central spin model'' 
\cite{prostamp1,stamp1,prostamp2,prostamp3,prostamp4} 
(where a 2-level system
couples to a bath of spins), quantum Brownian motion \cite{schmid,leggettE} 
and the damped oscillator \cite{leggettF,weisslivre} (where a single
oscillator couples to an oscillator bath). Only the last two of these 
have a classical analogue. 
In both classical and quantum physics, an enormous
amount about the physics of relaxation and decay has been learnt from
such models for the study of decoherence. However, significantly absent
from these studies is the generalisation to two coupled quantum systems.
In classical physics such a generalisation introduces resonant coupling,
``beats'', and correlates the relaxation. 
We shall see here that the quantum
generalisation has all this, as well as quantum interference between the
2 systems. Such phenomena are of course well-known in the context of many
examples, but the advantage of Eq. (\ref{htotal}) is that it allows us
to understand how things work when the 2 systems are strongly coupled
(either to each other or to their environment).

(ii) The model Eq. (\ref{htotal}) is particularly interesting when one
or both spins describe the low-energy dynamics of a macroscopic system.
In other papers under preparation \cite{dube1}, we have 
applied the results gained
herein to understand the tunneling dynamics of coupled Josephson
junctions in a SQUID
\cite{ycchen} and
coupled nanomagnets. 

Perhaps the most fundamental way in which Eq. (\ref{htotal}) can be
applied is to discuss the quantum measurement process
\cite{zurek}. We are aware of
no previous work which attempts to discuss both a measured system and
the measuring apparatus on an entirely quantum level, whilst both are
coupled to their environment. Previous studies (in particular those by
Leggett et al. \cite{leggettB,leggettC,leggettD}) 
deal exclusively with one or the other of these systems. In this paper 
we will already see some startling results relevant to this problem; 
we will devote an entire future paper to them \cite{dube2}. 

(iii) Models of the type Eq. (\ref{htotal}) are used widely throughout
physics, particularly in condensed matter physics. Some disparate
examples should suffice to illustrate this. A set of 2 coupled Anderson
(or Kondo) impurities are described by Eq. (\ref{htotal}) \cite{chakra}, 
with the
oscillators representing low-energy conduction electron excitations. A
pair of tunneling defects coupled to phonons or electrons \cite{sols}, 
or a pair of
chromophores in a large molecule, coupled to phonons or
polaritons \cite{chromo}, is almost trivially described 
by (\ref{htotal}), as are
nuclear spins in a metal \cite{vagner}. The problem of coupled 
2-level systems is also crucial to the understanding of many 
glasses \cite{glass}.
On a more macroscopic level, we have already
mentioned coupled SQUID's, nanomagnets, and the measurement problem. 
On a more microscopic level,
(\ref{htotal}) is an obvious generalisation  of Wilson's nucleus-meson
\cite{wilson} 
model to 2 nucleons. All these examples except for the nuclear spins
involve strong coupling to the environment. There are of course many
examples where this coupling is weak (perhaps the best known is that of
a pair of 2-level atoms, coupled to the EM field). Such weak-coupling
examples are well understood, and do not require the methods used below
for their solution. 

In this paper, we shall be concerned principally with the application of
Eq. (\ref{htotal}) to macroscopic systems at low temperature. This will
lead us to study a special case of (\ref{htotal}), in which 
$H = H_0 +H _{int}^{P}$, where
\begin{equation}
H_{0} =  - \frac{1}{2} (\Delta_{1} \hat{\tau}_{1}^{x} + 
                         \Delta_{2} \hat{\tau}_{2}^{x}) 
+ \frac{1}{2} K_{zz} \hat{\tau}_{1}^{z} \hat{\tau}_{2}^{z} + 
\frac{1}{2} \sum_{{\bf k}=1}^{N} m_{{\bf k}} 
( \dot{{\bf x}}_{{\bf k}}^{2} 
+ \omega_{{\bf k}}^{2} {\bf x_{k}}^{2})
\label{hpisces}
\end{equation}
\begin{equation}
H_{int}^{P} = \frac{1}{2} \sum_{{\bf k}=1}^{N} 
( c_{{\bf k}}^{(1)} e^{i {\bf k \cdot R}_{1}}
\hat{\tau}_{1}^{z} + c_{{\bf k}}^{(2)} e^{i {\bf k \cdot R}_{2}} 
\hat{\tau}_{2}^{z})
{\bf x_{k}}
\label{hintpisces}
\end{equation}
where the N oscillator bath modes $\{ {\bf x_{k}} \}$ are 
assumed for simplicity (but without real loss of generality) 
to be momentum eigenstates. 
The Pauli matrices
$\hat{\mbox{\boldmath $\tau$}}_{1}$ and 
$\hat{\mbox{\boldmath $\tau$}}_{2}$ represent the 2 
different spins. The two spins are coupled through a direct 
interaction term $K_{zz}$ giving ferro- or antiferromagnetic correlations, 
depending on whether $K_{zz}$ is negative or positive. 
In the absence of the coupling to the environmental sea, the
spins have their levels split by ``tunneling'' matrix elements
$\Delta_{1}$ and $\Delta_{2}$. We work in the basis in which
$ | \uparrow \rangle$ and $ | \downarrow \rangle$, eigenstates of
$ \hat{ {\bf \tau}}^{z}$, are degenerate until split by the off-diagonal
tunneling. 
 
Henceforth we will refer to Hamiltonians like (1.1) as
$H_{PISCES}(\mbox{\boldmath $\tau$}_{1},
\mbox{\boldmath $\tau$}_{2} ; \{ {\bf x_{k}} \} )$, 
where ``PISCES'' is an abbreviation for ``Pair
of Interacting Spins Coupled to an Environmental Sea''. They are an obvious
generalisation of the well-known spin-boson Hamiltonian, which has the form
\begin{equation} 
 H= -\frac{1}{2} \Delta \hat{\tau}^{x} + \epsilon \hat{\tau}^{z}
 + \frac{1}{2} \sum_{{\bf k}=1}^{N} m_{{\bf k}} ( \dot{{\bf x}}_{{\bf k}}^{2}
 + \omega_{{\bf k}}^{2} {\bf x_{k}}^{2})
 + \frac{1}{2} \hat{\tau}^{z} \sum_{{\bf k}=1}^{N} c_{{\bf k}}{\bf x_{k}}
 \label{hspinboson}
 \end{equation}   
whose dynamics was discussed in detail be Leggett et al \cite{leggettB}.

The task of the present paper is to solve for the 
the 2-spin {\it reduced} density matrix $\hat{\rho} ( 
\hat{\mbox{\boldmath $\tau$}}_{1},
\hat{\mbox{\boldmath $\tau$}}_{2}; 
\hat{\mbox{\boldmath $\tau$}}_{1}', 
\hat{\mbox{\boldmath $\tau$}}_{2}';t)$,
produced by integrating out the bath oscillators; this describes the dynamics
of the 2 spins. We emphasize here that such a study is an essential
preliminary to any attempt to understand the dynamics of any of the 
strongly-coupled physical
systems listed above; we devote the whole of the present paper to elucidating 
this dynamics. Although the application of these results is reserved mostly
to later papers, we will comment on this herein, where possible.

We may (very briefly) summarize our results by remarking that
there are essentially three regimes of interest for the PISCES model,
depending on how strong is the coupling of each spin to the bath, and also
how high is the temperature $T$. These are
 
 (i) The ``Strong Coupling'' regime, where the 2 spins are sufficiently
 correlated so they lock together, and essentially behave as a single
 spin, with a quite different tunneling matrix element and a different
 coupling to the bath.

 (ii) The ``Correlated Relaxation'' regime, where the 2 spins tunnel
 incoherently, but nevertheless their motion is correlated.

 (iii) The ``Mutual Coherence'' regime, where the motion of one or both
 spins still shows (damped) oscillations, which  
 are partially correlated.

 There is also a $4^{th}$ solvable regime which we have ignored - this is the
 perturbative regime, where the 2 spins are so weakly coupled to the bath
 and to each other that their dynamics can be understood by perturbation
 expansion in the interaction. What we have done in the present paper is to
 calculate the 2-spin density matrix in {\it analytic} form in the above 3
 regimes. We have concentrated on the case where the spins are coupled
 ``Ohmically'' to the bath, since this is usually the physically interesting
 case. In this Ohmic case, the results can be presented in the form of a 
 ``phase diagram''; see section V.

 {\bf Organisation of the paper :} In section II we discuss in more detail
 the physical origin of the PISCES model, in particular for the case where
 the 2 spins describe the low-energy dynamics of 2 macroscopic systems.
 In section III we discuss
 a simple ``toy''
 model, where 2 spins couple by a direct interaction
 $J_{0} \hat{\tau}_{1}^{z} \hat{\tau}_{2}^{z}$ only. This model 
 demonstrates correlations between the 2 spins in the absence of 
 any dissipative environment; it also makes clear the physical
 meaning of the various density matrix elements. In section IV we set up the
 formal solution for $\hat{\rho}$ for the full PISCES model, using influence
 functional methods. Section V is the most important in the paper; therein 
 we give the detailed analytic results for
 $\hat{\rho}$ in the various regimes described above, and discuss their
 physical significance.
 We summarise our results in section VI. The more complicated details of
 the calculations are relegated to a series of Appendices.
  
Those readers whose primary interest is the results for the PISCES dynamics
should go directly to section V, which can mostly be read independently of the 
rest of the paper.

\section{The Model}
\resetcounters

In this section we summarize the model, and sketch the way in which the 
truncated PISCES Hamiltonian can arise from a more microscopic model. No 
attempt is made at a detailed derivation, since this would require that
we specify a particular physical system - only the basic ideas are
outlined. 

\subsection{Effective Hamiltonians and Lagrangians}

To see how a PISCES Hamiltonian can arise, we start with a very
simple model, with a total  
Lagrangian given by 
\begin{equation}
L=L_{0}+L_{int}+L_{B}
\label{tlagrangian}
\end{equation}
\begin{equation}
L_{B}= \frac{1}{2} \sum_{{\bf k}} m_{{\bf k}}( \dot{x}_{{\bf k}}^{2} - 
\omega_{{\bf k}}^{2} {\bf x_{k}}^{2})
\label{blagrangian}
\end{equation}

\begin{equation}
L_{0} = \frac{1}{2} (M_{1}\dot{q}_{1}^{2} + M_{2}\dot{q}_{2}^{2})
- (V_{1}(q_{1}) + V_{2}(q_{2}))
\label{pbarelagrangian}
\end{equation}
\begin{equation}
L_{int} = - \sum_{{\bf k}} \left[ (q_{1}c_{{\bf k}}^{(1)} +
q_{2}c_{{\bf k}}^{(2)} ){\bf x_{k}}
      + \frac{1}{2} (\frac{{\bf x_{k}}^{2}}{m_{{\bf k}} \omega_{{\bf k}}^{2}}) 
( |c_{{\bf k}}^{(1)}|^{2}q_{1}^{2} + |c_{{\bf k}}^{(2)}|^{2}q_{2}^{2} ) \right] 
\label{pintlagrangian}
\end{equation} 
where it is assumed that $V_{1}(q_{1})$ and $V_{2}(q_{2})$ each describe
1-dimensional 2-well potentials, which are each symmetric
under the interchange $q_{\alpha} \rightarrow -q_{\alpha}$, with minima at
$q_{1}=\pm q_{01}/2$, $q_{2}=\pm q_{02}/2$. 
Note that in (\ref{pintlagrangian}), the effective 
couplings $c_{{\bf k}}^{(1)}$ and $c_{{\bf k}}^{(2)}$ are in 
general complex and out of 
phase with each other. Thus, for example, if the two systems in $L_{0}$ 
in (\ref{pbarelagrangian}) represent two systems at different positions 
$ {\bf R}_{1} $ 
and $ {\bf R}_{2} $, a commonly used form for the $c_{{\bf k}}^{\alpha}$ 
will be 
$ c_{{\bf k}}^{\alpha} = c_{{\bf k}} e^{i{\bf k \cdot R}_{\alpha}}$, 
where ${\bf k}$ is a 
wave vector and $\alpha =1,2$. 
The form we have chosen for the coupling between the environmental modes 
$\{ x_{k} \}$ and the coordinates $q_{\alpha}$ is linear in both, as in the 
usual Feynman-Vernon/Caldeira-Leggett \cite{vernon,leggettA} scheme- 
we discuss the validity and generality of this choice below. 

Included in (\ref{pintlagrangian}) 
are two separate counterterms for each potential, quadratic in $q_{\alpha}$. 
The purpose is, as usual, to restore the bare potential 
$V(q_{1},q_{2})=V_{1}(q_{1})+V_{2}(q_{2})$,  after renormalisation by the 
coupling to the environmental bosons, to its original value so that
we may treat it as the physical potential. In Eq. (\ref{blagrangian}), we 
have not yet truncated the problem to a ``PISCES'' problem, so that no direct
interaction terms appear. This is why the 
counterterm involves 2 separate contributions.
If in (\ref{pbarelagrangian}) there had been a term 
$V_{12}(q_{1},q_{2})$ involving
$q_{1}$ and $q_{2}$ together in a direct interaction, then the counterterm 
would no longer be separable either. Thus, in the form 
(\ref{pbarelagrangian})
the PISCES model only involves indirect interaction between system 1 and 
system 2, mediated by the oscillator bath. 

The Lagrangian (\ref{tlagrangian}) is a simple generalization of the single 
2-well model studied by Leggett et al. \cite{leggettB}, which they truncated
to the spin-boson model. The manoeuvres required for the truncation of 
(\ref{tlagrangian}) are the same; we require simply that the separations
$\Omega_{0}^{(1)}$ and $\Omega_{0}^{(2)}$, of the 2 lowest levels in wells 
1 and 2 respectively, from the higher levels in these wells, be much greater
than any energy scale (such as $T$) that we are interested in. If we impose
an ultraviolet cutoff $\omega_{c}$ on the bath modes, such that 
$\Omega_{0}^{(1)}, \Omega_{0}^{(2)} \gg \omega_{c} \gg \Delta_{1}, \Delta_{2},
 T $, then we may truncate (\ref{tlagrangian}) to the form
\begin{equation}
H=H_{0}+H_{int}+H_{B}
\label{thamiltonian}
\end{equation}
\begin{equation}
H_{B}=\frac{1}{2} \sum_{{\bf k}} m_{{\bf k}} ( \dot{{\bf x}}_{{\bf k}}^{2} 
+ \omega_{{\bf k}}^{2} {\bf x_{k}}^{2}) 
\label{bhamiltonian}
\end{equation}
\begin{equation}
H_{0} = - \frac{1}{2} (\Delta_{1} \hat{\tau}_{1}^{x} +
                               \Delta_{2} \hat{\tau}_{2}^{x})
\label{peffhamiltonian}
\end{equation}
\begin{equation}
H_{int} = \frac{q_{01}}{2} \hat{\tau}_{1}^{z} \sum_{{\bf k}} c_{{\bf k}}^{(1)}
{\bf x_{k}} 
  + \frac{q_{02}}{2} \hat{\tau}_{2}^{z} \sum_{{\bf k}} c_{{\bf k}}^{(2)}
{\bf x_{k}}
\label{pinthamiltonian}
\end{equation}

This is just the PISCES model of Eq. (\ref{hpisces}) and (\ref{hintpisces})
( in (\ref{hpisces}) and (\ref{hintpisces}), the factors $q_{01}$ and $q_{02}$
are absorbed into the couplings $c_{k}^{(1)}$ and $c_{k}^{(2)}$). 
 
The splittings $\Delta_{1}$ and $\Delta_{2}$ can be calculated from 
$L_{0}^{P}$ using WKB-instanton methods. Although this procedure is quite 
complicated and technically interesting for a spin-boson problem in which the 
wells
are not all degenerate, we will not discuss it here. We will simply assume 
that $\Delta_{1}$ and $\Delta_{2}$ are given quantities, in the spirit of the
low-energy effective Hamiltonian, and that they are, as usual, exponentially
small compared to the energy scale of $V(q_{1},q_{2})$ in 
(\ref{pbarelagrangian}) (ie., exponentially smaller than $\Omega_{0}^{(1)}$
and $\Omega_{0}^{(2)}$ respectively).

$H_{PISCES}$ as written in (\ref{hpisces}) is clearly not the most general
model of this kind.
A much wider range of direct couplings is possible; instead of
$K_{zz} \hat{\tau}_{1}^{z} \hat{\tau}_{2}^{z}$ we could use
\begin{equation}
H_{int}^{dir} = \frac{1}{2} \sum_{\mu \nu}
K_{\mu \nu} \hat{\tau}_{1}^{\mu} \hat{\tau}_{2}^{\nu}
\label{direct}
\end{equation}
We could also use more complicated indirect couplings
(ie., couplings to the bath)
like $\frac{1}{2}  \hat{\tau}^{\mu}_{\alpha} \sum_{{\bf k}}
c_{{\bf k} \mu}^{(\alpha)}
e^{i{\bf k \cdot R}_{\alpha}} {\bf x_{k}}$,
with $\mu = x,y,z$ and $\alpha=1,2$.
In the present paper we will stick to the diagonal coupling in
(\ref{hintpisces}) and keep only the direct coupling in
$\hat{\tau}^{z}_{1} \hat{\tau}^{z}_{2}$.
Our reasoning is as follows. Just as for the single spin-boson problem, we
expect diagonal couplings to dominate the low-energy dynamics of the
combined system, since the spins spend almost all their time in a diagonal
state (only a fraction $\Delta/\Omega_0$ of their time is in a non-diagonal
state, when the relevant system is tunneling under the barrier). In
certain cases the diagonal couplings can be zero (usually for symmetry
reasons), and then one must include non-diagonal
couplings
like $\frac{1}{2} \hat{\tau}^{\bot}_{\alpha}
\sum_{{\bf k}} (c_{{\bf k} \bot}^{\alpha}
e^{i {\bf k \cdot R}_{\alpha}} {\bf x_{k}} + H.c.)$.
Here we deal only with the generic case of dominant
diagonal couplings; we shall see in any case that the physics
generated by (\ref{hintpisces}) is complicated enough as it is.
 
 Our reason for keeping the longitudinal direct coupling in (\ref{direct})
 is connected
 to our choice of diagonal couplings between the bath and the
 systems. The interaction in (\ref{hintpisces}) will couple
 $\hat{\tau}^{z}_{1}$ and $\hat{\tau}^{z}_{2}$ to produce a
 longitudinal direct coupling of the form $ \frac{1}{2} {\cal J}
 \hat{\tau}^{z}_{1} \hat{\tau}^{z}_{2}$. In fact, one may quite
 generally observe that in a field-theoretical context, any direct
 interaction
 can be viewed as the result of a coupling between $\hat{{\bf \tau}}_{1}$ and
 $\hat{{\bf \tau}}_{2}$ through the high-frequency or ``fast'' modes of some
 dynamic field; ``fast'' in this context simply means much faster than the
 low-energy scales of interest, so that the interaction may be treated as
 quasi-instantaneous. In our case, integrating out these fast modes, those
 with
 a frequency greater than $\Omega_{0}$ then produces the
 longitudinal static interaction in (\ref{hpisces}).
  
  One further omission from (\ref{hpisces}) is an applied bias acting on one
  or both of the spins, of the form $\epsilon_{j} \hat{\tau}_{j}^{z}$, for
  example. This important extra ingredient will be discussed elsewhere
  \cite{dube1,dube2}- our present interest 
  is in the dynamic interaction  
 between the spins, generated through the bath, and  most of the present
  paper
  is devoted to unravelling its effects.

\subsection{The Oscillator Bath Representation}

The ``high-energy'' Lagrangian of (\ref{tlagrangian}) to (\ref{pintlagrangian})
already uses an oscillator bath representation of the environment, ie., it
already involves a low-energy truncation of some more detailed microscopic
representation of the environment.
We briefly recall the arguments that
justify this representation \cite{vernon,leggettA}, and when they fail.

One begins by assuming a large number
$N$ of degrees of freedom in the environment, with couplings $f_{ij}$ between
the environmental excitations $|i\rangle$ and $|j\rangle$ of order $1/N$. 
One also assumes that the coupling $V_{j}$ of $|j\rangle$ to an external 
``system'' obeys $V_{j} \sim O(N^{-1/2})$. In the PISCES model, as well as the 
spin-boson model, the $V_{j}$ are just the $c_{{\bf k}}$. 
There are 
of course many canonical examples of such environments, such as $^{3}He$ 
(normal, 
superfluid, solid, or quantum gas), $^{4}He$ superfluid and solid, metals, 
semiconductors, magnets, superconductors, simple insulators, etc. Notice that 
any {\it delocalised} degrees of freedom will {\it formally} satisfy these 
requirements, simply because their wave-functions will be normalised inside
a box of size $\sim N$.

If we treat the couplings $V_{j}$
perturbatively, up to $2^{nd}$-order, it is
always possible to represent the environment as a bath of oscillators 
(c.f.,  Feynman and Vernon \cite{vernon}, pp. 153-159). 
We are then bound to end up with an effective Lagrangian of form
\begin{equation}
L_{eff}=L_{0}+L_{int}+L_{B}
\label{leff}
\end{equation}
where $L_{B}$ takes the form (\ref{blagrangian}), $L_{0}(Q,\dot{Q})$ is the 
Lagrangian of the system and
\begin{equation}
L_{int} = - \sum_{{\bf k}} ( F_{{\bf k}}(Q) {\bf x_{k}} + 
G_{{\bf k}}(\dot{Q}) \dot{{\bf x}}_{{\bf k}})
\label{feynvern}
\end{equation}
(or some form related to this by a canonical transformation 
\cite{leggettprb}). 
It is often assumed that the excursions of the system are small, and then  
$L_{int}$ is written as $L_{int}=-\sum c_{{\bf k}}{\bf x_{k}}Q$, 
as in the PISCES and spin-boson problems. 
However in many practical situations one obviously has to go
to more general couplings (eg., the coupling 
$L_{int}= - \sum (V_{{\bf q}} e^{i{\bf q \cdot r}} b_{{\bf q}}^{+} 
+ H.c. $), arising from a potential 
probe $V(r)= \sum V_{{\bf q}} e^{i{\bf q \cdot r}}$; 
here $b_{{\bf q}}^{+}$ is an 
environmental boson operator, which could equally describe fermion 
particle-hole pairs via 
$b_{{\bf q}}^{+} = \sum_{{\bf k}} a_{{\bf k+q}}^{+} a_{{\bf k}}$), 
and one is led to (\ref{feynvern}).

Thus the question of the validity of the oscillator bath representation of the
environment boils down to the applicability of the 2nd-order perturbative
treatment of the $V_j$. It is important to understand that even if there 
appears to be some breakdown of the perturbative expansion, it may easily
be curable by a suitable canonical transformation.
Consider, eg., the Luttinger liquid with
back scattering; although the $\{ V_{j} \}$ are formally 
$ \sim O(N^{-1/2})$,  
they are infrared divergent
(in Caldeira-Leggett language, they are subohmic). 
Nevertheless,
a canonical transformation to a quite new set of environmental modes
\cite{proko1} reveals that 
this subohmic form is an artefact
of the original bosonisation- the coupling to the new modes is a 
conventional ohmic one. Another salutary example is provided by the coupling
of solitons to their environment.

However, an environment which {\it cannot} in general be mapped 
to an oscillator bath
is an environment of uncoupled (or very weakly coupled) spins, or 2-level
systems. 
In this case one finds that 
the couplings of the N individual spins $\{ \sigma_{k} \}$ to the central
quantum system are not $ \sim O(N^{-1/2})$, but independent of $N$.
In some cases they are also very strong, so that instead of 
the 
environmental dynamics being only weakly affected by the quantum system of
interest, they are in fact slaved to it 
\cite{prostamp1,prostamp3,prostamp4}.

{}From this very brief discussion (for a much more lengthy one see ref.
\cite{note1}), we see that provided we can ignore the effects of some ``spin
bath'' environment (eg., nuclear or paramagnetic spins, or local defects of
some kind), the oscillator bath description should be a good representation 
of the low-energy environmental degrees of freedom.

\section{Density Matrix for two Coupled Spins (no environment)}
\resetcounters

Before dealing with the full PISCES model, we summarize the results for the
dynamics of a much simpler problem, in which the 2 spins couple to each 
other by a static longitudinal interaction $J_{0} \hat{\tau}_{1}^{z}
\hat{\tau}_{2}^{z}$, and no environment is present.
In a certain sense the PISCES model can be viewed as a 
dissipative version of this ``toy model''. 

\subsection{Single Spin}

First recall that a single spin in a bias field $\epsilon$, with  
Hamiltonian
\begin{equation}
H=- \frac{\Delta_{0}}{2} \hat{\tau}_{x} + \frac{\epsilon}{2} 
\hat{\tau}_{z}
\label{2whamiltonian}
\end{equation}
and with eigenvalues
\begin{equation}
E_{\pm} = \pm |E| =
\pm \frac{1}{2} \sqrt{ \Delta^{2} + \epsilon^{2} }
\end{equation}
and eigenfunctions
\begin{equation}
\psi_{\pm} = A_{\pm} \, \left[ \, (E_{\pm}+\epsilon)| \up \rangle -
\Delta_{0} | \down \rangle \, \right] \, ,
\end{equation}
\begin{equation}
A_{\pm} = \left[ \frac{1}{(E_{\pm}+\epsilon)^{2}+(\Delta_{0})^{2}} \right]^{1/2}
\end{equation}
has a ``1-spin'' density matrix 
\begin{equation}
\rho_{\alpha \beta} (t) = c_{\alpha}(t)c_{\beta}^{*}(t)
\label{2wdenformal}
\end{equation}
in which the wave-function $\Psi (t)$ of the system, at time $t$, is expanded
as $\Psi (t) = \sum_{\alpha} c_{\alpha}(t) \phi_{\alpha}(t)$, with
$\phi_{1}=| \uparrow \rangle$ and $\phi_{2} = | \downarrow \rangle$. 
Assuming an initial state $| \uparrow \rangle$, $\rho(t)$ then evolves as 
\begin{equation}
\rho(t) = \left(
\begin{array}{cc}
1-\frac{\Delta_{0}^{2}}{E^{2}} \sin^{2} Et & -i \frac{\Delta_{0}}{E} \sin Et \\
i \frac{\Delta_{0}}{E} \sin^{2} Et & \frac{\Delta_{0}^{2}}{E^{2}} \sin Et 
\end{array} \right)
\label{2wbiasdensityt}
\end{equation}
In the equivalent 2-well problem, the 
initial ``wave-packet'' $| \uparrow \rangle $
partially oscillates between the 2 wells- the diagonal elements give the  
occupation probability of the wells, and the off-diagonal elements
describe oscillatory quantum interference between 
them, which is suppressed by
the bias; when $\epsilon \gg \Delta_{0}$, the particle 
stays in one well, in the absence of any coupling to
the environment.

\subsection{Two Spins}

Now consider 2 coupled spins, with a Hamiltonian containing
a simple longitudinal interaction :
\begin{equation}
H = -\frac{1}{2} ( \Delta_{1} \hat{\tau}_{1}^{x} + \Delta_{2} 
\hat{\tau}_{2}^{x} ) + J_{0} \hat{\tau}_{1}^{z} \hat{\tau}_{2}^{z}
\label{coupledspinsh}
\end{equation}
The coupling creates a ``dynamical bias'' whereby each
spin tends to force the other spin into one or other of its wells. 
The 2-spin density matrix 
takes the form
\begin{equation}
 \rho_{\tau_{1} \tau_{2} ; \tau_{1}' \tau_{2}'}(t)
 = A_{\tau_{1} \tau_{2}}(t) A_{\tau_{1}' \tau_{2}'}^{*}(t)
\end{equation}
(c.f., Eq (\ref{2wdenformal})),in which we order the states as 
$ \{ | \tau_{1} \tau_{2} \rangle \} = \{ | \uparrow \uparrow \rangle , 
 | \uparrow \downarrow \rangle , | \downarrow \uparrow \rangle , 
 | \downarrow \downarrow \rangle \}$;
 the amplitudes $A_{\tau_{1} \tau_{2}}(t)$ are given by
\begin{displaymath}
A_{\uparrow \uparrow} = \frac{1}{2} \left( [ \cos \Omega_{-} t + \cos 
\Omega_{+} t ] + iJ_{0} [ \frac{1}{\Omega_{+}} \sin \Omega_{+} t + 
\frac{1}{\Omega_{-}} \sin \Omega_{-} t ] \right) 
\end{displaymath}
\begin{displaymath}
A_{ \uparrow \downarrow} = \frac{i}{4} \left( \frac{\Delta_{1}- \Delta_{2}}
{\Omega_{-}} \sin \Omega_{-} t - \frac{\Delta_{1}+\Delta_{2}}{\Omega_{+}}
\sin \Omega_{+} t \right)
\end{displaymath}
\begin{displaymath}
A_{ \downarrow \uparrow} = -\frac{i}{4} \left( \frac{\Delta_{1}- \Delta_{2}}
{\Omega_{-}} \sin \Omega_{-} t + \frac{\Delta_{1}+\Delta_{2}}{\Omega_{+}}
\sin \Omega_{+} t \right) 
\end{displaymath}
\begin{equation}
A_{\downarrow \downarrow} = \frac{1}{2} \left( [ \cos \Omega_{-} t - \cos
\Omega_{+} t ] + iJ_{0} [ \frac{1}{\Omega_{+}} \sin \Omega_{+} t -
\frac{1}{\Omega_{-}} \sin \Omega_{-} t ] \right) 
\label{amplitudes}
\end{equation}
where at $t=0$ we have assumed an initial state $| \tau_{1}' \tau_{2}' 
\rangle = | \uparrow \uparrow \rangle$; and the eigenfrequencies $\Omega_{\pm}$
are given by
\begin{equation}
\Omega_{\pm}^{2} = J_{0}^{2} + \frac{1}{4} (\Delta_{1} \pm \Delta_{2})^2
\label{freefrequencies}
\end{equation}
so that when $J_{0}=0$, the amplitudes in (\ref{amplitudes}) reduce to 
\begin{displaymath}
A^{(0)}_{\uparrow \uparrow}= \cos \frac{\Delta_{1} t}{2} \cos \frac{\Delta_{2} t}{2}
\end{displaymath}
\begin{displaymath}
A^{(0)}_{\uparrow \downarrow}= -i \cos \frac{\Delta_{1} t}{2} \sin \frac{\Delta_{2} t}
{2}
\end{displaymath}
\begin{displaymath}
A^{(0)}_{\downarrow \uparrow}= -i \sin \frac{\Delta_{1} t}{2} \cos \frac{\Delta_{2} t}
{2}
\end{displaymath}
\begin{equation}
A^{(0)}_{\downarrow \downarrow} = \sin \frac{\Delta_{1} t}{2} \sin \frac{\Delta_{2} t}
{2}
\label{ampjzero}
\end{equation}

{}From this we obtain the probability $P_{\tau_{1}
\tau_{2}}(t) = | A_{\tau_{1} \tau_{2}}(t)|^2$ of finding the 2-spin system
in state $| \tau_{1} \tau_{2} \rangle $ at time $t$ (assuming a state
$| \uparrow \uparrow \rangle$ when $t=0$), and the interference between
these states. 
When $J_{0}=0$, the 2 spins oscillate independently,
with product wave function $| \tau_{1} \rangle | \tau_{2} \rangle$, and
$\rho(t)$ has oscillation components corresponding to the sum and 
difference frequencies $(\Delta_{1} \pm \Delta_{2})$ 
(see (\ref{ampjzero})). When $J_{0}$ is
finite, we still have sum and difference frequencies $\Omega_{\pm}$ in 
$\rho(t)$, but the behaviour is more complicated because the 
wave-function is no longer a product wave-function - the coupling 
``entangles'' the 2 spins. In fact the eigenfunctions are now
\begin{displaymath}
\Psi_{++} = \left( \frac{\Omega_{+} - J_{0}}{2\Omega_{+}} \right)^{1/2} 
\left( \frac{\Delta_{1} + \Delta_{2}}{2(\Omega_{+} -J_{0})} | \uparrow 
\uparrow \rangle + | \uparrow \downarrow \rangle + | \downarrow \uparrow 
\rangle + \frac{\Delta_{1}+\Delta_{2}}{2(\Omega_{+} -J_{0})} | \downarrow 
\downarrow \rangle \right)
\end{displaymath}
\begin{displaymath}
\Psi_{+-} = \left( \frac{\Omega_{-} - J_{0}}{2\Omega_{-}} \right)^{1/2}
\left( \frac{\Delta_{1} - \Delta_{2}}{2(\Omega_{-} -J_{0})} | \uparrow
\uparrow \rangle + | \uparrow \downarrow \rangle + | \downarrow \uparrow
\rangle + \frac{\Delta_{1}+\Delta_{2}}{2(\Omega_{-} -J_{0})} | \downarrow
\downarrow \rangle \right)
\end{displaymath}
\begin{displaymath}
\Psi_{-+} = \left( \frac{\Omega_{-} + J_{0}}{2\Omega_{-}} \right)^{1/2}
\left( \frac{\Delta_{1} - \Delta_{2}}{2(\Omega_{-} +J_{0})} | \uparrow
\uparrow \rangle + | \uparrow \downarrow \rangle + | \downarrow \uparrow
\rangle + \frac{\Delta_{1}-\Delta_{2}}{2(\Omega_{-} +J_{0})} | \downarrow
\downarrow \rangle \right)
\end{displaymath}
\begin{equation}
\Psi_{--} = \left( \frac{\Omega_{+} + J_{0}}{2\Omega_{+}} \right)^{1/2}
\left( \frac{\Delta_{1} + \Delta_{2}}{2(\Omega_{+} +J_{0})} | \uparrow
\uparrow \rangle + | \uparrow \downarrow \rangle + | \downarrow \uparrow
\rangle + \frac{\Delta_{1}+\Delta_{2}}{2(\Omega_{+} +J_{0})} | \downarrow
\downarrow \rangle \right)
\label{2cspineigenf}
\end{equation}
with energies
\begin{displaymath}
E_{++} = -E_{--} =  \Omega_{+}
\end{displaymath}
\begin{equation}
E_{+-} = - E_{-+} = \Omega_{-}
\label{2cspinenerg}
\end{equation}
and the way in which the entanglement occurs depends on the magnitude of $J_{0}$
and its sign. For $J_{0} < 0$ we have (FM) ferromagnetic coupling; 
when $-J_{0} \gg \Delta_{1}, \Delta_{2}$, the ground state
$| \Psi_{--} \rangle$
tends toward the superposition $2^{-1/2} ( | \uparrow \uparrow 
\rangle + | \downarrow \downarrow \rangle )$. For 
antiferromagnetic (AFM) coupling, the states $| \uparrow \downarrow \rangle$
and $| \downarrow \uparrow \rangle$ are emphasized and for $J_{0} \gg \Delta_{1}, 
\Delta_{2} $, the ground state, which is now $| \Psi_{++} \rangle$ 
 tends to the superposition 
$2^{-1/2} ( | \uparrow \downarrow \rangle + | \downarrow \uparrow \rangle)$.

\begin{figure}[t]
\epsfbox[50 200 550 760]{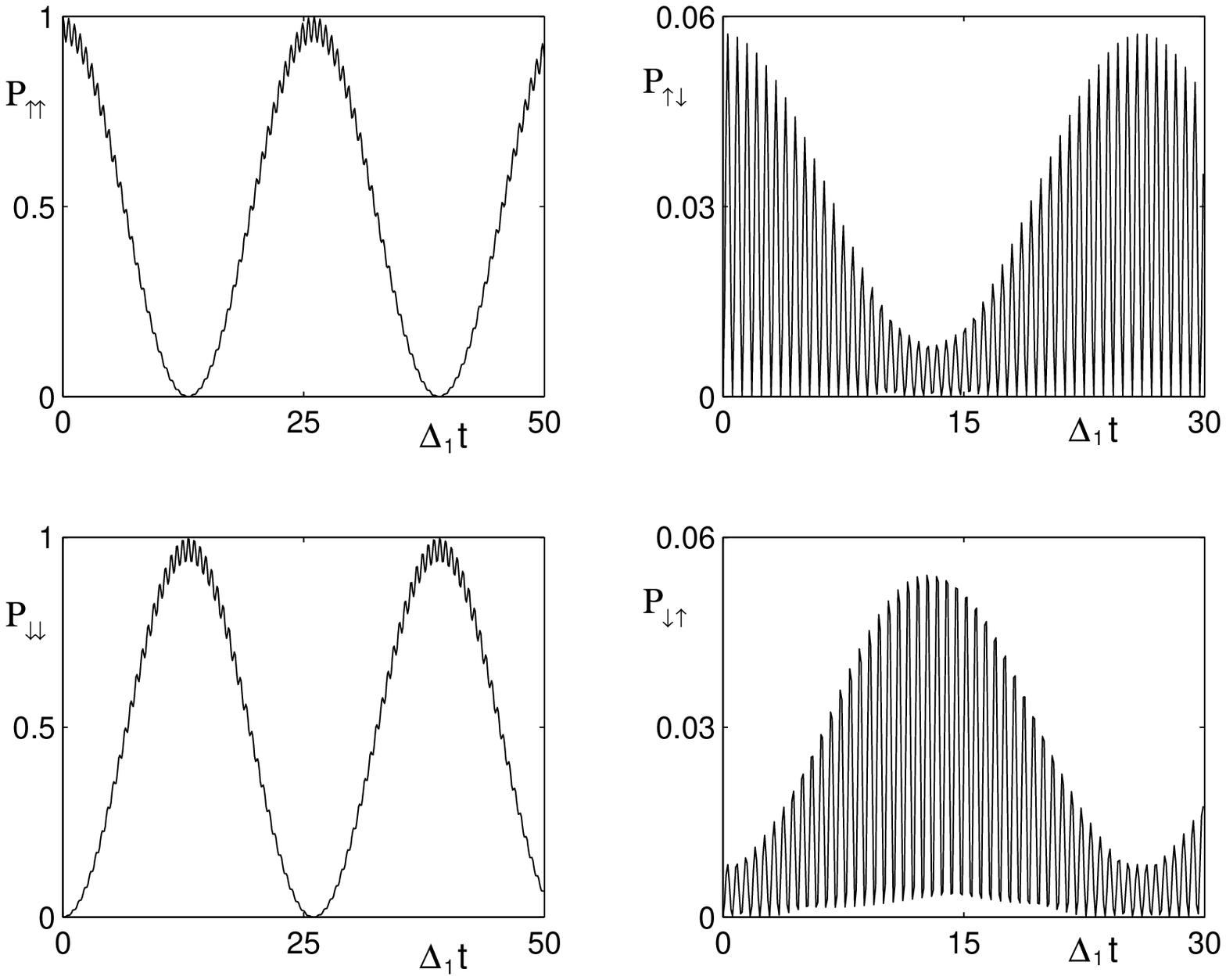}
\caption{}
\label{probjl}
\end{figure}

\begin{figure}[t]
\epsfbox[50 200 650 730]{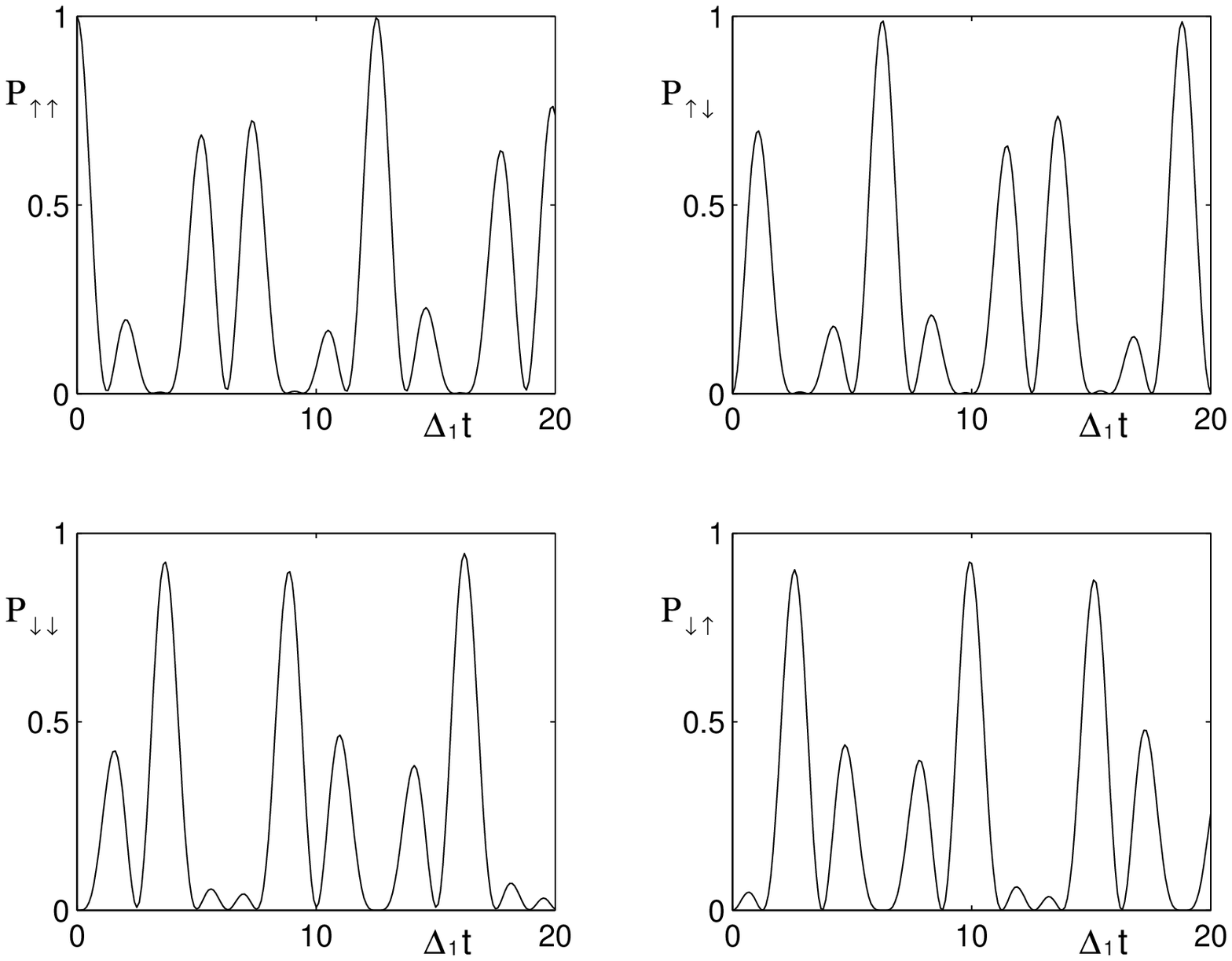}
\caption{}
\label{probjs}
\end{figure}

Consider now the probability $P^{0}_{\tau_{1} 
\tau_{2}}(t)$, for a system to start in $| \uparrow \uparrow \rangle$ at 
$t=0$, and finish at time $t$ in state $| \tau_{1} \tau_{2} \rangle$. 
In Figs.(\ref{probjl}) and
(\ref{probjs}) we plot these probabilities
for both strong-coupling ( $|J_{0}| \gg \Delta_{1} ,\Delta_{2}$ ) and 
weak-coupling ($|J_{0}| \ll \Delta_{1}, \Delta_{2}$). Consider first the 
strong-coupling regime. For both FM and AFM coupling, if the initial state 
is $| \uparrow \uparrow \rangle$, we see that the system
essentially oscillates between $| \uparrow \uparrow \rangle$ 
and $ | \downarrow \downarrow \rangle$ states with an effective frequency
$\bar{\Delta}=\Delta_{1} \Delta_{2}/(2 |J_{0}|)$. The oscillations between 
$| \up \down \rangle$ and $| \down \up \rangle$ are very fast, 
frequency $\sim J_{0}$, but their amplitude is reduced by a factor 
$\Delta/J_{0}$.
Notice the AFM system cannot relax to some 
combination of the lower-energy states $| \uparrow \downarrow \rangle $ and
$| \downarrow \uparrow \rangle$ (there is no environment), a problem which 
does not arise (for this initial state) in the FM case, where  
the initial state $| \uparrow \uparrow \rangle$ is 
already a superposition of the 2 low-lying eigenstates $|\Psi_{-+}
 \rangle$ and $| \Psi_{--}\rangle$ in (\ref{2cspineigenf}). 
This situation is precisely analogous to the strongly biased 2-well system,
where an initial state $| \uparrow \rangle$ in the biased system is stuck at 
high energy. 
In the same way, there is  
little coherence between the ($| \uparrow \uparrow \rangle$, $| \downarrow
 \downarrow \rangle$) manifold and the ($| \uparrow \downarrow \rangle$,
$| \downarrow \uparrow \rangle$) manifold when $|J_{0}|$ is large, i.e., 
$J_{0}$ suppresses both coherence between singlet and triplet states, 
and the off-diagonal elements in $\rho(t)$.

In the opposite limit of weak coupling the density matrix $\rho(t)$ 
differs little from $\rho^{(0)}(t)$, and there is mixing between
all four states. 
{}From Eq. (\ref{2cspinenerg}) we see that for small $|J_0|$,  
the displacement of the
energy levels $ \sim O(J_{0}^{2})$ unless $\Delta_{1} = \Delta_{2}$, in which case
the 2 central levels separate linearly with $J_{0}$. 
In this limit
\begin{equation}
\Psi_{++} = \Psi_{++}^{0} + \frac{1}{2} \frac{J_{0}}{\Delta_{1} + \Delta_{2}} \left[
| \uparrow \uparrow \rangle - | \uparrow \downarrow \rangle - | \downarrow 
\uparrow \rangle + | \downarrow \downarrow \rangle \right)
\end{equation}
\begin{equation}
\Psi_{+-} = \Psi_{+-}^{0} + \frac{1}{2} \frac{J_{0}}{\Delta_{1} - \Delta_{2}} \left[
| \uparrow \uparrow \rangle - | \uparrow \downarrow \rangle - | \downarrow 
\uparrow \rangle + | \downarrow \downarrow \rangle \right]  
\end{equation}
\begin{equation}
\Psi_{-+} = \Psi_{-+}^{0} - \frac{1}{2} \frac{J_{0}}{\Delta_{1} - \Delta_{2}} \left[
| \uparrow \uparrow \rangle - | \uparrow \downarrow \rangle - | \downarrow 
\uparrow \rangle + | \downarrow \downarrow \rangle \right]  
\end{equation}
\begin{equation}
\Psi_{--} = \Psi_{--}^{0} - \frac{1}{2} \frac{J_{0}}{\Delta_{1} + \Delta_{2}} 
\left[
| \uparrow \uparrow \rangle - | \uparrow \downarrow \rangle - | \downarrow 
\uparrow \rangle + | \downarrow \downarrow \rangle \right)  
\end{equation}
 
This concludes our analysis of this coupled 2-spin system.

\section{Density Matrix for the PISCES Model}
\resetcounters

In this section we set up a formal expression for the 2-spin density matrix
for the PISCES system. This is done by first integrating out the oscillator
bath modes to produce an expression in terms of an influence functional; 
then we show how the resulting expression can be summed over instanton paths.
Some of the derivation is an obvious generalization of the work from
Leggett et al. \cite{leggettB}, whereas other parts are somewhat less 
trivial. To help readers who are familiar with the spin-boson model, we have
adopted a notation which coincides fairly closely with that of Leggett et al.
\cite{leggettB}, with obvious generalizations.

\subsection{The Influence Functional}

In order to calculate the reduced density matrix for the system of interest,
we will use the well known technique of integrating out the environmental modes
in a path integral formalism. Thus, for some general Lagrangian of the form
(\ref{tlagrangian}), with an associated action
\begin{equation}
S[Q,\{ {\bf x_{k}} \} ] = S_{0}[Q] + S_{env}[ \{ {\bf x_{k}} \} ] 
+ S_{int}[Q, \{ {\bf x_{k}} \} ] \, ,
\end{equation}
(here we denote the environmental coordinates by $ \{ x_{k} \}$), the
reduced density matrix $\rho$ propagates according to 
\begin{equation}
\rho(Q_{f},Q_{f}';t) = \int dQ_{i}  \, dQ_{i}' \, 
J_{\rho}(Q_{f},Q_{f}',t;Q_{i},Q_{i}',0) \, \rho(Q_{i},Q_{i}';0)
\label{densityt}
\end{equation}
We write the propagator $J$ as a weighted double path integral: 
\begin{equation}
J_{\rho}(Q_{f},Q_{f}',t;Q_{i},Q_{i}',0) = 
\int_{q(0)=Q_{i}}^{q(t)=Q_{f}} D[q] \int_{q'(0)=Q_{i}'}^{q'(t)=Q_{f}'}
 D[q'] \, e^{i(S_0[q]-S_0[q'])} F[q,q'] 
\end{equation}
a double integral over the paths $q(t)$ and $q'(t)$, starting at $t=0$
from the positions $Q_{i}$ and $Q_{i}'$ and ending at time $t$ in
$Q_{f}$ and
$Q_{f}'$ respectively.
$F[q,q']$ is the famous ``influence functional'',
whose general properties are discussed by Feynman
and Vernon \cite{vernon}. It incorporates the interaction between the
paths $q$ and $q'$ with an environment.
In the case of a system coupled
independently to a set of independent external ``environmental
coordinates'',
we have
\begin{equation}
F[q,q'] = \prod_{{\bf k}} F_{{\bf k}}[q,q']
\end{equation}
with $F_{{\bf k}}$ the functional of the ${\bf k}^{th}$ external system
(which
in this case is just the ${\bf k}^{th}$ oscillator).

Here we are interested in the representation of the
PISCES problem either via a 4 well representation, or as eigenstates of
$\tau^{z}_{(1)} \tau^{z}_{(2)}$ (which in the truncated approximation
are equivalent).
The path $q$ of the combined 2-spin system is written as a double path
$q = (q_{1},q_{2})$, one for each spin (and likewise for the path $q'$ as
well as for $Q=(Q_{1},Q_{2})$).
The general form of the kernel $J_{\rho}$ in equation
(\ref{densityt}) is then
\begin{equation}
J_{\rho} = \int D[q_{1}] \int D[q_{2}] \int D[q_{1}'] \int D[q_{2}']
\, A_{1}[q_{1}] \, A_{2}[q_{2}] \, A_{1}^{*}[q_{1}] \, A_{2}^{*}[q_{2}]
        F[q_{1},q_{1}',q_{2},q_{2}']
\label{suppropag}
\end{equation}
where the boundary conditions are defined by
\begin{equation}
\int D[q_{\alpha}] \equiv
\int_{q_{\alpha}(0)=Q_{\alpha,i}}^{q_{\alpha}(t)=Q_{\alpha,f}}
D[q_{\alpha}]
\end{equation}
with $\alpha=1,2$.
The factors
$A_{1}[q_{1}] = i\Delta_{1}$/2, $A_{2}[q_{2}] = i\Delta_{2}/2$ are the
transition amplitude associated with a single flip (instanton)
of the relevant spin (1 or 2). In an instanton calculation
$\Delta \sim \Omega_{0} e^{-S_{cl}}$, where $S_{cl}$ is the action of
the path that minimises Lagrange's equations in imaginary time, and
$\Omega_{0}$ is the preexponential factor, of order the small oscillation
frequency of the system.
$F[q,q'] = F[q_{1},q_{2},q_{1}',q_{2}']$ is again the influence functional,
produced
by integrating out the oscillator environment.

To determine the form of $F[q,q']$ we need the correlation 
functions of the boson bath. To illustrate the general case let us first 
consider a particular one, in which the oscillator modes 
$\{ {\bf x_{k}} \}$
can be classified by momentum quantum number {\bf k}, and the two spins 
are considered to be at 
2 positions ${\bf R}_{1}$ and ${\bf R}_{2}$. Then defining 
${\bf x}({\bf r},t) = \sum_{{\bf k}} e^{-i {\bf k} \cdot {\bf r}} 
{\bf x}_{{\bf k}}(t)$ and
${\bf R} = {\bf R}_{1} - {\bf R}_{2}$, we require the correlation function
\begin{equation}
J_{\alpha \beta}({\bf R},\tau ,T ) = \langle x({\bf R}_{\alpha},\tau ) 
x({\bf R}_{\beta},\tau ) \rangle_{T}
\end{equation}
which is conveniently written in the matrix form
\begin{equation}
J_{\alpha \beta}({\bf R},\omega ) = \left( \begin{array}{cc}
J_{1}(\omega ) & J_{12}({\bf R},\omega ) \\
J_{12}^{*}({\bf R},\omega ) &  J_{2}(\omega ) \end{array} \right) 
\end{equation}
with $J_{\alpha \beta}({\bf R},\omega ) = \int d\tau e^{-i \omega \tau} 
J_{\alpha \beta}({\bf R}, 
\tau )$; as before, the indices $\alpha, \beta = 1,2$, and label the two
different spins. The diagonal terms $J_{\alpha}=J_{\alpha \alpha}$ are given 
directly from $H_{PISCES}$ (Eq. (\ref{hpisces}) and (\ref{hintpisces})) as
\begin{equation}
J_{\alpha}(\omega )=\frac{\pi}{2}
\sum_{{\bf k}}\frac{|c_{{\bf k}}^{(\alpha)}|^{2}}
{m_{{\bf k}}\omega_{{\bf k}} } \delta ( \omega - \omega_{{\bf k}} )
\end{equation}
whereas the inter-spin spectral function is given by
\begin{equation}
J_{12}({\bf R},\omega) = \frac{\pi}{2} 
\sum_{{\bf k}} \frac{c_{{\bf k}}^{(1)}
                 c_{{\bf k}}^{(2)*} }{m_{{\bf k}} \omega_{{\bf k}} }
                 D({\bf R},\omega_{{\bf k}})
                 \delta ( \omega - \omega_{{\bf k}} ) 
\end{equation}
where $D({\bf R},\omega_{{\bf k}}) = \sum_{{\bf k}} 
e^{i {\bf k} \cdot {\bf R}} D_{{\bf k}}
(\omega)$ is just the propagator for the bath modes (eg., in a phonon bath
it is the usual phonon propagator).
The diagonal spectral functions $J_{\alpha}(\omega)$ are of course
equivalent to those used by Caldeira and Leggett.
$J_{12}$ is similar, but note that (a) in general the function 
$J_{12}({\bf R},\omega)$ is
not a separable function of ${\bf R}$ and $\omega$, and (b) it is complex 
(it is a retarded correlation function).

The retardation effects begin to be important once $\omega_{{\bf k}} \geq
v_{{\bf k}}/|{\bf R}|$, where $v_{{\bf k}}$ is the propagation velocity of 
the $k^{th}$ bath mode. For distances $|{\bf R}| \ll v_{{\bf k}}/
\omega_{{\bf k}}$, we can ignore retardation effects, and 
$J_{12}({\bf R},\omega)$ will be separable. In this paper we will always work
in this limit. Typically we will be interested in interactions mediated by
either phonons or electrons. In the case of an electronic bath, one has 
typically the Ohmic form
\begin{equation}
J_{\alpha \beta }({\bf R},\omega ) = \omega \eta_{\alpha \beta} 
e^{- \omega/\omega_{c}}
\end{equation}
\begin{equation}
\eta_{12}({\bf R}) \sim (\eta_{1} \eta_{2})^{1/2} {\cal V}_{e}({\bf R})
\end{equation}
where $\eta_{1}$ and $\eta_{2}$ are local friction coefficients of the kind
discussed by Caldeira and Leggett \cite{leggettA}, and ${\cal V}_{e}({\bf R})$ 
has the form ${\cal V}_{e}({\bf R}) \sim \sin^{2}(k_{f} R)/
(k_{f}R)^{2}$, in 3 dimensions. The upper cut-off $\omega_{c}$ in an electron
bath must satisfy $\omega_{c} \ll \Omega_{0}^{(1)} , \Omega_{0}^{(2)}$, as
discussed in section II. Typical electronic velocities are $\sim$ the Fermi
velocity $v_{f} \sim 10^{6} m \, s^{-1}$; this means that if, say,
$\Omega_{0} \sim 1 K$, then retardation effects do not become important until
$R \sim 50 \, \mu m$.

For a phonon bath, one usually has the superOhmic form
\begin{equation}
J_{\alpha \beta}({\bf R},\omega ) = \bar{g}_{\alpha \beta}({\bf R}) 
\Theta_{D} \left( 
\frac{\omega}{\Theta_{D}} \right)^{m} e^{- \omega/\omega_{c}} 
\end{equation} 
\begin{equation} 
\bar{g}_{12}({\bf R})\sim (\bar{g}_{1} \bar{g}_{2})^{1/2} {\cal V}_{\phi}
({\bf R}) 
\end{equation} 
where $m \geq 3$ in 3 dimensions, and ${\cal V}_{\phi}(R) \sim (a_{0}/R)^{3}$, 
with $a_{0}$ a lattice constant, and $\bar{g}_{1}$ and $\bar{g}_{2}$ again being
local coupling constants. For a phonon velocity $ \sim 5 \times 10^{3} 
m \, s^{-1}$, and $\Omega_{0} \sim 1 K$ again, we now find that 
retardation effects
become important for distances $R \sim \, 250 nm$ or greater.
For distances $|{\bf R}|$ much less than these limits, these forms can be
used. Again, a more precise specification 
of $J_{\alpha \beta}({\bf R},\omega )$ can
be given once we fix our physical problem. 
 
Returning now to the influence functional, we introduce the time 
correlation functions
\begin{equation}
\Phi_{\alpha} (\tau -s) = \int^{\infty}_{0} d \omega \, 
J_{\alpha}(\omega ) \,
             \sin\omega (\tau -s)
\vspace{2mm}
\end{equation}
\begin{equation}
\Gamma_{\alpha} (\tau -s) = \int^{\infty}_{0} d \omega \, 
J_{\alpha}(\omega ) \,
             \cos\omega (\tau -s) \coth(\omega /2T)
\vspace{2mm}
\end{equation}
for the diagonal elements $(\alpha=1,2)$, and 
\begin{equation}
\Phi_{12} ({\bf R},\tau -s) = \int^{\infty}_{0} d \omega \, J_{12}({\bf R},
\omega ) \, \sin\omega (\tau -s)
\vspace{2mm}
\end{equation}
\begin{equation}
\Gamma_{12} ({\bf R},\tau -s) = \int^{\infty}_{0} d \omega \, J_{12}({\bf R}, 
\omega ) \, \cos\omega (\tau -s) \coth(\omega /2T)
\vspace{2mm}
\end{equation}
for the off-diagonal elements. The general form of the influence functional is
now
\begin{equation}
F[q_{1},q_{2},q_{1}',q_{2}'] = F_{1}[q_{1},q_{1}'] \; F_{2}[q_{2},q_{2}'] \;
                     F_{12}[q_{1},q_{2},q_{1}',q_{2}'] \, ,
\end{equation}
with $F_{1}$ and $F_{2}$ being the influence
functionals for systems 1 and 2 alone, and  $F_{12}$ the
part representing their interaction through the bath. 
Each path $q_{\alpha}(t)$, with $\alpha=1,2$, is
composed of a series of jumps between 
wells, which we assume to be located at $\pm q_{0\alpha}/2$. We can then
define the functions \cite{leggettB}:
\begin{equation}
\xi_{\alpha} (\tau ) = q_{0\alpha}^{-1}[q_{\alpha}(\tau )-q_{\alpha}'(\tau )]
\label{param1}
\end{equation}
\begin{equation}
\chi_{\alpha} (\tau ) = q_{0\alpha}^{-1}[q_{\alpha}(\tau )+q_{\alpha}'(\tau )]
\label{param2}
\end{equation}
so that the single spin functionals are given by
\begin{equation}
F_{\alpha}[\tau,\tau'] = 
\exp \left( \frac{q_{0\alpha}^{2}}{\pi} 
\int_{0}^{t} dt \int_{0}^{\tau}ds 
( i\Phi_{\alpha}(\tau -s)  \xi_{\alpha}(\tau )\chi_{\alpha}(s) 
- \Gamma_{\alpha}(\tau -s)   \xi_{\alpha}(\tau )\xi_{\alpha}(s)  )
\right)
\label{complete}
\end{equation}
whereas the interaction functional 
$F_{12}$ is
\begin{eqnarray}
F_{12} &=& \exp \left( \frac{q_{01}q_{02}}{\pi} \int_{0}^{t} dt
\int_{0}^{\tau}ds 
( i \Phi_{12}({\bf R},\tau -s) [ \xi_{1}(\tau )
\chi_{2}(s) + \xi_{2}(\tau ) \chi_{1}(s)] \right. \nonumber \\
&-& \left. \frac{}{} \Gamma_{12}({\bf R},\tau -s) [\xi_{1}(\tau ) \xi_{2}(s) 
+ \xi_{2}(\tau) \xi_{1}(s) ] ) \right)
\label{completeint}
\end{eqnarray}

\begin{figure}
\epsfxsize=6.5in
\epsfbox{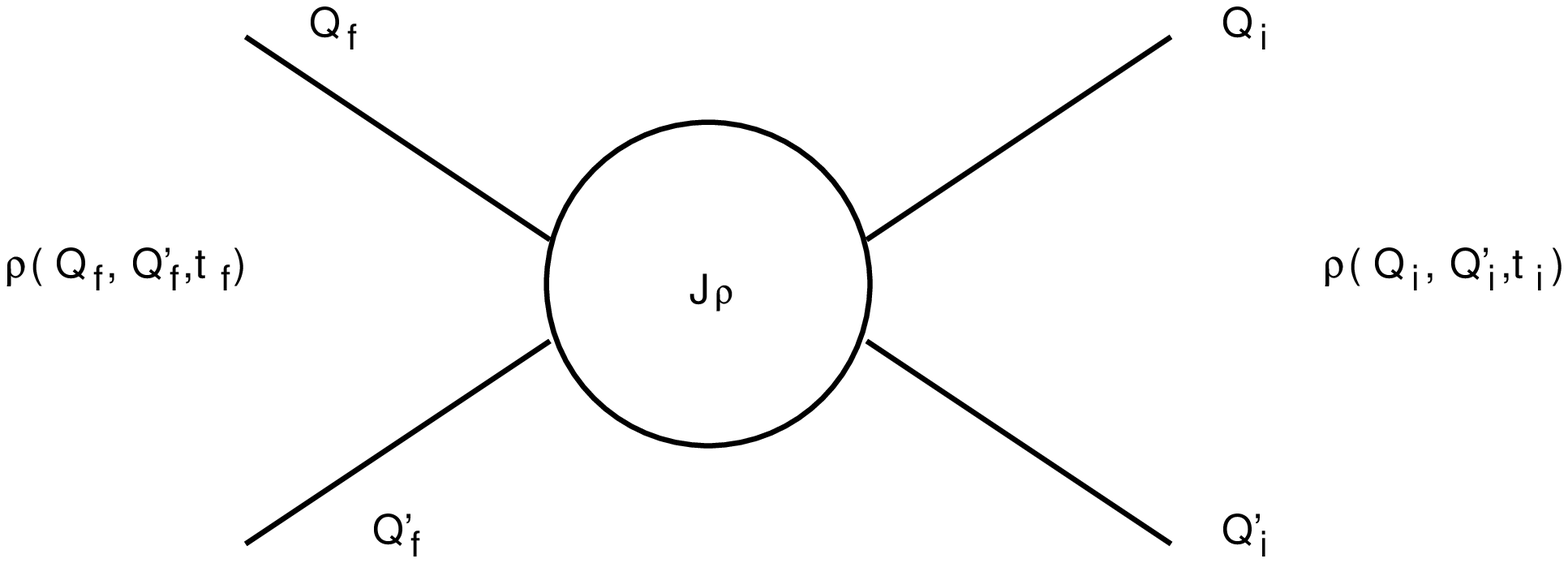}
\caption{}
\label{sysenvint}
\end{figure}

We see that in both the single spin functional $F_{\alpha}$ and in the 
interaction functional $F_{12}$, there is a purely reactive ``phase'' 
correction $\Phi$, 
and a purely dissipative damping term $\Gamma$. The damping 
operates when paths
depart from each other, ie., when both
$\xi_{1}(\tau)$ and $\xi_{2}(s)$ (or vice-versa) are non-zero.
We may interpret these equations diagramatically 
(Fig. (\ref{sysenvint})). 
For both the
spin-boson and PISCES problems Eq. (\ref{complete}) and
(\ref{completeint}) 
massively simplify,
because $\xi_{\alpha}((\tau)$ and $\chi_{\alpha}(\tau)$ may only take one of
the 3 values $0, \pm 1$; each time the spin $\alpha$ tunnels, both
$\xi_{\alpha}$ and $\chi_{\alpha}$ change by $\pm 1$. Notice that these
equations give $F_{\alpha \beta}$ solely in terms of the sum and difference
variables between pairs of paths; the paths themselves have now disappeared 
from the formalism.

\subsection{Path Integral for the Density Matrix}

The formal calculation of the density matrix elements (and hence all physical
properties of the system) can be carried out starting from
(\ref{densityt}), (\ref{suppropag}), (\ref{complete}) and (\ref{completeint}).
In order to do this, we must integrate over 4 separate paths
($q_{1}(t),q_{2}(t),q_{1}'(t),q_{2}'(t)$, or, equivalently 
$ \xi_{1}(t),\xi_{2}(t),\chi_{1}(t),\chi_{2}(t)$ ).

One way of handling the path integral for $J$ would be to write it in terms of
a single path over the 16 possible states of $q_{1},q_{2},q_{1}'$ and 
$q_{2}'$. Here we shall use a procedure analogous to that employed by 
Leggett et al. \cite{leggettB} for the single spin-boson problem. Taking
advantage of the fact that the $\chi_{\alpha}$ and $\xi_{\alpha}$ are not
independent (when $\chi_{\alpha}=0$, $\xi_{\alpha} = \pm 1$ and vice-versa),
we write $F_{\alpha \beta}$ as an integral over 2 paths, one for each spin, 
in which we distinguish only between diagonal ``sojourn'' states
($\xi_{\alpha} =0$, but $\chi_{\alpha} = \pm 1$) and off-diagonal 
``blip'' states ($\xi_{\alpha} = \pm 1$ whilst $\chi_{\alpha} = 0$). 
This has the diagrammatic representation shown in Fig. (\ref{typpath}).
 Each such path must contain an even number of transitions, with action
$-i \Delta_{\alpha}/2$ for each transition. The transitions occur at times
$t_{j} \, (j=1,2,...2n_{1})$ for spin 1, and times
$u_{k} \, (k=1,2,...2n_{2})$ for spin 2. We may then follow the standard
manoeuvre of introducing a set of ``charges'' which label the states (blip or
sojourn) for each path; calling these charges 
$\eta_{1j},\zeta_{1j}$ (for spin 1) and 
$\eta_{2k},\zeta_{2k}$ (for spin 2), we allow them to have values $\pm 1$
according to
\begin{equation}
\chi_{1}(\tau) = \sum_{j=0}^{n_{1}} \, \eta_{1j} \, [ \,
\theta( \tau - t_{2j} ) - \theta( \tau - t_{2j+1} ) \, ]
\label{charges1}
\end{equation}
\begin{equation}
\xi_{1}(\tau) = \sum_{j=0}^{n_{1}} \, \zeta_{1j} \,
[ \, \theta( \tau - t_{2j-1} ) - \theta( \tau - t_{2j} ) \, ]
\end{equation}
\begin{equation}
\chi_{2}(\tau) = \sum_{k=0}^{n_{2}} \, \eta_{2k} \,
[ \,  \theta( \tau - u_{2k} ) - \theta( \tau - u_{2k+1} ) \, ]
\end{equation}
\begin{equation}
\xi_{2}(\tau) = \sum_{k=0}^{n_{2}} \, \zeta_{2k} \,
[ \,  \theta( \tau - u_{2k-1} ) - \theta( \tau - u_{2k} ) \, ] \, .
\label{charges4}
\end{equation}
 
The sum over the 2 paths now becomes a sum over
all possible arrangements of the charges $\eta_{\alpha j}$ 
and $\zeta_{\alpha k}$, these charges
living on a set of discrete times between $0$ and $t$ (with, however,
certain restrictions on the paths, and on the charges, depending on which
density matrix element is being computed). 
 
The explicit expansion of the density matrix $\rho$ in terms of sums over the
various charge and blip ordering configurations is rather messy; to make 
the paper easier to follow, the detailed path integral
calculations are confined to a series of appendices. In Appendix A the 
detailed structure of the complete influence functional is given, along with
explicit expressions for the blip-blip and blip-sojourn interactions, both for
bath-mediated interactions between different states of the same spin, and also
for interspin interactions. The results can be summarised as follows. The
single spin functionals $F_{\alpha}$ are those derived by Leggett et al.
\cite{leggettB}; 
they can often be treated in the ``dilute blip approximation'', in
which only the ``blip self-energy'' terms (ie., between 2 times in the same 
blip), and interaction between a blip and the sojourn immediately preceding it
are included. This works, provided the interactions with the bath are 
sufficiently strong that blips are rare; essentially it assumes that the
environmental decoherence is strong enough to severely suppress the 
``off-diagonal'' paths in the density matrix, in which the damping term
$\Gamma_{\alpha}(\tau-s)$ in Eq. (4.23) comes in. Consequently, during the 
waiting time between two such off-diagonal excursions or ``blips'', the 
system completely ``forgets'' the effects of the previous blip. 

The dilute blip approximation is much more problematic when an interaction 
term in the Hamiltonian is trying to force the density matrix 
into off-diagonal states.
This
point is rather crucial when we come to the interaction functional
$F_{12}$, since the main effect of the bath turns out to be the introduction
of an extra interaction 
$\bar{\epsilon}({\bf R}) \hat{\tau}_{1}^{z} \hat{\tau}_{2}^{z}$ 
between the 2 spins, with $\bar{\epsilon}({\bf R})$ given by
\begin{equation}
\bar{\epsilon}({\bf R}) \equiv \frac{q_{01}q_{02}}{\pi } \int_{0}^{\infty}
d\omega \, \frac{J_{12}({\bf R}, \omega )}{\omega } \, .
\label{bias}
\end{equation}
For Ohmic dissipation, the explicit form of $\bar{\epsilon}({\bf R})$ is
\begin{equation}
\bar{\epsilon}({\bf R}) =  \alpha_{12}({\bf R})\omega_{c}
\end{equation}
where $\alpha_{12}({\bf R}) = \eta_{12}({\bf R}) q_{01} q_{02} / 2 \pi \hbar$. 
For superohmic dissipation, we get
\begin{equation}
\bar{\epsilon}({\bf R}) = 2\Gamma(s) \left( \frac{\tilde{\omega}}{\omega_{c}}
\right)^{s} \bar{g}_{12}({\bf R}) \omega_{c}
\end{equation}
In the subohmic case, $s<1$ so that the integral diverges and the model becomes
ill-defined.
It is important to notice that the magnitude of this bias depends on the
upper cut-off frequency of the bath - this is simply because the total
strength of the interaction involves the high-energy (adiabatic) modes.
Now, this {\it bath-induced} interaction simply adds to the 
original high frequency ``direct'' interaction
to give the total adiabatic interaction between the two spins as
\begin{equation}
{\cal J}({\bf R}) \equiv K_{zz}({\bf R}) + \bar{\epsilon}({\bf R})
\end{equation}
The big problem for the theory is now that ${\cal J}({\bf R})$ plays the role
of a ``mutual bias'' between the 2 spins, just as $J_{0}$ did in our toy model
of section III; it turns out that most of its effects come from interactions
between blips of one system and sojourns of the other. Unless both systems are
overdamped (or both locked together, so that blips on one cannot overlap with
sojourns of the other), then it is clear that we cannot introduce any kind of
dilute blip approximation on a single system, independently of the
other, to handle the interspin term $F_{12}$. 
Nevertheless we will show, in the next section (and in Appendices C and D) 
how it is possible to evaluate $F_{12}$ by generalising this approximation.
We now turn to the detailed behaviour of the density matrix.

\section{Dynamics of the PISCES System}
\resetcounters

In this section, we obtain the dynamics of the 2-spin reduced density matrix.
We explicitly calculate the 4 diagonal matrix elements
$P_{\tau_{1} \tau_{2}}(t) \equiv \rho(\tau_{1} \tau_{2},t;\up \up, 0)$, 
which are
the probabilities for the system to end up in a state $| \tau_{1} \tau_{2}
\rangle$ after a time $t$, having started in a state $|\up \up \rangle$.
We will concentrate in what follows on the most interesting Ohmic case, where 
the 2 spins are coupled Ohmically to the sea of oscillators. Since it is
conventional to express the strength of this Ohmic coupling by a dimensionless
constant $\alpha$, we will henceforth use the index $\beta=1,2$ to label the
2-spins (we do not use $\beta =1/kT$ in what follows). We will use temperature
and frequency
units in which $k_{B}=\hbar=1$, so that the thermal energy $T$ and the
tunneling splitting energies are $\Delta_{\beta}$.

For this Ohmic PISCES problem, we will concentrate on those parameter regimes
in which analytic results can be obtained for $P_{\tau_{1} \tau_{2}}(t)$. 
Having achieved these, it is possible to build up a fairly complete picture of
the behaviour of $P_{\tau_{1} \tau_{2}}(t)$., in all temperature and coupling
range of interest. 
We can identify 4 regimes in the dynamics, as shown
in the ``phase diagram '' of Fig. (\ref{pphase}).

These 4 regimes are, in the Ohmic case, defined as follows:

(i) The ``Locked Phase'' (${\cal J} \gg T, \Delta^{*}_{\beta}/\alpha_{\beta}$):
In this regime, the effective coupling ${\cal J}$ is so strong that the 2 
spins lock together, in either the states $| \up \up \rangle$ or
$| \down \up \rangle$ depending on the sign of ${\cal J}$. This state is very
similar to the strongly-coupled state described in section 3; we shall see
that the combined ``locked spin'' oscillates between $|\up \up \rangle$ and
$| \down \down \rangle$ (for FM coupling), or between $|\up \down \rangle$ and
$| \down \up \rangle$ (for AFM coupling), at a renormalised frequency
$\Delta_{c} = \Delta_{1} \Delta_{2} / | {\cal J}|$. However these oscillations 
are damped - in fact the locked spin now behaves like a single spin-boson
system, with a new coupling $\alpha_{c}=\alpha_{1}+\alpha_{2}
\pm 2 \alpha_{12}$ to
the oscillator bath, the $+$ ($-$) corresponding to ferromagnetic 
(antiferromagnetic) coupling between the spins. 

(ii) The ``Mutual Coherence'' phase ($\Delta_{\beta}/\alpha_{\beta} \gg T \gg
{\cal J}$; ${\cal J} > \Delta_{\beta}$): 
Here, the thermal energy overcomes the mutual coupling; 
nevertheless if the dissipative couplings $\alpha_{\beta}$'s are sufficiently
small ($\alpha_{\beta} \ll 1$), it is possible for the energy scale
$\Delta_{\beta}^{*}/\alpha_{\beta}$ to dominate even if 
$\Delta_{\beta} < {\cal J}$. In this case, even though we are dealing with a 
strong coupling, and the bath dissipation is still important, some
coherence in the motion of each spin is maintained - moreover, the small
${\cal J}$ causes ``mutual coherence'' between the two spins, ie., their
damped oscillations are correlated to some extent. 

(iii) The ``Correlated Relaxation'' or High-T phase ($T \gg \Delta_{\beta}/
\alpha_{\beta}, {\cal J}$). In this regime, the bath causes each spin to relax
incoherently; however, the relaxation of the two spins is still correlated 
(indeed each spin relaxes in the time-dependent bias 
generated by the other). 
 
(iv) Finally, and much less interesting, the ``perturbative regime'' 
(${\cal J} \ll \Delta_{\beta}^{*}$), in which the total coupling is so 
weak that the 2 spins relax almost independently; all correlations can
be handled perturbatively, and only weakly affect the behaviour that one
calculates from the standard spin-boson model.

We notice that the locked and correlated relaxation phases always exist; but
the mutual coherence phase is much more delicate, and only appears if the
coupling of each spin to the bath is weak (ie., $\alpha_{\beta} \ll 1$). 
To get a better intuitive feeling for the energy scales and the different 
regimes, it is useful to imagine how the behaviour of the PISCES system 
changes as one moves along various paths in the phase diagram. Imagine first
that we start off in the lower right half of Fig. (\ref{pphase}), at a very 
low $T$ (much less than any other energy). Then temperature fluctuations are 
irrelevent, even if each spin is strongly coupled to the bath; the effect 
of the induced interaction ${\cal J}$ is simply to lock the 2 spins into
one. This ``spin complex'' also couples to the bath, but now with a 
quite different
coupling; and it can also tunnel between 2 orientations,
but at a much lower frequency. Typically its motion at finite $T$ will be
completely overdamped, and it will localize as $T$ tends to zero. However if we now raise $T$, the thermal fluctuations increase in importance, and 
eventually compete with ${\cal J}$; the spins start to unlock from each 
other. This crossover regime cannot be handled analytically, but further 
increase in $T$ causes complete unlocking- we enter the correlated 
relaxation phase, where the thermal fluctuations are strong enough to 
destroy any coherent dynamics of the system, but the relaxtion of each is
still correlated with that of the other. If $T \gg {\cal J}$, even these 
correlations become irrelevant- the 2 spins relax independently.

One may make a similar traversal, this time from correlated to locked phase,
this time by starting at high $T$ but weak coupling (top left in the Figure),
and then increasing the coupling. Much more interesting is to go to low $T$
and then increase the coupling. Two situations are then possible. One is 
where the coupling $\alpha$ to the bath is not small. Then there is no
mutual coherence regime, and we find ourselves initially in the lower right
quadrant of the Figure. Even with large $\alpha$, one may still envisage a 
very small ${\cal J}$ (eg., if the 2 spins are very far apart). Then each 
behaves independently (with weak perurbative interactive corrections), as 
a spin-boson system. Increasing the coupling (ie., increasing their distance
apart) increases the correlations, until they eventually lock; the crossover
cannot be handled analytically.

On the other hand if $\alpha \ll 1$, so each spin is weakly coupled to the 
bath, one can get a situation in which their mutual coupling, although 
weak, still exceeds the renormalised tunneling splitting $\Delta^{*}$. In 
this case the $T$-fluctuations are not strong enough to upset their
mutual correlations, or to destroy coherent behaviour on the part of each 
spin, even if $T > {\cal J}$, provided $\alpha T < \Delta^{*}$. The spin is 
simply so weakly coupled to the bath that it hardly sees the $T$-
fluctuations, but ${\cal J}$ causes the oscillations of the 2 spins to 
correlate- we get ``damped beats''. Further increase in $T$ decoheres these 
completely, and further increase in ${\cal J}$ locks the 2 spins together.
This mutual coherence phase is clearly going to be favoured if we have 
2 spins close to each other (to make ${\cal J}$ big enough), yet very weakly
coupled to the bath.

The difficulty of setting up a PISCES model which allows a mutual coherence
phase is connected with the following important general feature of the 
results. In the problem of a {\it single} spin coupled to an oscillator bath,
it is known that one may always get some coherent behaviour provided $T$ is
low enough, and also $\alpha < 1/2$. However in the present case of 2 spins,
coherent behaviour is much harder to find; even if $\alpha$ is very small, 
coherence is destroyed if we ``switch on'' the coupling between the 2 spins, 
once ${\cal J}$ reaches the very small energy scale $\Delta^{*}/\alpha$.
This is because the ``dynamic bias'' on one spin, coming from the other, 
removes the ``resonance condition'' between initial and final states which is 
necessary for coherent oscillations. It shows, in essence, that 
mutual interactions {\it increase decoherence effects}. This enhanced 
decoherence is particularly interesting in the context of the measurement
operation, or indeed for any attempt to look for macroscopic quantum 
coherence.

A more formal understanding of these energy scales comes out of a 
renormalisation group treatment, to which we proceed first. After this the 
analytic results for each regime are presented.

\subsection{Energy Scales From the Renormalisation Group}

Although a renormalisation group analysis cannot give us the dynamics
of the PISCES model, it gives considerable insight into the 
energy scales mentioned above. We shall use the 
``poor-man'' renormalisation procedure first applied to the Kondo problem by
Anderson et al. \cite{anderson}, and then later to the single spin-boson
problem \cite{bmoore}. Cardy \cite{cardy}  devised a method
to treat the multisite problem, which was then applied to the case of two
Anderson impurities \cite{chakra} and two tunneling impurities in
metal \cite{sols}. As shown in Appendix B, in the case where both spins
are identical (ie., where
$\Delta_{1}=\Delta_{2}=\Delta$, and $\alpha_{1}=\alpha_{2}=\alpha$) the 
partition function for this problem is identical to that considered by 
Chakravarty and Hirsch \cite{chakra}, and we carry over directly their 
scaling equations for a set of fugacities 
\begin{eqnarray}
\frac{dy}{d \ln \tau_{c}} &=& y(1-\alpha) + y(y_{F}+y_{AF}) \nonumber
\\
\frac{dy_{F}}{d \ln \tau_{c}} &=& y_{F}(1-\alpha/2-\alpha_{12}/2) +
y^{2}
\label{fugacities} \\
\frac{dy_{AF}}{d \ln \tau_{c}} &=& y_{AF}(1-\alpha/2+\alpha_{12}/2) +
y^{2}
\nonumber
\end{eqnarray}
where the dimensionless fugacity $y = \Delta/\omega_c$ and 
$y_{F}$ and $y_{AF}$ are the fugacities associated with
simultaneous transitions of the 2 spins.
Since $\alpha \gg \alpha_{12}$, the behaviour of the fugacities is
determined
solely by the single spin dissipation coefficients.
These coefficients are
also renormalised; their scaling is
\begin{eqnarray}
\frac{d \alpha}{d \ln \tau_{c}} &=& -\alpha
(2y^{2}+y_{F}^{2}+y_{AF}^{2})-
\alpha_{12} (y_{F}^{2}-y_{AF}^{2}) \nonumber \\
\frac{d \alpha_{12}}{d \ln \tau_{c}} &=& -\alpha_{12} (2y^{2}+y_{F}^{2}+
y_{AF}^{2})- \alpha (y_{F}^{2}-y_{AF}^{2})
\label{kays}
\end{eqnarray}
Finally, there is also a scaling of the coupling energy
\begin{equation}
\frac{d (\tau {\cal J})}{d \ln \tau_{c}} = (1-4y^{2})(\tau {\cal J})
\label{sbreaking}
\end{equation}

The renormalisation group analysis assumes that the dimensionless
quantities
$y$, $\tau_{c} T$ and $\tau_{c} {\cal J}$ are all much less than
one. The scaling has to be stopped if one of the parameters renormalises
to the strong coupling regime. The three important scaling parameters
are thus $1/\Delta^{*}$,
a renormalised tunneling matrix element, $1/T$ and $1/{\cal J}$.
Let us first set ${\cal J}=0$. The scaling
equations are then very similar to the case of the single spin-boson system.
At $T=0$, the scaling of $y$ is determined by $\alpha$. For $\alpha >1$, $y$
decreases and since $y_{F}$ and $y_{AF}$ grow as $y^{2}$, this decrease
cannot be compensated and the scaling leads to $y=0$, the well-known
localisation phenomenon \cite{leggettB}. 
The dynamics will then be determined by $y_{F}$ or
$y_{AF}$. Depending on the strongest of the two, only the ferromagnetic or
antiferromagnetic states will be occupied.
For $\alpha < 1$, then the scaling can stop at some
$\tau = 1/\Delta^{*}$, a renormalised tunneling matrix element
determined self-consistently as
\begin{equation}
\Delta^{*}_{\beta} = \Delta_{\beta} \left(
\frac{\Delta_{\beta}}{\Omega_{0}}
\right)^{\alpha/(1-\alpha)}
\end{equation}
In this case, the behaviour is determined by the single tunneling fugacity, the
two spins are still correlated, but behave nearly independently.
At finite temperature, the value of the renormalised tunneling matrix element
is determined by the ratio between the energy scales $1/\Delta^{*}$ and
$1/T$. If $T$ is the largest energy scale, the scaling must be terminated at
$\tau_{c} \sim 1/T$ and the resulting tunneling matrix element is
$ \Delta^{*}(T) = \Delta_{0} ( T/\Omega_{0} )^{\alpha/(1-\alpha)}$.

In the presence of an interaction, the symmetry between the levels is broken
and Eq. (\ref{sbreaking}) indicates that this difference
will {\em grow} with the scaling,
even if the fugacity grows as well (since $y \ll 1$ by assumption). The time
it takes for $\tau_{c} {\cal J}$ to grow to the strong coupling phase is roughly
$\tau_{{\cal J}} \sim 1/ {\cal J}$. If both $1/\Delta_{\beta}^{*}$ and $1/T$
are smaller than $1/{\cal J}$, the scaling stops before the strong coupling
phase is attained and the system behaves like two single spins, although
coupled by the interaction ${\cal J}$. However, if ${\cal J}$ is the largest
energy scale, the system rapidly gets into the strong coupling phase and the
RG equations break down. It is clear however that the new strongly coupled
phase is made of the two spins locked together by ${\cal J}$. The thermal
fluctuations (controlled by $T$) and the tendency to individual dynamics
(determined by $\Delta_{\beta}^{*}$) cannot compensate this strong coupling.

\subsection{Locked Phase}

Here ${\cal J}$ is so
strong that a prohibitively high energy cost is incurred 
if a blip on one path overlaps with a sojourn on the other.
In Appendix B this is discussed in more detail, and we demonstrate that
for a strongly FM-coupled system which starts in initial state 
$|\up \up \rangle$,
the resulting dynamics is is equivalent to that of 
a {\it single} spin- boson system, with a new tunneling
matrix element $\Delta_{c} = \Delta_{1} \Delta_{2} / | {\cal J} |$,
and with this ``spin complex'' now coupled
to the environment with a new Ohmic coupling 
$\alpha_{c}= \alpha_{1} + \alpha_{2} +2 \alpha_{12}$.
The new oscillation 
frequency is not surprising- we have already seen it in our toy model in the 
strong coupling regime. As for the new coupling to the bath, it is
simply equivalent to the coefficient $K_{F,F}$ coming from the
renormalisation group analysis (see Appendix B). 
Physically, in the case of FM coupling, the
system oscillates between $| \up \up \rangle$ and $| \down \down \rangle$ at
a frequency $\Delta_{c}$, if we ignore the environmental
dissipation. Mixing
with the high-energy states $| \up \down \rangle$ and $|\down \up \rangle$
is negligible (with ${\cal J} \gg T$, even thermally-excited transitions,
via absorption of bath quanta, will be exponentially rare), and so the 
system reduces to the single spin-boson problem.

A system coupled by a strong AFM interaction will have the
same effective tunneling matrix element $\Delta_c$, but will now be
coupled to the environment by a dissipation coefficient
$\alpha_c = \alpha_1 + \alpha_2 - 2 \alpha_{12}$. Once again, the new
coupling can be related to a the RG analysis, in this case, it is simply
the coefficient $K_{AF,AF}$ of Appendix B. 

Notice that the case of 
an AFM-coupled system which begins in the state
$|\up \up \rangle$ is quite different, since this is then a {\it high} 
energy state. Without the bath,
the system is frozen in this state (c.f.,
section 3). The coupling to the bath allows relaxation to the 
low-energy manifold (now composed of the states 
$ | \up \down \rangle$ and $| \down \up \rangle$). 
Thus this case needs to be treated slightly differently.
Intuitively one expects that the system will first try to relax to an
AFM state before simultaneous transitions take place (since
a transition from $| \up \up \rangle$ to $| \down \down \rangle$ still
takes the system to a high-energy state); this turns out to be correct. 

Let us begin with the FM-coupled case.
The equivalence to the unbiased spin-boson system means that the
dilute-blip approximation can be used, and we will simply take over the
results already derived by Leggett et al \cite{leggettB}. Thus, defining
the Laplace transform of $P_{\up \up}(t)$ as 
\begin{equation}
P_{\up \up}(\lambda)=\int_{0}^{\infty} dt e^{-\lambda t} P_{\up \up}(t)
\end{equation}
one finds in this locked phase
\begin{equation}
P_{\up \up}(\lambda ) = \frac{1}{2\lambda} +
\frac{1}{ \lambda +  f(\lambda )} \, .
\label{puufullambda}
\end{equation}
The function $f(\lambda)$ is defined in terms of an effective tunneling matrix
element $\Delta_{c}^{eff}$, viz., 
\begin{equation}
\Delta_{c}^{eff} \equiv [ \Gamma (1-2\alpha_{c} ) \cos (\pi \alpha_{c} )
]^{ \frac{1}{2(1- \alpha_{c} )} } \Delta_{c}^{*}
\end{equation}
\begin{equation}
\Delta_{c}^{*} \equiv \Delta_{c} \left( \Delta_{c} / \Omega_{o}
\right)^{ \frac{\alpha_{c} }{1- \alpha_{c} } }
\label{deltacr}
\end{equation}
\begin{equation} 
\Delta_{c} = \frac{\Delta_{1} \Delta_{2}}{{\cal J}} 
\end{equation}

The function $f(\lambda)$, in the limits appropriate 
to the truncation procedure
($T, \Delta_{c} \ll \Omega_{0}$, as well as $\omega_{c}t \gg 1$) 
is given by 
\begin{equation}
f(\lambda,T)= \Delta_{c}^{eff} \left( 
\frac{2\pi T}{\Delta^{eff}_{c}}
\right)^{2\alpha_{c}-1}
 \frac{\Gamma(\alpha_{c}+\lambda/2\pi T)}{\Gamma(1-\alpha_{c} +
\lambda/2\pi T)}
\label{fclambda}
\end{equation}
where $\Gamma(x)$ is the Gamma function and where, as noted above,
$\alpha_{c}=\alpha_{1}+\alpha_{2}+2 \alpha_{12}$.
The corresponding behaviour of
$P_{\up \up}(t)$ is well known:

(i) At $T=0$ and for $\alpha_{c} < 1/2$, the system exhibits 
damped oscillations
of frequency $\sim \Delta_{eff}$. For $1/2< \alpha_{c}<1$, 
the dynamics is an exponential decay superposed on an incoherent 
background and for $\alpha_{c}>1$, the system is 
localised, that is $P_{\up \up}(t)=1$ for all times. 
(ii) At finite $T$, this oscillatory behaviour persists as long as 
$T < \Delta_{c}/ \pi \alpha_{c}$. For $T > \Delta_{c}/ \pi \alpha_{c}$, 
the dynamics is
overdamped, and the system relaxes exponentially to $P_{\up \up} = 1/2$.

The various regimes are shown in Fig. 
(\ref{spbosphase}). For
the present problem the system is most likely to be in the overdamped regime;
because $\Delta_{c}^{eff}$ is so small, extremely small values of $\alpha_{c}$ 
and/or $T$
will be required to see underdamped oscillations. In the overdamped regime,
the damping rate is 
\begin{equation}
\Gamma_c = \frac{\Delta_{c}^{2}}{2\Omega_{0}}
\left[ \frac{2\pi T}{\Omega_{0}} \right]^{2\alpha_{c}-1}
\frac{\Gamma^{2}(\alpha_{c})}{\Gamma(2\alpha_{c})}
\end{equation}
and, provided $T > \Delta_{c}/\pi \alpha_{c}$, one has the simple result 
\begin{equation}
P_{\up \up}(t) = \frac{1}{2} + \frac{1}{2} e^{- \Gamma_{c}t}
\end{equation}
If $T < \Delta_{c}/\alpha_{c}$, 
numerical methods must be employed if one is interested in the exact 
behaviour of the system. Notice that when 
$\alpha_{c} > 1/2$, an increase in temperature increases $\Gamma_{c}$, 
corresponding to faster relaxation. However, if $\alpha_{c} < 1/2$, 
the relaxation rate $\Gamma_{c}$ (which is now much
smaller) decreases with increasing $T$. Thus the thermal environment plays a
different role on either side of the ```Toulouse line'' $\alpha_{c}=1/2$; for
$\alpha_{c} > 1/2$, it actively retards relaxation. In the limit 
$\alpha_{c} \ll 1$, we can establish an upper bound on the relaxation rate
$\Gamma_{c} < \Delta_{c}/2\pi$.

Obviously, the same discussion will go through if the coupling is 
antiferromagnetic but with an initial $| \up \down \rangle$ or 
$| \down \up \rangle$
configuration. 

Consider now the case of an AFM coupling with an initial
$| \up \up \rangle$ state. As mentioned before, the system must first
relax to a low energy state before any simultaneous transitions
can take place. In the present case of strong coupling this is obvious-
the energy gap between the upper ``FM-ordered'' doublet and the lower
AFM doublet is so large that this is by far the fastest 
transition, and so we do not attempt to derive this result. The 
initial relaxation is then 
simply determined by the
relaxation of a single spin-boson system in the presence of a 
{\em static} bias ${\cal J} \gg  T$
\begin{equation}
\Gamma_{AFM} = \alpha_{\beta} (\Delta_{\beta}^{2}/{\cal J})
({\cal J}/\Omega_0)^{2 \alpha_{\beta}}
\label{ratecoupledanti}
\end{equation}
where $\beta$ represent the spin with the fastest relaxation rate.
Once the system has relaxed to the low-energy states $\{ \up \down \}$ or
$\{ \down \up \}$, the analysis for the coupled dynamics can be applied.

It has to be said that the locked phase is the one that is most likely to be 
encountered in physical applications. For microscopic models such as the
problem of 2 coupled magnetic impurities, it represents the ordered state;
for macroscopic models such as 2 coupled nanomagnets or 2 coupled SQUID's,
it will prevail unless the mutual coupling is very weak (ie., they are
very far apart, since usually $\alpha$ is not small for such systems).
Many quantum measurement schemes actually {\it require} that a
strong correlation be established between apparatus and measured system
at some point, thereby freezing the system dynamics (even if the system
is otherwise able to display coherent tunneling). 

\subsection{Correlated Relaxation Phase}

As the coupling ${\cal J}$ is reduced, or $T$ is increased, temperature
fluctuations break up the ``locking of blips'' which allowed the solution of
the locked phase. When $T \gg {\cal J}$, we may use instead an approximation
in which blips of one path almost always overlap with with sojourns of the
other. It is important that this approach only works {\em outside} of the
perturbative regime - but it is the strong-coupling regime we are interested
in here.

Again, the detailed evaluation of the path integrals, etc., 
is given in Appendix D. There we show that one can   
ignore both  
interactions between blips on the same path, and between blips on the 2
separate paths. 
This requires ${\cal J} \gg \Delta_{\beta}$, 
and thus places us directly outside the range of a perturbative
treatment. Our results therefore cannot be necessarily extended to the limit
${\cal J} \rightarrow 0$, especially if one is interested in the
coherent properties of the coupled system in this limit. However, provided 
that the two conditions $T \gg {\cal J}$ and ${\cal J} \gg \Delta_{\beta}$, 
then our results are valid for any values of the dissipation coefficients 
$\alpha_{\beta}$.

In fact one finds that the same path summation technique also works for the 
mutual coherence phase, and so we first give a general expression for 
the Laplace transform
of $P_{\up \up}(t)$ (again we 
assume an initial state $|\up \up \rangle$). As shown in Appendix D, 
$P_{\up \up}(\lambda)$ is given in terms of the 2 functions
\begin{eqnarray}
g_{\beta}(\lambda) &=& \Delta_{\beta}^{2} \int_{0}^{\infty} dt \,
e^{-\lambda t - Q_{2}^{(\beta)}(t) } \cos[Q_{1}^{(\beta)}(t)] \,
\cos [ {\cal J} t ] \nonumber \\
&=& \frac{1}{2} (f_{\beta}(\lambda + i{\cal J}) +
f_{\beta}(\lambda-i{\cal J}))
\\
h_{\beta}(\lambda) &=& \Delta_{\beta}^{2} \int_{0}^{\infty} dt \,
e^{-\lambda t - Q_{2}^{(\beta)}(t) } \sin[Q_{1}^{(\beta)}(t)] \,
\sin [ {\cal J} t] \nonumber \\
&=& \frac{\tan \pi \alpha}{2i}
(f_{\beta}(\lambda - i{\cal J}) - f_{\beta}(\lambda +i{\cal J}))
\label{definitions}
\end{eqnarray}
where $f_{\beta}(\lambda)$ is equivalent to Eq.(\ref{fclambda}) provided
that the appropriate changes of parameters are made. We then have
\begin{eqnarray}
\label{puulambda}
P_{\up \up}(\lambda) &=& \frac{1}{\lambda} -\frac{1}{4} \left[ g_{1}
\left(
 1+ \frac{h_{1}}{g_{1}} \right) + g_{2} \left( 1+ \frac{h_{2}}{g_{2}}
\right)
\right]  \frac{1}{\lambda} \, \frac{1}{\lambda + g_{1} +g_{2} }
\nonumber \\
& - &  \frac{1}{4} \left[ g_{1} \left( 1+ \frac{h_{1}}{g_{1}} \right) +
g_{2} \left( 1+ \frac{h_{2}}{g_{2}} \right) \right]
\frac{1}{\lambda^{2} + \lambda(g_{1}+g_{2}) +g_{1}g_{2}
(1-\frac{h_{1}h_{2}}{
g_{1}g_{2} })  } \\
&-& \frac{1}{2}g_{1}g_{2} \left[ 1 - \frac{h_{1}h_{2}}{g_{1}g_{2}}
\right]
\frac{1}{\lambda} \, \frac{1}{\lambda^{2} + \lambda(g_{1}+g_{2})
 +g_{1}g_{2} (1-\frac{h_{1}h_{2}}{g_{1}g_{2}}) } \, . \nonumber
\end{eqnarray}
Analogous expressions for the other probabilities appear in Appendix D.
The dynamics is controlled by
the poles of $P_{\tau_{1} \tau_{2}}(\lambda)$. There are two types of
denominator; the ``global denominator'', 
\begin{equation}
\lambda + g_{1}(\lambda)+g_{2}(\lambda)
\label{denom1}
\end{equation}
corresponds to the overall system dynamics, 
whereas the ''individual spin'' denominator, 
\begin{equation}
(\lambda+g_{1}(\lambda))(\lambda+g_{2}(\lambda))
-h_{1}(\lambda)h_{2}(\lambda)
\label{denom2}
\end{equation}
corresponds to the individual spin dynamics.
Both spins are correlated through the terms
in $g_{\beta}/h_{\beta}$,
which essentially behave as $\tanh {\cal J} /2T$ at long
times. This correlation is present both in the denominators (which
give the decay rates and oscillation frequency) and in their coefficients
(controlling the weight of each term) which results in a very complex
behaviour.

On the other hand the limit of $\lambda P_{\tau_{1} \tau_{2}}(\lambda)$ as
$\lambda \rightarrow 0$ yields the universal long time probabilities 
\begin{equation}
P_{\up \up}(t \rightarrow \infty) = P_{\down \down}(t \rightarrow
\infty) =
\frac{1}{4} [ 1-\tanh {\cal J}/2T ]
\label{probinfini1}
\end{equation}
\begin{equation}
P_{\up \down}(t \rightarrow \infty) = P_{\down \up}(t \rightarrow
\infty)=
\frac{1}{4} [ 1+\tanh {\cal J}/2T ] \, .
\label{probinfini2}
\end{equation}
These results are of course nothing but
the thermodynamic equilibrium expectation values, once all 
dynamic relaxation
to equilibrium has ceased.

Let us now deal with the regime $T \gg {\cal J}$, and also assume that
$T > \Delta_{\beta}/\alpha_{\beta}$; this is the overdamped correlated 
relaxation regime (cf. Fig. (\ref{spbosphase})). 
Formally this regime is very easy to deal
with, since we may set $\lambda=0$ in $f_{\beta}(\lambda)$, and the explicit
dependence on $h_{1}$ and $h_{2}$ in $P_{\tau_{1} \tau_{2}}(\lambda)$ 
can be eliminated using \cite{leggettB}
$h_{\beta}(0)/g_{\beta}(0) = \tanh ({\cal J}/ 2T)$. 
The Laplace inversion results in a sum of decaying
exponentials for each $P_{\tau_{1} \tau_{2}}(t)$: 

\begin{equation}
P_{\up \up}(t) = A_{-} + A_{+} e^{ - ( \Gamma_{1} + \Gamma_{2} )t }
        + B_{+} e^{-R_{+}t} + B_{-} e^{-R_{-}t}
\label{ppp}
\end{equation}
\begin{equation}
P_{\down \down}(t) = A_{-} + A_{+} e^{ - ( \Gamma_{1} + \Gamma_{2} )t }
- B_{+} e^{-R_{+}t} - B_{-} e^{-R_{-}t}
\vspace{2mm}
\end{equation}
\begin{equation}
P_{\up \down}(t) = A_{+}[1-e^{ - ( \Gamma_{1} + \Gamma_{2} )t }]
+ C_{+} [ e^{-R_{+}t} - e^{-R_{-}t} ]
\vspace{2mm}
\end{equation}
\begin{equation}
P_{\down \up}(t) = A_{+}[1-e^{ - ( \Gamma_{1} + \Gamma_{2} )t }]
- C_{+} [ e^{-R_{+}t} - e^{-R_{-}t} ] \, ,
\vspace{2mm}
\end{equation}
with the different constants defined as
\begin{equation}
A_{\pm} = \frac{1}{4} [1 \pm \tanh({\cal J} /2T)]
\label{constA}
\end{equation}
\begin{equation}
B_{\pm} =  \frac{1}{4} \left[ 1 \pm  \frac{\Gamma_{1}+\Gamma_{2}}{\sqrt{R_{12}}}
 \tanh({\cal J} /2T) \right]
\end{equation}
\begin{equation}
R_{\pm} = \frac{1}{2}(\Gamma_{1}+\Gamma_{2}) \pm \frac{1}{2} \sqrt{R_{12}}
\end{equation}
\begin{equation}
R_{12}(T) = \Gamma_{1}^{2} + \Gamma_{2}^{2} -2\Gamma_{1}\Gamma_{2} \left[ 1-2
\tanh^{2} ({\cal J} /2T) \right]
\end{equation}
\begin{equation}
C_{+}=A_{+} \frac{\Gamma_{1}-\Gamma_{2}}{\sqrt{R_{12}}}
\end{equation}
The relaxation rates $\Gamma_{1}$ and $\Gamma_{2}$ are simply the functions 
$g_{1}(0)$ and $g_{2}(0)$, the relaxation rates of a spin-boson 
system in the presence of a bias. These are \cite{leggettB} :
\begin{equation}
\Gamma_{\beta} = \frac{\Delta_{\beta}^{2}}{2\Omega_{0}}
\left[ \frac{2\pi T}{\Omega_{0}} \right]^{2\alpha_{\beta}-1}
\frac{\cosh({\cal J}/2T)}{\Gamma(2\alpha_{\beta})}
| \Gamma(\alpha_{\beta}+i{\cal J}/2\pi T) |^{2}
\label{decayratesover}
\end{equation}
Eq.(\ref{ratecoupledanti}) corresponds to the limit ${\cal J} \gg T$ of this
equation. In the other limiting case, ${\cal J} \ll T$ and for 
$\alpha_{\beta} \sim O(1)$, 
\begin{equation}
\Gamma_{\beta} = \frac{\Delta_{\beta}^{2}}{2\Omega_{0}}
\left[ \frac{2\pi T}{\Omega_{0}} \right]^{2\alpha_{\beta}-1}
\frac{\Gamma^{2}(\alpha_{\beta})}{\Gamma(2\alpha_{\beta})} + O({\cal J}/T)^{2}
\; \; \; \; \; \; ({\cal J}/T \ll 1)
\end{equation}
However, if $\alpha_{\beta} \ll 1$ as well, then it is
\begin{equation}
\Gamma_{\beta} = \frac{\Delta_{\beta}^{2}}{2\Omega_{0}}
\left[ \frac{2\pi T}{\Omega_{0}} \right]^{2\alpha_{\beta}-1}
\frac{2 \alpha_{\beta}}{\alpha_{\beta}^{2} + ({\cal J}/2 \pi T)^{2}}
\; \; \; \; \; \; ({\cal J}/T, \alpha_{\beta} \ll 1)
\end{equation}
Thus  one can get different behaviour depending on the ratio between
$\alpha_{j}$ and ${\cal J}/T$

The relaxation times appearing in $P_{\tau_{1} \tau_{2}}(t)$ if
$T \gg {\cal J}$ 
can be quite different, and their contributions will in general appear in
different proportions. 
Fig.(\ref{relaxfig}) illustrates this by plotting the 4 probabilities.

Let us now look more closely at the structure of $P_{\tau_{1} \tau_{2}}(t)$. 
If ${\cal J}=0$, so that the two system are
completely uncorrelated, it simply decomposes into a product of
the individual spin probabilities $P_{\tau_{1} \tau_{2}}(t)=P_{\tau_{1}}(t)
P_{\tau_{1}}(t)$, with $A_{\pm}=B_{\pm}=C_{+}=1/4$, $R_{+}=\Gamma_{1}$ and
$R_{-}=\Gamma_{2}$. 
Turning on the interaction ${\cal J}$ causes the individual probabilities to 
merge together to produce a term corresponding to the relaxation of the system 
as a whole
(with relaxation rate $\Gamma_{1}+\Gamma_{2}$, and weight 
$A_{\pm}$), mixed in with correlated individual relaxation 
(of rates $R_{\pm}$, and weight
$B_{\pm}$ or $C_{+}$). The mixing of the relaxation rate is independent of the 
sign of 
${\cal J}$; as ${\cal J}$ increase, $R_{+}$ tends to $\Gamma_{1}+\Gamma_{2}$
while $R_{-} \rightarrow 0$, indicating a reduction of the individual dynamics
of the spins. However, the coefficients $A_{\pm}$,
$B_{\pm}$ and $C_{+}$ do depend on the sign of the interaction, as we would 
expect - we have started the system in the state $| \up \up \rangle$.
Notice that the correlations of the two spins is only weakly felt
in the relaxation rates $R_{\pm}$; it represents a correction of order
${\cal J}/T << 1$. The relaxation at long time will reflect this
correlation more strongly, mostly through the terms $A_{\pm}$.

It is important to remember that all these expressions assume ${\cal J} \ll T$ 
(and can thus be expanded in powers of ${\cal J}/T$). Using them for 
${\cal J} \sim T$ would be incorrect- one is then in the crossover 
region
between the correlated and the locked phase. In this crossover, we
cannot solve the problem analytically - the assumption of non-overlapping 
blips is breaking down, but they are still not fully correlated (as in the
locked phase). Physically, when ${\cal J} \sim T$ the spins are trying to lock,
but thermal fluctuations are preventing them.

\subsection{Mutual Coherence Phase}

If $\alpha_{\beta} \geq O(1)$, then we have exhausted all possible phases
of the PISCES system; either the 2 spins are locked, or they exhibit correlated
overdamped relaxation, depending on the ratio of ${\cal J}$ to $T$.

However if $\alpha_{\beta} \ll 1$, a new possibility emerges; 
for temperatures such that 
${\cal J} \ll T \ll \Delta_{\beta}/\alpha_{\beta}$, the
thermal fluctuations are strong enough to prevent the locking of the
blips but still not capable of destroying coherence. As a result,  
the dynamics is composed of underdamped oscillations. In the PISCES
problem the oscillations of the 2 systems will be correlated, and we can 
even expect decaying ``beats'', between the 2 spins.

Because we are in the non-perturbative limit ${\cal J} \gg {\Delta}$
(even though $\alpha_{\beta} \ll 1$), the
oscillation frequencies are roughly equal to ${\cal J}$, 
with corrections of order $O((\Delta_{\beta}/{\cal J})^2)$ and 
$O( \Delta_{\beta}^{2} (\alpha_{\beta} T)^2 / {\cal J}^4)$. There are 2
characteristic relaxation rates in the system. There is a ``global'' decay 
at a rate 
$\Gamma_{glob} \sim \alpha_{\beta} T (\Delta_{\beta} / {\cal J})^2$, and 
a much faster damping of the mutual oscillations,at a rate
$\Gamma_{osc} \sim \alpha_{\beta} T \ll \Gamma_{glob}$. We give more 
detailed expressions below. 

As noted above, expressions like (\ref{puulambda}) for
$P_{\tau_{1} \tau_{2}}(\lambda)$
are still valid for this regime, but an expansion
of $f_{\beta}(\lambda,T)$ in (\ref{fclambda}) (and hence of $g_{\beta}
(\lambda)$ and $h_{\beta}(\lambda)$) in powers of $\lambda$ is invalid -
instead we must expand in $(\alpha_{\beta}+\lambda/2 \pi T) \ll 1$.
This expansion gives
\begin{equation}
f_{\beta}(\lambda)= \Delta_{\beta}^{* 2}
\left( \frac{2\pi T}{\Delta_{\beta}^{*}} \right)^{2\alpha_{\beta}}
\frac{1}{\Gamma(1-2\alpha_{\beta})} \frac{1}{2\pi T \alpha_{\beta} +
\lambda}
\end{equation}
In general the denominators
in $P_{\tau_{1} \tau_{2}}$
are of polynomial degree 3 or 6, which renders an analytic
inverse Laplace transformation hopeless.
We can however study 2 cases. 
One is the case where both spins are equivalent, ie., $\Delta_{1}=\Delta_{2}
= \Delta$ and $\alpha_{1}=\alpha_{2}=\alpha$; this is the Equivalent Spin
case. We can also examine what happens if one spin is overdamped, and the
other is underdamped; this is the Overdamped plus Underdamped case. We 
re-emphasize that in real physical situations it will be quite difficult to 
see mutual coherence. One possibility for macroscopic systems involves
2 conducting nanomagnetic particles imbedded in a {\it semiconducting} 
substrate; the low electronic density so reduces $\alpha$ and the mutual 
RKKY interactions that the equivalent spin phase is a real possibility.
The ``overdamped plus underdamped'' system is interesting in view of the 
possibility of coupling an overdamped measuring apparatus to an underdamped 
quantum system, without destroying the coherence properties of the latter.

We start by defining  
three energy scales, viz., $a \equiv \pi \alpha T$, $({\cal J}^{2} +
\bar{\Delta}^2)^{1/2}$ and $\bar{\Delta}$, where 
$\bar{\Delta}$ is defined as
\begin{equation}
\bar{\Delta}^{2} =
\frac{\Delta^{*2}}{\Gamma(1-2\alpha)}
\left( \frac{2\pi T}{\Delta^{*}}
\right)^{2\alpha}
\label{dbar}
\end{equation}
Of these 3 energy scales, $a$ is the smallest. 
It is now convenient to define (by analogy with the biased spin-
boson model \cite{weissA}):
\begin{eqnarray}
\gamma_{R,s}(a_{\beta},{\cal J}, \bar{\Delta}_{\beta}^{2}) &=& 2 a_{\beta}
\frac{\bar{\Delta}_{\beta}^{2}}{
{\cal J}^{2} + \bar{\Delta}_{\beta}^{2}}  \nonumber \\
\gamma_{s}(a_{\beta},{\cal J}, \bar{\Delta}_{\beta}^{2}) &=& 2a_{\beta} -
\frac{1}{2} \gamma_{R,s} =
\frac{a_{\beta}}{{\cal J}^{2} + \bar{\Delta}_{\beta}^{2}} ( 2{\cal
J}^{2} +
\bar{\Delta}_{\beta}^{2})
  \label{mutualsingle}  \\
\nu^{2}_{s}(a_{\beta},{\cal J},\bar{\Delta}_{\beta}^{2}) &=& {\cal J}^{2} +
\bar{\Delta}_{\beta}^{2} +4a_{\beta}^{2}
-2 \gamma_{R,s} \gamma_{s} - \gamma_{s}^{2} \nonumber \\
& = & {\cal J}^{2} + \bar{\Delta}^{2}_{\beta} - \frac{a^{2}_{\beta}
\bar{\Delta}_{\beta}^{4}}{
({\cal J}^{2} + \bar{\Delta}^{2}_{\beta})^{2}} \left[ 1 +
4 \frac{{\cal J}^{2}}{
\bar{\Delta}^{2}_{\beta}} \right] \nonumber
\end{eqnarray}
with $a_{\beta} \equiv \pi \alpha_{\beta} T$ and $\bar{\Delta}_{\beta}$
defined as in Eq. (\ref{dbar}) with $\Delta$ replaced by
$\Delta_{\beta}$. In the limits considered in this section, 
$\gamma_{R,s} \sim 2 a_{\beta} \bar{\Delta}_{\beta}^{2} /{\cal J}^{2}$,
$\gamma_{s} \sim 2 a_{\beta}$ and 
$\nu^{2}_{s} \sim {\cal J}^{2}$.

\subsubsection{Equivalent Spin Case}

For this case, defined by $T \gg {\cal J}$ and
$\Delta/\alpha \gg T$ for both spins, the equation for the roots of the
denominator in $P_{\tau_{1} \tau_{2}}(\lambda)$ decouples into 2 equations of
polynomial degree 3, which we can handle by perturbation in 
$a = \pi \alpha T$. 
The resulting forms for the ``FM- correlated matrix elements'' 
$P_{\up \up}(t)$ and $P_{\down \down}(t)$ are then 
found to be (cf. App. E):
\begin{eqnarray}
\label{oscilluu}
P_{\up \up}(t) &=& A_{-} - C_{2}e^{-\Gamma_{R}t} + C_{3}e^{-\Gamma
t}\cos(\mu t)
+C_{4}e^{-\Gamma t}\sin(\mu t) + \nonumber \\
& & \sum_{i=\pm} C_{5i}^{+}e^{-\gamma_{Ri}t}
+ C_{6i}^{+}e^{-\gamma_{i} t}\cos(\nu_{i} t) +C_{7i}^{+}e^{-\gamma_i t}
\sin(\nu_{i} t)
\end{eqnarray}
\begin{eqnarray}
\label{oscilldd}
P_{\down \down}(t) &=& A_{-} - C_{2}e^{-\Gamma_{R}t} + 
C_{3}e^{-\Gamma t}\cos(\mu t) +C_{4}e^{-\Gamma t}\sin(\mu t) + \nonumber \\
& & \sum_{i=\pm} C_{5i}^{-}e^{-\gamma_{Ri}t}
+ C_{6i}^{-}e^{-\gamma_{i} t}\cos(\nu_{i} t) +C_{7i}^{-}e^{-\gamma t}
\sin(\nu_{i} t)
\end{eqnarray}
where $\Gamma_{R} = \gamma_{R,s}(a,{\cal J},2 \bar{\Delta}^{2})$,
$\Gamma= \gamma_{s}(a,{\cal J},2\bar{\Delta}^{2})$,
$\mu^{2}=\nu^{2}_{s}(a,{\cal J},2\bar{\Delta}^{2})$, and  
$A_{\pm}$ is defined by Eq. (\ref{constA}).
If we now introduce
${\cal J}_{\mp} = 4{\cal J}A_{\mp}$ and
$\bar{\Delta}^{2}_{\pm} = 4 \bar{\Delta}^{2} A_{\pm}$,  
then we have 
$\gamma_{R \pm} =  \gamma_{R,s}(a,{\cal J}_{\mp},\bar{\Delta}_{\pm}^{2}$),
$\gamma_{\pm} = \gamma_{s}(a,{\cal J}_{\mp},\bar{\Delta}_{\pm}^{2})$ and the
frequencies 
$\nu_{\pm}^{2} = \nu_{s}(a,{\cal J}_{\mp},\bar{\Delta}_{\pm}^{2}$).
The explicit expressions for the 3 characteristic frequencies of the system 
are then:
\begin{equation}
\mu^{2} =  {\cal J}^{2} +2 \bar{\Delta}^{2} -
4 \frac{a^{2} \bar{\Delta}^{4}}{({\cal J}^{2} + 2\bar{\Delta}^{2})^{2}}
\left[ 1 + 2 \frac{{\cal J}^{2}}{ \bar{\Delta}^{2}} \right]
\label{muf}
\end{equation}
\begin{equation}
\nu_{\pm}^{2} = {\cal J}^{2} + \bar{\Delta}^{2}
- \frac{a^{2} \bar{\Delta}^{4}}{({\cal J}^{2} + \bar{\Delta}^{2})^{2}}
4 A_{\pm} \left[ 4 A_{\pm} + 16 A_{\mp}
\frac{{\cal J}^{2}}{ \bar{\Delta}^{2}} \right]
\label{nuf}
\end{equation}
The frequency $\mu$ pertains to the combined system,
whereas the frequency pair $\nu_{\pm}$ describes correlated oscillations of
the individual spins (``beats'').
The various constant $C_i$'s in Eq. (\ref{oscilluu}) 
and Eq. (\ref{oscilldd}) 
are complicated functions of the frequencies and decay rates and
given in Appendix E. 

On the other hand, in this equivalent spin limit, 
the AFM correlated probabilities $P_{\up \down}(t)$ and $P_{\down \up}(t)$
have a much simpler form:
\begin{equation}
P_{\up \down}(t) = P_{\down \up}(t) = A_{+} + C_{2}e^{-\Gamma_{R}t} -
C_{3}e^{-\Gamma t}\cos(\mu t) - C_{4}e^{-\Gamma t}\sin(\mu t)
\label{oscillud}
\end{equation}

Let us now try to understand some of the details of this result.
The appearance of oscillations in the density matrix at 3 different 
frequencies reflects the fact that the 2 spins are neither in a
completely correlated nor in a completely uncorrelated state. The total
dynamics then results from a compromise between global and individual 
dynamics. The global dynamics can be eliminated if one considers the
probability of a spin being in a given state, independently of the state
of the other. For example, the quantity
$[P_{\up \up}(t) + P_{\up \down}(t)]$ describes 
the dynamics of
the first spin, with the second one ``averaged out''. 
Notice, incidentally,  that it is not possible to recover the oscillation 
frequencies of our ``toy-model'', Eq. (\ref{muf}) and (\ref{nuf}), in the
limit $\alpha = 0$ since paths with overlapping blips do not contribute
to $P_{\tau_1 \tau_2}$ in the uncorrelated phase. The lower limit on
$\alpha$ is set by ${\cal J} \sim \Delta /\alpha$. 

To leading order in $T/{\cal J}$ and $\Delta/{\cal J}$, 
the decay rates $\Gamma_R$ and $\gamma_{R,i}$ are given by
$\gamma_{R,i} = \Gamma_R / 2 \sim 2 \pi \alpha T \bar{\Delta}^2 / {\cal
J}$, which simply corresponds to the limit $\alpha T \ll {\cal J}$
(always with ${\cal J} \gg \Delta$) of the individual decay rate
$\Gamma_{\beta}$ given in Eq. (\ref{decayratesover}). 
The same is true of the decay rates $\Gamma$ and $\gamma_i$, the decay
rates associated to the oscillatory terms.  
However, they are strongly corrected by a term of order 
$O(({\cal J}/\Delta)^2)$, so that the oscillations are 
damped out very quickly compared to the general decay of the probability
function. 

We illustrate this case with Fig. (\ref{mutualfig}), showing a special
case of the Fourier
transform of $P_{\up \up}(t)$, namely $Im[ -\lambda P_{\up \up}(i\lambda)]$,
Without the environment, the oscillation frequencies would show up as
divergences in this function. With the environment, the divergences are
smeared out by the decay rates $\Gamma$ and $\gamma_{\pm}$. At this scale,
the differences between $\mu$ and $\nu_{\pm}$ are clear, but not the
difference between $\nu_{+}$ and $\nu_{-}$. Taking a larger value of
$\alpha$ (resulting in a higher $a=\pi \alpha T$)
increases the decay rates associated with
the oscillations and the width of the individual resonances
becomes large enough for the two to overlap.

\subsubsection{Overdamped plus Underdamped Case}

We can also imagine two different systems labelled by $1$ and 
$2$, such that 
$\Delta_{2}/\alpha_{2} > T$ 
whereas $\Delta_{1}/\alpha_{1} < T$. 
In a single spin analysis, the spin-$1$ 
would be underdamped with the spin-$2$ overdamped. 
The limit ${\cal J} \gg
\Delta_{1},\Delta_{2}$ 
assuring the non-overlap of the blips is implicit.

To establish the dynamical behaviour, let us first look at the
two denominators involved in the Laplace transform. In Appendix F, we
show that the global denominator $(\lambda + g_1 + g_2)$ yields decay rates
and oscillation frequencies given by 
$\Gamma_{2} +\bar{\gamma}_{R}$,
$\Gamma_{2}+\bar{\gamma}$ and $\bar{\mu}^{2}$, where 
$\Gamma_{2}$ is the decay rate of the overdamped spin, 
Eq. (\ref{decayratesover}) and
$\bar{\gamma}_{R}$, $\bar{\gamma}$ and $\bar{\mu}^{2}$ are expressed with
respect to Eq. (\ref{mutualsingle}) as 
\begin{eqnarray}
\bar{\gamma}_{R} &=& \gamma_{R,s}(\bar{a},{\cal J},
\bar{\Delta}_{1}^{2}) \nonumber \\
\bar{\gamma} &=& \gamma_{s}(\bar{a},{\cal J},
\bar{\Delta}_{1}^{2}) \\
\bar{\mu}^{2} &=& \nu^{2}_{s}(\bar{a},{\cal J},
\bar{\Delta}_{1}^{2}) \nonumber
\end{eqnarray}
where $\bar{a} \equiv a_{1}-\Gamma_{2}/2$.

To treat the second denominator $[(\lambda + g_1)(\lambda + g_2) - h_1h_2]$,
we simply drop 
the terms $h_{1}(\lambda) h_{2}(\lambda)$. We can do this because 
$h_{1} \sim {\cal J}/ T$, whether
the spin is over- or underdamped. Thus as far as the decay rates and
oscillation frequencies are concerned, these terms will only be a small
correction of order $({\cal J}/2T)^{2}
(\Delta_{1}/ {\cal J})^{2} \ll 1$. We are then left with the
decay rates and oscillation frequencies of a single over- or underdamped
spin
in a {\it static} bias ${\cal J}$. These are $\Gamma_{2}$, as given by
Eq. (\ref{decayratesover}),
$\gamma_{1,R} = \gamma_{R,s}(a_{1},{\cal J},\bar{\Delta}_{1})$,
$\gamma_{1} = \gamma_{s}(a_{1},{\cal J},\bar{\Delta}_{1})$ 
and the frequency $\nu_{1}^{2} = 
\nu_{s}^{2}(a_{1},{\cal J},\bar{\Delta}_{1})$, defined in
Eq. (\ref{mutualsingle}).
Notice however, that the correlations brought by 
$h_{1}$, $h_{2}$ in the numerators of $P_{\tau_1 \tau_2}(\lambda)$
must be kept to determine the weight of the relaxation and oscillation terms
as well as the long time limit of the probabilities. 

With the decay rates and
oscillation frequency identified, it is straightforward 
to perform the inverse Laplace
transform to obtain the probability of occupation $P_{\up \up}(t)$ as
\begin{eqnarray}
P_{\up \up}(t) &=& A_{-} + C_{2} e^{-\Gamma_{2}-\bar{\gamma}_{R}}
+ C_{3} e^{-\Gamma_{2}-\bar{\gamma}} \cos (\bar{\mu} t)
+ C_{4} e^{-\Gamma_{2}-\bar{\gamma}} \sin (\bar{\mu} t) \nonumber
\\
&+&  C_{5} e^{-\Gamma_{2} t} + C_{6} e^{-\gamma_{1,R} t}
+ C_{7} e^{-\gamma_{1} t} \cos (\nu_{1} t)
+ C_{8} e^{-\gamma_{1} t} \sin (\nu_{1} t)
\end{eqnarray}

Again, oscillations associated to the global and the individual dynamics 
of the spins are present in $P_{\up \up}$. Notice that the  
corrections to the frequency $\bar{\mu}$ are ``tunable'' if one has
very good control over all the parameters. In fact,
with, $\Gamma_{2} = a_{1}$, the frequency $\bar{\mu}^{2}$ is 
simply equal to ${\cal J}^{2} + \bar{\Delta}^{2}$.
On the other hand, once again most of the effect of the environment is felt 
as a damping of the oscillatory terms, 
much faster than the the global decay
(with the oscillations of frequency $\bar{\mu}$ being damped faster 
than those of frequency $\nu_{1}$).

This case is illustrated of Fig. (\ref{overpunderfig}). The coherence
peaks are still present, but the line-width caused by the environment
impedes their resolution.
 
\section{Conclusion}

We have spoken very little in this article about the possible physical 
applications of our results, preferring first to gain a general 
understanding of how the model behaves. We have in fact already used these 
results to do a number of detailed calculations for 2 particular problems.
The first concerns the dynamics of a pair of nanomagnets, each coupled to 
a background conducting or semiconducting substrate. To properly treat this 
we need to add to what we have done here the presence of an external bias
acting on each nanomagnet. It is straightforward to make this 
generalisation, and also to calculate from microscopic theory what are the 
parameters entering the PISCES Hamiltonian as functions of nanomagnet size,
distance between nanomagnets, electron density, etc.; and one may also 
incorporate the effect of nuclear spins and dipolar interactions. This 
problem is not only of academic interest; it is also relevant to future 
information storage technology.

The second problem we have looked at describes a fairly general class of 
physical systems in which we have a macroscopic measuring apparatus coupled
to a central ``measured'' system; and in which at low energies the dynamics of 
both truncates to 2-level systems. The purpose of this work is to go beyond
the usual idealised models of the measurement process, and see how the 3
important partners in the measurement (system, apparatus, and environment)
work together. Again one finds that the PISCES model (with added biases)
can be used to derive the full dynamics, with some rather surprising results.
Both of these investigations will be the subject of detailed future papers;
and they confirm that the results we have here can be used directly to 
give a realistic description of the full quantum dynamics of coupled 
macroscopic systems (as well as a large number of microscopic ones).

\section{Acknowledgements}

We would like to thank E. Mueller, N.V. Prokof'ev, and R.F.Kiefl, 
for suggestions and comments.
We also acknowledge support from NSERC and the Canadian Institute of Advanced
Research (CIAR); and P.C.E.S. thanks the Max Planck Institute for support while
this paper was being finished.

\newpage

\appendix

\section{Reduced Density Matrix of the Two-Spin System}
\resetcounters

In this appendix, we derive the form of the reduced 
density matrix of the combined two-spin system. 
We will assume the system starts in the polarised
state $| \uparrow \uparrow \rangle$, so both $\eta_{10}$ and $\eta_{20}$ must be
$+1$. Now consider the density matrix element
$\rho_{\tau_{1} \tau_{2}, \uparrow \uparrow}(t)$, i.e., the probability
$P_{\tau_{1} \tau_{2}}(t)$ that at time $t$ the system is in the
state $\tau_{1} \tau_{2}$ where $\tau_{1}$ and $\tau_{2}$ 
represent either $\uparrow$ or $\downarrow$.
Using the decomposition of the influence functional into individual
functionals and the interaction $F_{12}$, we express
the complete summation for the probabilities as
\begin{eqnarray}
P_{\tau_{1} \tau_{2}}(t) &=& \sum_{n_{1} n_{2}}
(-1)^{n_{1}+n_{2}+(d^{(1)}_{\alpha \beta}+d^{(2)}_{\alpha \beta})/2}
\left( \frac{\Delta_{1}}{2} \right)^{2n_{1}+d^{(1)}_{\alpha \beta}}
\left( \frac{\Delta_{2}}{2} \right)^{2n_{2}+d^{(2)}_{\alpha \beta}}
\int_{0}^{t} D\{t_{2n_{1}}\} \int_{0}^{t} D\{u_{2n_{2}}\}
 \sum_{ \{ \zeta_{1j},\eta_{1j} \} }
\sum_{ \{ \zeta_{2j},\eta_{2j} \} } \nonumber \\
& & \tilde{F}^{(1)}_{n_{1}} (\{t_{j}\}, \{\zeta_{1j}\},\{\eta_{1j}\})
\tilde{F}^{(2)}_{n_{2}} (\{u_{2k}\}, \{\zeta_{2k}\},\{\eta_{2k}\})
\tilde{G}_{n_{1}n_{2}}
( \{ t_{1j} \}, \{ u_{2k} \}, \{ \zeta_{1j} \}, \{ \eta_{1j} \},
\{ \zeta_{2k} \} , \{ \eta_{2k} \} )
\label{somme}
\end{eqnarray}
We first explain each term of this expression. The factors
$d^{(i)}_{\alpha \beta}$ are different for each probability calculated and
reflect the constraint on the last sojourn of each paths. If we are interested
in calculating $P_{\up \up}(t)$, then at time $t$, the two paths must be in
the state $[ \up \up ]$ after having performed $2n_{1}$ and $2n_{2}$ jumps.
We thus have the constraints  $\eta_{1n_{1}} \equiv
\eta_{2n_{2}} \equiv +1$ and $d^{(1)}_{\up \up}=d^{(2)}_{\up \up}=0$.
However, for $P_{\downarrow \downarrow}(t)$, the paths must end in the state
$[\downarrow \downarrow]$. This means that at least one blip must be present in
the paths considered. They will thus be composed of $n_{1}+1$ and $n_{2}+1$
blips, both $n_{1}$ and $n_{2}$ ranging from zero to infinity, with the
constraints $\eta_{1n_{1}+1} \equiv \eta_{2n_{2}+1} \equiv -1$ and
$d^{(1)}_{\down \down}=d^{(2)}_{\down \down}=2$.
Finally, of course, the paths for $P_{\uparrow \downarrow}(t)$ and
$P_{\downarrow \uparrow}(t)$ will be formed by a mix of these two cases.
For $P_{\uparrow \downarrow}$, the constraints are $\eta_{1n_{1}}=+1$ and
$\eta_{2n_{2}+1} = -1$ with $d^{(1)}_{\down \down}=0$ and
$d^{(2)}_{\down \down}=2$,
while for $P_{\downarrow \uparrow}$, the constraints are $\eta_{1n_{1}+1} = -1$
and $\eta_{2n_{2}}=+1$ and  $d^{(1)}_{\down \down}=2$,
$d^{(2)}_{\down \down}=0$,
 
The time integrations are defined as
\begin{equation}
\int_{0}^{t} D\{t_{2n}\} = \int_{0}^{t} dt_{2n} \int_{0}^{t_{2n}} dt_{2n-1}
\cdots \int_{0}^{t_{3}} dt_{2} \int_{0}^{t_{2}} dt_{1}
\label{contract}
\end{equation}
ie., an
integration over the times at which the jumps occur. Notice that the blips
of a given path are well ordered, ie., they occur successively;
however, nothing specifies the order of the blips of one path with respect to
those of the other, and so 
all the different different configurations must be included.
For example, consider a path relevant to the calculation of $P_{\uparrow
\uparrow}(t)$ with
$n_{1}=n_{2} = 1$; there are four transition times $t_{1}$, $t_{2}$, $u_{1}$
and $u_{2}$, which can be arranged in 6 different configurations,
as represented in Fig. \ref{pathupup}, always with the restriction
$t_{1}<t_{2}<t$ and $u_{1}<u_{2}<t$.  In general, there will be
\begin{equation}
\label{nombre}
\left( \begin{array}{c}
                2n_{1}+2n_{2} \\
                2n_{1}
       \end{array} \right)
 \equiv  \frac{ (2n_{1}+2n_{2})! }{2n_{1}! \, 2n_{2}!}
\end{equation}
such configurations. This is simply a
reflection of the fact that we cannot
describe the two paths with respect to a single time where the all the jumps
would be well ordered
 
The terms $\tilde{F}^{(\alpha)}_{n_{\alpha}} (\{t_{j}\}, 
\{\zeta_{\alpha \, j}\},\{\eta_{\alpha \, j}\})$
are the single-spin functionals which can be decomposed as
\begin{equation}
\tilde{F}_{n_{\alpha}}^{(\alpha)} (\{ t_{j} \} , \{ \zeta_{\alpha \, j} \} , 
\{ \eta_{\alpha \, j} \} ) =
\tilde{F}_{n_{\alpha}}^{(\alpha ; se)} \tilde{F}_{n_{\alpha}}^{(\alpha ; b-b)} 
\tilde{F}_{n_{\alpha}}^{(\alpha ; b-s)}
\label{1spincomplete}
\end{equation}
where
\begin{equation}
\tilde{F}_{n_{\alpha}}^{(\alpha ; se)} =\exp \left[ 
-\frac{q_{0}^{2}}{\pi} \sum_{j=1}^{n_{\alpha}}
Q_{2}^{(\alpha)}(t_{2j}-t_{2j-1}) \right]
\label{f1}
\vspace{2mm}
\end{equation}
\begin{equation}
\tilde{F}^{(\alpha ; b-b)}_{n_{\alpha}} =\exp 
\left[- \frac{q_{0}^{2}}{\pi}\sum_{j>k}^{n_{\alpha}}
\zeta_{\alpha \, j} \, \zeta_{\alpha \, k} \, \Lambda_{jk}^{(\alpha)} \right]
\vspace{2mm}
\end{equation}
\begin{equation}
\tilde{F}^{(\alpha ; b-s)}_{n_{\alpha}} =\exp 
\left[ i\frac{q_{0}^{2}}{\pi} \sum_{j>k}^{n_{\alpha}}
\zeta_{\alpha \, j} \, \eta_{\alpha \, k} \, X_{jk}^{(\alpha)} \right]
\vspace{2mm}
\end{equation}
in which the kernel matrices are
\begin{equation}
\Lambda_{jk}^{(\alpha)} \equiv Q_{2}^{(\alpha)}(t_{2j}-t_{2k-1}) + 
Q_{2}^{(\alpha)}(t_{2j-1}-t_{2k}) -Q_{2}^{(\alpha)}(t_{2j}-t_{2k}) - 
Q_{2}^{(\alpha)}(t_{2j-1}-t_{2k-1})
\label{lambdajk}
\end{equation}
\begin{equation}
X_{jk}^{(\alpha)} = Q_{1}^{(\alpha)}(t_{2j}-t_{2k+1}) +Q_{1}^{(\alpha)}
(t_{2j-1}-t_{2k}) -Q_{1}^{(\alpha)}(t_{2j}-t_{2k})
         -Q_{1}^{(\alpha)}(t_{2j-1}-t_{2k+1})
\vspace{4mm}
\label{xjk}
\end{equation}
and are defined in terms of the correlation functions
\begin{equation}
Q_{2}^{(\alpha)}(t) \equiv \int_{0}^{\infty} d\omega \, 
\frac{J_{\alpha \alpha}(\omega)}{\omega^{2}} \,
[\, 1-\cos \omega t \, ] \coth (\omega /2T ) \, .
\vspace{4mm}
\label{cul2}
\end{equation}
\begin{equation}
Q_{1}^{(\alpha)}(t) \equiv \int^{\infty }_{0} d\omega \, 
\frac{J_{\alpha \alpha}(\omega )}{\omega^{2}} \,
                \sin \omega t
\vspace{4mm}
\label{cul1}
\end{equation}
The correlation $Q_{1}^{(\alpha)}(t)$ refers to the reactive (phase) term 
$\Phi_{\alpha}(t)$ in (\ref{complete}), and $Q_{2}^{(\alpha)}(t)$ refers
to the dissipative term $\Gamma_{\alpha}(t)$.
 
As discussed at great length by Leggett et al., (\ref{somme}) can be
analysed in many regimes using the ``dilute blip'' approximation, in which the
interactions $\Lambda_{jk}^{(\alpha)}$ are dropped entirely and only the term
$\tilde{F}_{n_{\alpha}}^{(\alpha ; se)}$ and the term in 
$X_{jk}^{(\alpha)}$ corresponding to the interaction
of a blip with the sojourn immediately preceding it are kept.
Thus the single-spin influence functionals 
$\tilde{F}^{(\alpha)}_{n_{\alpha}}$ in
(\ref{somme}) can be evaluated without worrying about the precise ordering
of the charges in the configuration. 

Now let us turn to the interaction
functional $F_{12}$ in (\ref{completeint}).  One finds that $F_{12}$ is a sum
over all possible double paths integrals of the form
\begin{equation}
\int_{0}^{t} D\{t_{2n_{1}}\} \int_{0}^{t} D\{u_{2n_{2}}\}
\sum_{ \{ \zeta_{1j},\eta_{1j} \} }
\sum_{ \{ \zeta_{2j},\eta_{2j} \} }
\tilde{G}_{n_{1}n_{2}}
( \{ t_{1j} \}, \{ u_{2j} \}, \{ \zeta_{1j} \}, \{ \eta_{1j} \},
\{ \zeta_{2j} \} , \{ \eta_{2j} \} )
\label{somme2}
\end{equation}
with weightings $ (-i \Delta_{1} /2)^{2n_{1}} (-i \Delta_{2}/2)^{2n_{2}} $
( here we have chosen a set of path corresponding to the calculation of
$P_{\uparrow \uparrow}(t)$; (c.f., Eq (\ref{somme})). The integrand
$\tilde{G}_{n_{1}n_{2}}$ has the form
\begin{equation}
\tilde{G}_{n_{1}n_{2}}({\bf R}) = G^{(b-b)}_{n_{1}n_{2}}({\bf R})
 G^{(b-s)}_{n_{1}n_{2}}({\bf R})
\label{indcomplete}
\end{equation}
where
\begin{equation}
G^{(b-b)}_{n_{1}n_{2}}({\bf R}) = \exp \left[ -\frac{q_{01}q_{02}}{\pi}
\sum_{j,k}^{n_{1},n_{2}} \zeta_{1j} \zeta_{2k} \Lambda_{jk}({\bf R}) \right]
\end{equation}
in which the blip-blip interaction matrix is
\begin{equation}
\Lambda_{jk}^{(12)}({\bf R}) = [ Q_{2}^{(12)}({\bf R},|t_{2j}-u_{2k-1}|)
+ Q_{2}^{(12)}({\bf R},|t_{2j-1}-u_{2k}|) -
Q_{2}^{(12)}({\bf R},|t_{2j-1}-u_{2k-1}|) -
Q_{2}^{(12)}({\bf R},|t_{2j}-u_{2k}|) ]
\label{lambdajkint}
\end{equation}
Notice the difference from (\ref{lambdajk}). The
$G_{n_{1}n_{2}}^{(b-s)}({\bf R})$ term is
\begin{eqnarray}
G_{n_{1}n_{2}}^{(b-s)}({\bf R}) &=& \exp \left[ i \frac{q_{01}q_{02}}{\pi}
\sum_{j}^{n_{1}} \sum_{k}^{n_{2}} X_{jk}^{(12)} ({\bf R}) 
(\zeta_{1j} \eta_{2k} +
\eta_{1j} \zeta_{2k} ) \right] \times \nonumber \\
& & \exp \left[ -i K_{zz}({\bf R}) \left[ 
\sum_{j=1}^{n_{1}} \zeta_{1j} \eta_{2P}
(t_{2j}-t_{2j-1}) + \sum_{k=1}^{n_{2}} \zeta_{2k} \eta_{1P} (u_{2k}-u_{2k-1})
 \right] \right]
\end{eqnarray}
where
\begin{equation}
X_{jk}^{(12)}({\bf R}) = Q_{1}^{(12)}({\bf R},|t_{2j}-u_{2k-1}|)
+ Q_{1}^{(12)}({\bf R},|t_{2j-1}-u_{2k}|) -
Q_{1}^{(12)}({\bf R},|t_{2j-1}-u_{2k-1}|) -
Q_{1}^{(12)}({\bf R},|t_{2j}-u_{2k}|)
\label{xjkint}
\end{equation}
$K_{zz}$ is the direct interaction and couples
blips and sojourns belonging to different systems, but only 
while they overlap in time. A term directly coupling blips with earlier
sojourns would come from retardation effects, and these are ignored here.
This restriction is indicated by 
the index $P$ appearing in the charge of the sojourn.
The correlation functions have the obvious definitions
\begin{equation}
Q_{1}^{(12)}(t) = \int_{0}^{\infty} d\omega \,
\frac{J_{12}({\bf R},\omega )}{\omega^{2}} \, \sin \omega t
\end{equation}
\begin{equation}
Q_{2}^{(12)}(t) \equiv \int_{0}^{\infty} d\omega \,
\frac{J_{12}({\bf R},\omega )}{\omega^{2}} \,[1-\cos\omega t]
\coth(\omega/2T)
\end{equation}
We notice that there is no term in (\ref{indcomplete}) corresponding to the
term $F_{n}^{(1)}$ in (\ref{1spincomplete}), i.e., no ``self-energy''
term in the interaction functional corresponding to the the blip self-energy
in the single-spin functional $F_{j}$; $\tilde{G}_{n_{1}n_{2}}$ contains only
spin-spin interaction terms. This point is crucial - if we try to make an
approximation analogous to the non-interacting blip approximation, so useful
for the single spin-boson problem, we would be throwing away all spin-spin
interaction terms! Thus the PISCES problem is intrinsically more
difficult than the single-spin problem, since we must in some way deal with
these interactions.

Consider now the structure of the interaction influence 
functional. We need only consider three explicit 
configurations- the main characteristics of the analysis depend on 
whether or not the blips of a path overlap the sojourns of the other (well
separated blips and sojourns give standard results).
First, if the blip and sojourn are not overlapping,
the interaction is
identical to what was found before. An example of this case is the
interaction of the first spin-2 blip (charge $\zeta_{21}$) with the first
spin-1 sojourn (charge $\eta_{10}$) in Fig. \ref{pathupup}-a
The result is:
\begin{equation}
\exp  i \, \eta_{10} \, \zeta_{21} \, [ \, Q_{1}^{(12)}({\bf R}, u_{1}-t_{0})
+ Q_{1}^{(12)}({\bf R}, u_{2}-t_{1}) - Q_{1}^{(12)}({\bf R}, u_{2}-t_{0}) -
Q_{1}^{(12)}({\bf R}, u_{1}-t_{1}) \, ]
\end{equation}
However, if a blip occurs entirely within the time interval of a sojourn,
complications arise. As an example, consider the interaction of the first
spin-1 blip (charge $\zeta_{11}$) with the first spin-2 sojourn 
(charge $\eta_{20}$)
in Fig. \ref{pathupup}-b
The time integration is:
\begin{equation}
\int_{t_{1}}^{t_{2}} d\tau \int_{t_{0}}^{\tau} ds  \, \sin \omega (\tau-s) =
\frac{1}{\omega}(t_{2}-t_{1}) - \frac{1}{\omega^{2}} [\, \sin\omega(t_{2}-t_{0})-\sin\omega(t_{1}-t_{0}) \, ]
\end{equation}
which gives the result
\begin{equation}
\exp i \, \zeta_{11} \, \eta_{20} \, [ \, (\, - Q_{1}^{(12)}({\bf R},
t_{1}-t_{0})
 + Q_{1}^{(12)}({\bf R}, t_{2}-t_{0}) \, )
 - (K_{zz}({\bf R}) + \bar{\epsilon}({\bf R})) (t_{2}-t_{1})  \, ] \, .
\label{biasedcase}
\end{equation}
where $\bar{\epsilon}({\bf R})$ is given by
\begin{equation}
\bar{\epsilon}({\bf R}) \equiv \frac{q_{01}q_{02}}{\pi } \int_{0}^{\infty}
d\omega \, \frac{J_{12}({\bf R}, \omega )}{\omega } \, .
\end{equation}
This function is discussed in the main text.
 
The $3^{rd}$ case involves overlapping blips, and the total interaction
appears as well.
If the blips are overlapped as in Fig. \ref{pathupup}-b
it is convenient to consider
together the $\zeta_{11} - \eta_{20}$
and the $\zeta_{21} - \eta_{11}$ interactions. The corresponding
factor in the influence functional is then
\begin{eqnarray}
& & \exp i  \zeta_{11}  \eta_{20}  [ \, Q_{1}^{(12)}({\bf r}, t_{2}-
u_{1}) + Q_{1}^{(12)}({\bf R}, t_{1}-t_{0}) - Q_{1}^{(12)}({\bf R}, t_{2}-t_{0}) ]
\times  \exp i  \zeta_{21}  \eta_{11}  Q_{1}^{(12)} ({\bf R}, u_{2}-t_{2})
  \nonumber \\
& \times & \exp  i {\cal J}
[  \zeta_{11}  \eta_{20}   (u_{1}-t_{1})
+ \zeta_{21}  \eta_{11}  (u_{2}-t_{2})  ]
\end{eqnarray}
Finally, if a blip completely overlaps another blip, as in
Fig. \ref{pathupup}-c, the result of the interaction with the two
neighbouring sojourns is
\begin{eqnarray}
& & \exp +i {\cal J} [  \zeta_{11}
\eta_{20}   (u_{1}-t_{1}) + \zeta_{11}  \eta_{21}  (t_{2}-u_{2})  ]
\nonumber \\
 \exp i  \zeta_{11}  \eta_{20} 
& & [ Q_{1}^{(12)}({\bf R}, t_{1}-t_{0}) + Q_{1}^{(12)}({\bf R}, t_{2}-u_{1})
- Q_{1}^{(12)}({\bf R}, t_{2}-t_{0})  ]
- i  \zeta_{11} \eta_{21}  Q_{1}^{(12)}({\bf R}, t_{2}-u_{2})]
\end{eqnarray}
All other cases in Fig. \ref{pathupup} are similar to these, provided
one interchanges the times $t_{j}$ and $u_{k}$.
 
This summarises the construction of the influence functional for any
configuration.

\section{Poor Man's scaling for the Ohmic PISCES problem}
\resetcounters

We start from the reduced partition function of the Ohmic problem,  
expressed as:
\begin{eqnarray}
Z/Z_{env} &=& \int D[q_{1}]D[q_{2}] e^{-S_{0}[q_{1}]-S_{0}[q_{2}] + {\cal J}
({\bf R}) \int_{0}^{1/T} ds q_{1}(s) q_{2}(s)} \nonumber \\
& & \exp  \frac{1}{2\pi} \int_{0}^{1/T} ds \int_{0}^{1/T} ds' [
\eta_{1} \dot{q}_{1}(s) \dot{q}_{1}(s') +
\eta_{2} \dot{q}_{2}(s) \dot{q}_{2}(s') +
\eta_{12}({\bf R}) (\dot{q}_{1}(s) \dot{q}_{2}(s')
+\dot{q}_{2}(s) \dot{q}_{1}(s'))]
\label{partition}
\end{eqnarray}
where $Z_{env}$ is the partition function of the environment,
$S_{0}[q_{\alpha}]$ is the action of the $\alpha$-th two-level system,
and the
remaining terms represent their interaction with the oscillator bath.
Notice that the total adiabatic interaction ${\cal J}({\bf R})
 q_{1}(s) q_{2}(s)$ is used, so that 
all interactions resulting from {\em static}
configurations are explicitly separated from those coming from 
changes in the configurations (ie., transitions).
Each term in the equation can easily be interpreted with respect to the
single spin-boson problem. 
 
The partition function is evaluated using the usual instanton
method. The trajectories are defined by a series of tunneling jumps (the
instantons), of duration smaller
than $\tau_{c}$, taking place at times $t_{j}$ and
connecting the different configurations of the system. We
refer to the different configurations by the label $z_{l}$, $l =
1,2,3,4$ such that
$z = ( \{ \up \up \}, \{ \up \down \}, \{ \down \down \}, \{ \down \up \} )$.
These are connected by the tunneling matrix elements
$\Delta(k,l) = \Delta(l,k)$; here 
$ \Delta(1,1) = \Delta_{1}$ corresponds to transitions between
$z_{1}$ or $z_{2}$ with
$z_{3}$ or $z_{4}$, $\Delta(2,2)=\Delta_{2}$
connects $z_{1},z_{4}$ to $z_{2},z_{3}$ and
we introduce $\Delta(1,3)=\Delta_{F}$ and $\Delta(2,4) = \Delta_{AF}$
connecting the ferro- and
antiferromagnetic configurations $z_{1} - z_{3}$ and $z_{2} - z_{4}$
respectively. Even if no such transitions are present in the original
problem, they will be generated through the renormalisation procedure.
We also define the dimensionless fugacities $y(k,l) = \tau_{c}
\Delta(k,l)$.
To each configuration is associated a {\em static} energy
$\epsilon_{l}$, and a corresponding dimensionless energy
${\cal M}_{l} \equiv \tau_{c} \epsilon_{l}$. For the PISCES problem considered
here,
${\cal M}_{1} = {\cal M}_{3} = \tau_c {\cal J}/2$ and
${\cal M}_{2} = {\cal M}_{4} = - \tau_c {\cal J}/2$. 
 
The boundary conditions on the trajectories are such that a path starting in
a configuration $z_{l}$ at $t=0$ must return to this configuration at the
time $t=1/T$. Paths starting from the 4 different configurations must be
considered.
The time integrals
in Eq.(\ref{partition}) can now be done, and the partition function is 
\begin{equation}
Z/Z_{env} = \sum_{n=0}^{\infty}  \tilde{Z}_{n} =
\sum_{n} \sum_{z_{l}} \prod_{l=1}^{n} \left[  y(l,l-1)
 \int_{0}^{1/T} \frac{dt_{l}}{\tau_{c}} \Theta (t_{l}-t_{l-1} - \tau_{c})
\right]
\exp \left[ - \sum_{l=0}^{n} \left( \frac{t_{l}-t_{l-1}}{\tau_{c}}
 \right) {\cal M}_{l} + U(\{ t_{l} \}) \right]
\end{equation}
Just as for the dynamical behaviour of the PISCES problem,
nothing specifies the ordering of the instantons and we must sum over all
the possible set of configuration $\{ z_{l} \}$.
The time integral accounts for all the possible positions of the instantons,
with the $\Theta$- function insuring that
none are within $\tau_{c}$ of each other. The effect of the bath is
contained in the interaction matrix $U(\{ t_{l} \})$, defined as
\begin{equation}
U(\{ t_{l} \}) = \sum_{l>j} \ln \left( \frac{t_{l}-t_{l-1}}{\tau_{c}}
 \right) [ K_{l,j-1}+K_{l-1,j}-K_{l,j}-K_{l-1,j-1} ]
\end{equation}
with $K_{l,j}$ representing the interaction of the $l^{th}$ instanton with 
all those at earlier times.
In terms of the parameters of the PISCES problem, these are
$ K_{\beta, 2} = \alpha_{\beta 2}$, $K_{\beta, F}=\alpha_{\beta} +
\alpha_{12}$, $K_{\beta, AF}=\alpha_{\beta}-\alpha_{12}$, and, $K_{F, F} =
\alpha_{1}+\alpha_{2}+2\alpha_{12}$, $K_{AF, AF}=\alpha_{1}+\alpha_{2}-2
\alpha_{12}$, $K_{F, AF}=\alpha_{1}-\alpha_{2}$.
This interaction is logarithmic, which is what allows the
use of the poor man's renormalisation group. If one suppresses the 
energy difference between configurations, this partition function is
completely equivalent to the one treated by Cardy \cite{cardy}. In the 
present case where 
symmetry between the levels is broken, the poor man's renormalisation
procedure requires the inequalities $y(k,l) \ll 1$, $T \ll \Omega_{0}$, and
${\cal J} \ll \Omega_{0}$. If we also assume the 2 spins are identical,
ie., that $\Delta=\Delta_1 = \Delta_2$ and $\alpha = \alpha_1 = \alpha_2$,
then the partition function
is equivalent to that derived by Chakravarty and Hirsch \cite{chakra} for the
2-impurity Anderson model, and we may simply take over their scaling 
equations, quoted in the text.

\section{Dynamics in the Locked Phase}
\resetcounters

When the indirect coupling is the largest energy scale, the two spins will tend
to tunnel simultaneously. The time interval spent in a ``unlocked'' state
( $\{ \up \down \}, \{ \down \up \}$ if the coupling is ferromagnetic
(or $\{ \up \up \}, \{ \down \down \}$ if the
coupling is antiferromagnetic ) will be $ \sim 1/|{\cal J}|$.
In terms of our path integral for the density matrix, this means that the
dominant paths will be those where the blips are overlapping.
This minimises the overlap between blips and sojourns, thus cancelling the
fast varying factor $\exp i {\cal J} (t_{j}-u_{k})$ in the influence
functional which would otherwise give this configuration a vanishing
contribution to the path integral.
Thus, we can restrict the summation to be only over the paths where the
beginning (end) of one blip is within ${\cal J}^{-1}$ of the beginning
(end) of the second blip.
Furthermore,
the paths $q_{1}(\tau)$ and $q_{1}(\tau')$ are nearly identical to
the paths $q_{2}(\tau)$ and $q_{2}(\tau')$ respectively, and this brings
a restriction on the values that the charges of the blips and
sojourn can take. In the ferromagnetic case, they must be equal. That
is $\zeta_{1j}=\zeta_{2j}=\pm 1$ and $\eta_{1j}=\eta_{2j}=\pm 1$.
However, in the antiferromagnetic case, they must be opposite ($\zeta_{1j}=
-\zeta_{2j}=\pm 1$ and $\eta_{1j}=-\eta_{2j}=\pm$).

With these paths, the time integration over the end-points of the blips, Eq.
(\ref{somme}) is greatly simplified. For strong bias,
such that ${\cal J}$ is
the shortest time scale of the problem, we can set $t_{j}-u_{k} =
t_{j}-t_{k}$ provided that $j \neq k$. In the influence functional, we can
also set $Q_{2}^{(12)}(t_{2j}-u_{2j})=Q_{2}^{(12)}(0)=0$ and
$Q_{1}^{(12)}(t_{2j}-u_{2j})=Q_{1}^{(12)}(0)=0$ since they will be of order
$1/|{\cal J}|$. We clearly cannot do the same with the factor
$\exp i {\cal J} (t_{k}-u_{k})$ but it is now easy to integrate over
it. Always in the limit of strong coupling, the integration 
over the the endpoints of the $n^{th}$ pair of overlapping blips with 
charge $\zeta_n$,
placed between sojourns of charge $\eta_{n-1}$ and $\eta_n$ will be
\begin{equation}
4 \lim_{1/{\cal J} \rightarrow 0} \int_{0}^{1/{\cal J}} du
e^{ \pm i\eta_{n-1}\zeta_{n} {\cal J} u}
\int_{0}^{1/{\cal J}} du'
e^{\pm i\eta_{n}\zeta_{n} {\cal J} u'} \sim
\left( \frac{\pm 1}{i|{\cal J}|} \right)^{2}
\frac{1}{\eta_{n-1}\zeta^{2}_{n}\eta_{n}}
\label{intecoupled}
\end{equation}
$u$ and $u'$ being short times over which the blips are not overlapping.
The signs $\pm$ correspond to ferro- or antiferromagnetic coupling
respectively and the factor 4 comes from the four possible configuration of 
two overlapping blips. In Eq. (\ref{intecoupled}), we have neglected a
factor of $O(1)$ coming from the terms $e^{i \zeta_n \eta_n} -1)$. 
Taking into account the $(n+1)^{th}$ blip brings a equivalent factor,
but with charges $\eta_{n}\zeta^{2}_{n+1}\eta_{n+2}$ so that for a chain
of $n$ overlapping blips, 
the total contribution from the time integration of the overlap is
\begin{equation}
\frac{(-1)^{n}2^{2n}}{{\cal J}^{2}} \frac{1}{\eta_{0} \eta_{n}}
\end{equation}
This shows that we now have a new effective tunneling matrix element equal
to $\Delta_{c} = \Delta_{1} \Delta_{2} / | {\cal J} |$ (compare
Fig. (\ref{probjl}); there, the problem is set up so that $J_{0}={\cal J}/2$).
Furthermore, within this approximation, the interaction
of two overlapping blips, as given by Eq. (\ref{lambdajkint}) becomes
\begin{equation}
\Lambda_{jk}({\bf R}) \rightarrow 2 Q_{2}^{(12)}(t_{2k}-t_{2k-1})
\end{equation}
which is simply a contribution to the self-energy of each blips.  
Similarly, the interaction of the blip of one spin with the sojourn of 
the other spin becomes also simply adds to the contribution of the 
individual spins.

Therefore, the whole system behaves like 
a single spin-boson system to which is associated a
tunneling matrix element $\Delta_{c}$ and coupled to the environment
through a new coefficient $ \alpha_{c} = \alpha_{1} + \alpha_{2}
+2 \alpha_{12}$ for ferromagnetic coupling (since $\zeta_{1j} \zeta_{2j}
= +1$) and $ \alpha_{c} = \alpha_{1} + \alpha_{2}
-2 \alpha_{12}$ for antiferromagnetic (in this case
$\zeta_{1j} \zeta_{2j} = -1$).

Now the dissipative dynamics for this system is very well known, and all
known results can be taken over directly. 
Assuming a ferromagnetic coupling, so
that the system oscillates between the state $\{ \up \up \}$ and
$\{ \down \down \}$, it is straightforward to use the dilute-blip
approximation and to obtain Eq. (\ref{puufullambda}),
the Laplace transform of $P_{\uparrow \uparrow}(t)$. 
The expression for $P_{\down \down}(\lambda)$ will be equivalent, but with the
$+$ sign replaced by a $-$ sign (coming from the constraint
$\eta_{0}=0,\eta_{n}=-1$.
\begin{equation}
P_{\down \down}(\lambda ) = \frac{1}{2\lambda} -
\frac{1}{ \lambda + f(\lambda )} \, .
\end{equation}
with $f(\lambda)$ given by Eq. (\ref{fclambda}).

Notice that if we had an 
antiferromagnetically coupled system, starting at $t=0$ in the position
$ \{ \up \down \}$ or $\{ \down \up \}$, the structure of the summation is
completely similar, one merely needs to replace $P_{\up \up}$ and $P_{\down
\down}$ by $P_{\up \down}$ and $P_{\down \up}$. The case of antiferromagnetic
coupling, but starting configuration $\{ \up \up \}$ is discussed in the main
body of the text.

\section{Dynamics in the Correlated Relaxation Phase}
\resetcounters

As long as $T \gg {\cal J}$, the renormalisation group analysis shows
that there is no great tendency of the 2 spins to lock together. In
terms of the path integral, this means that the oscillating factors
$ \exp (i \zeta_{\alpha j} \eta_{\beta k} {\cal J} (t_j -u_k ))$ do not
reduce the weight of paths with overlapping blips and sojourns as
efficiently as in the locked phase. On the other hand, in the limit 
${\cal J} \gg \Delta_{\beta}$, if the  blip
of one path overlaps with a blip of the other, a factor 
$ \Delta_1 \Delta_2 / {\cal J} \ll \Delta_{\beta}$ still appears in 
the path integral, which then tends to reduce 
drastically such overlaps. 
If furthermore, the blips on the individual paths are naturally dilute,
then the occurrence of overlapping blips will be extremely rare and we
can neglect them completely. The paths summation can then be done by
using only paths where a blip of one path only overlaps in time with a 
sojourn of the other path.
The two essential conditions of this approximation are thus 
$T \gg {\cal J}$, to insure that we are not in the locked phase, and
${\cal J} \gg \Delta_{\beta}$ so as to neglect paths with overlapping
blips. In addition the blips of each individual spin must be in a 
dilute states \cite{leggettB}. 
If $\alpha_{\beta} > 1$, this is so for all values of 
${\cal J}$ and $T$. 
On the other hand, if $\alpha_{\beta} \ll 1$, then
the blips are dilute provided that 
\begin{equation}
\left. \frac{dg_{\beta}}{d\lambda} \right|_{\lambda=0} \sim
\frac{\Delta^{*2}_{\beta}}{ (2 \pi \ \alpha_{\beta} T)^{2} + {\cal
J}^{2}}
\end{equation}
Notice that as long as ${\cal J} \gg \Delta_{\beta}$,
the ratio $\alpha_{\beta} T /\Delta_{\beta}$ can be
arbitrary, thus allowing the occurrence of the mutual coherence phase. 

We can 
now proceed to explain in detail the path summations that lead to 
the expression given in section V-C. Each paths are formed of $n_{1}$ blips
of the first path, and $n_{2}$ blips of the second path, such that no blips
are overlapping. This approximation corresponds to the dilute-blip
approximation for the single-spin influence functional and means that
we keep only the direct interaction (${\cal J}=K_{zz}+\bar{\epsilon}$)
in the interspin functional $\tilde{G}_{n_{1}n_{2}}$. All the other
terms vanish by the usual arguments of the dilute-blip approximation.

Blips of a given type are well ordered amongst themselves. What changes from
one configuration to the other is the ordering between blips of different type.
Therefore, the summation over the $\{\zeta_{1j}\}$ and $\{\zeta_{2k}\}$ 
in Eq. (\ref{somme}) is
independent of the configuration and can be performed immediately. It
brings a factor 
\begin{eqnarray}
& & 2^{n_{1}} \prod_{j=1}^{n_{1}} \cos \left[ 
\eta_{1 j-1}Q_{1}^{(1)}(t_{2j}-t_{2j-1})
- \eta_{2P} {\cal J}(t_{2j}-t_{2j-1}) \right] \nonumber \\
&\times & 2^{n_{2}} \prod_{k=1}^{n_{2}} \cos \left[ 
\eta_{2k-1} Q_{1}^{(2)}(u_{2k}-u_{2k-1})
- \eta_{1P} {\cal J}(u_{2k}-u_{2k-1}) \right]
\end{eqnarray}
where $\eta_{1P}$ and $\eta_{2P}$ are the charges of the sojourns
overlapping the blips $\zeta_{2k}$ and $\zeta_{1j}$ respectively.
 
With the charges of the blips removed from the problem, we will refer to
the blips by the charge of the sojourns immediately preceding them.
The fact that the blips are not overlapping then allows the use of the
Laplace transform to perform the summation, just as in the single
spin-boson problem \cite{leggettB}. 
To keep track of the different indices involved in the summation, it is
convenient to define a $2 \times 2$ matrix ${\bf g}^{(\alpha)}(\lambda)$
with
components $(g^{(\alpha)})_{\tau}^{\sigma}$ as \footnote{We thank E.
Mueller for pointing out this matrix method to us. The complete path
summation was first done using a more complicated combinatorial method.}
\begin{equation}
{\bf g}^{(\alpha)} = \left( \begin{array}{cc}
(g^{(\alpha)})_{+}^{+} & (g^{(\alpha)})_{-}^{+} \\
(g^{(\alpha)})_{+}^{-} & (g^{(\alpha)})_{-}^{-}
\end{array} \right)
\end{equation}
with the definition
\begin{equation}
(g^{(\alpha)}(\lambda))_{\tau}^{\sigma}  =
-\frac{\Delta_{\alpha}^{2}}{2\lambda}
\int_{0}^{\infty} dt \,
e^{-\lambda t} e^{- Q_{2}^{(\alpha)}(t)} \cos [ 
\tau {\cal J} t -  \sigma
Q_{1}^{(\alpha)}(t) ]
\end{equation}
The function $(g^{(\alpha)})_{\tau}^{\sigma}$ represents the
contribution of
a blip of type-$\alpha$. The index $\sigma$ refers to the charge of the
sojourn on the path $\alpha$ immediately preceding the blip while the
index
$\tau$ refers to the charge of the sojourn overlapping the blip.

The contribution of the blips to a given path will now be expressed as a
product of the components of ${\bf g}^{(\alpha)}(\lambda)$.
What remains to be done is the sum over the configurations and the set
of the $\{\eta_{1j}\}$ and $\{\eta_{2k}\}$. This must be performed
simultaneously since the summation over the $\eta$ gives a different
result
from one configuration to the other.

We will give a detailed explanation of the procedure to calculate
$P_{\up \up}(t)$ and simply state what needs to be done differently in
calculating the three other probabilities.
The basic building blocks of the summation are the chains of coupled
``clusters'' of blips. The clusters are defined as the ensemble of
successive
blips of the same type $\alpha$, successive meaning that they are all
overlapping the {\em same} sojourn of charge $\eta_{\beta,P}$. This
corresponds
to a single spin-boson system evolving in a {\em static} bias
$\eta_{\beta,P} {\cal J}$. All the different
clusters are
then linked together by the charge of the sojourn preceding the first
blip
of a given cluster. $P_{\up \up}(\lambda)$ then consists in a summation
over
all the possible chains of clusters, each cluster containing all the
possible
number of successive blips. This is represented schematically on Fig.
(\ref{chainfig}) and Fig. (\ref{clusterfig}).

The complete summation over a cluster is easy since it is equivalent to
the
spin boson problem. The total contribution of a cluster of type
$\alpha$, whose
preceding sojourn has a charge $\sigma$ and overlapping a sojourn of
charge $\tau$ is then
\begin{eqnarray}
(\tilde{g}^{(\alpha)})_{\tau}^{\sigma} &=& \sum_{ \{ \sigma_{j} \} }
\sum_{0}^{\infty} \left[ \prod_{j=0}^{n}
(g^{(\alpha)})_{\tau}^{\sigma_{j}}
\right] \nonumber \\
&=& (g^{(\alpha)})_{\tau}^{\sigma}
\frac{1}{1-(g^{(\alpha)})_{\tau}^{+} - (g^({\alpha}))_{\tau}^{-}}
\label{cluster}
\end{eqnarray}
Notice that the 
expression $(\tilde{g}^{(\alpha)})_{\tau}^{+} -
(\tilde{g}^{(\alpha)})_{\tau}^{-}$ is nothing but the function
$P(\lambda)=P_{\up}(\lambda)-P_{\down}(\lambda)$ obtained for the single
biased spin-boson system in a static bias $ \tau {\cal J}$
within the dilute-blip approximation.

The summation of the chains of clusters is now straightforward. Let us
refer
to the contribution of a chain of clusters beginning with a cluster of
type $\alpha$ and ending with a cluster of type $\beta$, with $n$
clusters of
type $\alpha$ as $C_{n}^{\alpha,\beta}$. The summation over the charges
of
the sojourn is included in this notation. The complete probability
summation
can be expressed as
\begin{equation}
P_{\up \up} = \frac{1}{\lambda} +
\frac{1}{\lambda} \sum_{n=1}^{\infty} C_{n}^{(1-1)} +
\frac{1}{\lambda} \sum_{n=1}^{\infty} C_{n}^{(2-2)} +
\frac{1}{\lambda} \sum_{n=1}^{\infty} C_{n}^{(1-2)} +
\frac{1}{\lambda} \sum_{n=1}^{\infty} C_{n}^{(2-1)}
\end{equation}
where the factor $1/\lambda$ comes from the definition of the Laplace
transformation. Let us consider in detail a particular contribution
$P^{\alpha-\beta} \equiv \sum C_{n}^{\alpha-\alpha}$.
It is composed of a sum over a product of
$2n-2$ successive clusters (n $\alpha$-clusters and $n-2$
$\beta$-clusters).
No clusters of the same type are adjacent. Before the summation over the
charges, the product is thus of the form
$(\tilde{g}^{(\alpha)})_{\tau_{1}}^{\sigma_{1}}
(\tilde{g}^{(\beta})_{\tau_{2}}^{\sigma_{2}} ...
(\tilde{g}^{(\alpha)})_{\tau_{j}}^{\sigma_{j}}
(\tilde{g}^{(\beta})_{\tau_{j+1}}^{\sigma_{j+1}}
... (\tilde{g}^{(\alpha)})_{\tau_{2n-2}}^{\sigma_{2n-2}}$.
However, since they are successive, there is the restriction
$\sigma_{j+1} = \tau_{j}$. It is then possible to write
\begin{equation}
P^{(\alpha-\beta)} = \sum_{n=0}^{\infty}
(\tilde{g}^{(\alpha)})_{+}^{+}
(\tilde{g}^{(\beta)})_{\sig_{1}}^{+}
\prod_{j=1}^{n}
\sum_{ \{ \sigma_{j} \} }
(\tilde{g}^{(\alpha)})_{\sigma_{2j}}^{\sigma_{2j-1}}
(\tilde{g}^{(\beta)})_{\sigma_{2j+1}}^{\sigma_{2j}}
\end{equation}
where for each $n$, there is a restriction $\sigma_{2n+1} \equiv 1$.
This is then nothing but 
a particular element of the multiplication of two matrices:
\begin{equation}
P^{(\alpha-\beta)} = (\tilde{g}^{(\alpha)})_{+}^{+}
\left[ \tilde{{\bf g}}^{(\beta)}
(1-\tilde{{\bf g}}^{(\alpha)} \tilde{{\bf g}}^{(\beta)})^{-1}
\right]_{+}^{+}
\end{equation}
The similar expression for a chain where the two limiting clusters are
of the
same type is
\begin{eqnarray}
P^{(\alpha-\alpha)} &=& (\tilde{g}^{(\alpha)})_{+}^{+}
\sum_{n=1}^{\infty}
\prod_{j=1}^{n}
\sum_{ \{ \sigma_{j} \} }
(\tilde{g}^{(\beta)})_{\sigma_{2j-1}}^{\sigma_{2j-2}}
(\tilde{g}^{(\alpha)})_{\sigma_{2j}}^{\sigma_{2j-1}}
\nonumber \\
& = & (\tilde{g}^{(\alpha)})_{+}^{+}
\left[
(1-\tilde{{\bf g}}^{(\beta)} \tilde{{\bf g}}^{(\alpha)})^{-1}
\right]_{+}^{+}
\end{eqnarray}
The complete probability $P_{\up \up}(\lambda)$ is thus given by
\begin{eqnarray}
P_{\up \up}(\lambda) &=& \frac{1}{\lambda} +
 \frac{1}{\lambda} (\tilde{g}^{(1)})_{+}^{+}  \left[
(1-\tilde{{\bf g}}^{(2)} \tilde{{\bf g}}^{(1)})^{-1} \right]_{+}^{+} +
 \frac{1}{\lambda} (\tilde{g}^{(2)})_{+}^{+}  \left[
(1-\tilde{{\bf g}}^{(1)} \tilde{{\bf g}}^{(2)})^{-1} \right]_{+}^{+} +
\nonumber \\
& &  (\tilde{g}^{(1)})_{+}^{+}
\left[ \tilde{{\bf g}}^{(2)}
(1-\tilde{{\bf g}}^{(1)} \tilde{{\bf g}}^{(2)})^{-1} \right]_{+}^{+} +
(\tilde{g}^{(2)})_{+}^{+}
\left[ \tilde{{\bf g}}^{(1)}
(1-\tilde{{\bf g}}^{(2)} \tilde{{\bf g}}^{(1)})^{-1} \right]_{+}^{+}
\label{puubiaslambda}
\end{eqnarray}

The other probabilities are then obtained in a straightforward manner, 
they simply correspond to taking different matrix elements in Eq.
(\ref{puubiaslambda}).
We obtain:
\begin{eqnarray}
P_{\down \down}(\lambda) &=&
 \frac{1}{\lambda} (\tilde{g}^{(1)})_{+}^{+}  \left[
\tilde{{\bf g}}^{(2)}
(1-\tilde{{\bf g}}^{(1)} \tilde{{\bf g}}^{(2)})^{-1} \right]_{-}^{+} +
 \frac{1}{\lambda} (\tilde{g}^{(2)})_{+}^{+}  \left[
\tilde{{\bf g}}^{(1)}
(1-\tilde{{\bf g}}^{(2)} \tilde{{\bf g}}^{(1)})^{-1} \right]_{-}^{+} +
\nonumber \\
& &  \frac{1}{\lambda} (\tilde{g}^{(1)})_{+}^{+}
\left[ \tilde{{\bf g}}^{(2)} \tilde{{\bf g}}^{(1)}
(1-\tilde{{\bf g}}^{(2)} \tilde{{\bf g}}^{(1)})^{-1} \right]_{-}^{+} +
\frac{1}{\lambda} (\tilde{g}^{(2)})_{+}^{+}
\left[ \tilde{{\bf g}}^{(1)} \tilde{{\bf g}}^{(2)}
(1-\tilde{{\bf g}}^{(1)} \tilde{{\bf g}}^{(2)})^{-1} \right]_{-}^{+}
\label{pddbiaslambda}
\end{eqnarray}
Notice that the matrices now involve $[ \ldots ]_{-}^{+}$, since the
final
sojourns must have a charge of $-1$. There is also the factor
$1/\lambda$
missing, since there must be at least one blip in each path, to go from
the
state $| \up \up \rangle$ to $| \down \down \rangle$.

Finally, for the last two remaining probabilities,
\begin{eqnarray}
P_{\up \down}(\lambda) &=&
 \frac{1}{\lambda} (\tilde{g}^{(1)})_{+}^{+}  \left[ \tilde{{\bf
g}}^{(2)}
(1-\tilde{{\bf g}}^{(1)} \tilde{{\bf g}}^{(2)})^{-1} \right]_{+}^{+} +
 \frac{1}{\lambda} (\tilde{g}^{(2)})_{+}^{+}  \left[
(1-\tilde{{\bf g}}^{(1)} \tilde{{\bf g}}^{(2)})^{-1} \right]_{+}^{+} +
\nonumber \\
& &  \frac{1}{\lambda} (\tilde{g}^{(1)})_{+}^{+}
\left[ \tilde{{\bf g}}^{(2)} \tilde{{\bf g}}^{(1)}
(1-\tilde{{\bf g}}^{(2)} \tilde{{\bf g}}^{(1)})^{-1} \right]_{-}^{+} +
\frac{1}{\lambda} (\tilde{g}^{(2)})_{+}^{+}
\left[ \tilde{{\bf g}}^{(1)}
(1-\tilde{{\bf g}}^{(2)} \tilde{{\bf g}}^{(1)})^{-1} \right]_{-}^{+}
\label{pudbiaslambda}
\end{eqnarray}
with $P_{\down \up}(\lambda)$ simply related to $P_{\up \down}(\lambda)$
by the
substitution $ (1) \leftrightarrow (2) $.

These are obviously fairly complicated expressions, despite their
compactness.
However, the matrices
${\bf g}^{(\alpha)}$ are $2 \times 2$ so that it is relatively
easy to expand. Notice first that the denominator of $P_{\tau_1 
\tau_2}(\lambda)$,
which determines to poles in $\lambda$ and consequently the dynamics is
given as
\begin{equation}
1- Tr (\tilde{{\bf g}}^{(1)} \tilde{{\bf g}}^{(2)} )
+ Det (\tilde{{\bf g}}^{(1)} \tilde{{\bf g}}^{(2)} )
\end{equation}
This denominator reduces to a product of polynomials of degree
2 in $\lambda$ and it is then possible to simplify
$P_{\tau_1 \tau_2} (\lambda)$ further.

At this point, we can revert to the notation of Leggett et al. and use
the functions $g_{\beta} (\lambda)$ and $h_{\beta} (\lambda$ defined by
Eq. (\ref{definitions}) to simplify $P_{\up \up} (\lambda)$ 
to Eq. (\ref{puulambda}) of section V.

Similarly, we obtain for $P{\down \down} (\lambda)$ 
\begin{eqnarray}
\label{pddlambda}
P_{\down \down}(\lambda) &=& -\frac{1}{4} \left[ g_{1} \left(
1+ \frac{h_{1}}{g_{1}} \right) + g_{2} \left( 1+ \frac{h_{2}}{g_{2}} \right)
\right]  \frac{1}{\lambda} \, \frac{1}{\lambda + g_{1} +g_{2} }  \nonumber \\
& + &  \frac{1}{4} \left[ g_{1} \left( 1+ \frac{h_{1}}{g_{1}} \right) +
g_{2} \left( 1+ \frac{h_{2}}{g_{2}} \right) \right] \,
\frac{1}{\lambda^{2} + \lambda(g_{1}+g_{2}) +g_{1}g_{2} (1-\frac{h_{1}h_{2}}{
g_{1}g_{2} })  } \\
& + & \frac{1}{2}g_{1}g_{2} \left[ 1 - \frac{h_{1}h_{2}}{g_{1}g_{2}} \right]
\frac{1}{\lambda} \, \frac{1}{\lambda^{2} + \lambda(g_{1}+g_{2})
+g_{1}g_{2} (1-\frac{h_{1}h_{2}}{g_{1}g_{2}}) } \, . \nonumber
\end{eqnarray}
 
While the last two remaining probabilities simplify to 
 
\begin{eqnarray}
\label{pudlambda}
P_{\up \down}(\lambda) &=& \frac{1}{4} \left[ g_{1} \left(
1+ \frac{h_{1}}{g_{1}} \right) + g_{2} \left( 1+ \frac{h_{2}}{g_{2}} \right)
\right]  \frac{1}{\lambda} \, \frac{1}{\lambda + g_{1} +g_{2} }   \\
& - &  \frac{1}{4} \left[ g_{1} \left( 1+ \frac{h_{1}}{g_{1}} \right) -
g_{2} \left( 1+ \frac{h_{2}}{g_{2}} \right) \right]
\frac{1}{\lambda^{2} + \lambda(g_{1}+g_{2}) +g_{1}g_{2} (1-\frac{h_{1}h_{2}}{
g_{1}g_{2} })  } \, . \nonumber
\end{eqnarray}
 
\begin{eqnarray}
\label{pdulambda}
P_{\down \up}(\lambda) &=& \frac{1}{4} \left[ g_{1} \left(
1+ \frac{h_{1}}{g_{1}} \right) + g_{2} \left( 1+ \frac{h_{2}}{g_{2}} \right)
\right]  \frac{1}{\lambda}  \frac{1}{\lambda + g_{1} +g_{2} }   \\
& + &  \frac{1}{4} \left[ g_{1} \left( 1+ \frac{h_{1}}{g_{1}} \right) -
g_{2} \left( 1+ \frac{h_{2}}{g_{2}} \right) \right]
\frac{1}{\lambda^{2} + \lambda(g_{1}+g_{2}) +g_{1}g_{2} (1-\frac{h_{1}h_{2}}{
g_{1}g_{2} })  } \, . \nonumber
\end{eqnarray}
For identical systems, $P_{\up \down}=P_{\down \up}$, as should be expected, 
but the form of these probabilities also become quite simple, setting
$g=g_{1}=g_{2}$ and $h=h_{1}=h_{2}$, the probabilities are 
\begin{equation}
P_{\up \down}=P_{\down \up} = \frac{1}{2 \lambda} (g+h) 
\frac{1}{\lambda + 2 g} \; \; \; \; \; \; \; \; \mbox{(Identical systems)}
\end{equation}

\section{Mutual Coherence Regime: Identical Systems}
\resetcounters

In this appendix, we discuss the more tedious details  pertaining to the
mutual coherence case. We start with the denominators in $P_{\tau_1
\tau_2} (\lambda)$, which yields the oscillation frequencies and decay
rates and then give the form of the coefficients 
$C_ij$ appearing in Eq. (\ref{oscilluu}), Eq. (\ref{oscilldd}) and 
Eq. (\ref{oscillud}). All the following expression are of course quite
messy, although straightforward to obtain. Furthermore, they are
obtained by working to lowest order in $a = \pi \alpha T$. 
It is nevertheless important to get an idea of the value of relaxation
rates and oscillation frequencies as well as the relative weight of the
different terms in the overall probability function. 

For two equivalent spins, the denominators of the Laplace transforms 
$P_{\tau_1 \tau_2} (\lambda)$, Eq. (\ref{denom1}) and Eq. (\ref{denom2})
are now
\begin{equation}
\lambda + 2 g(\lambda)
\end{equation}
and
\begin{equation}
(\lambda + g(\lambda) + h(\lambda))(\lambda + g(\lambda) -h(\lambda)
\end{equation}
With the appropriate forms of $g(\lambda)$ and $h(\lambda)$ inserted,
the equations for the poles of the Laplace transform are then
\begin{equation}
\lambda^3 + 4 a \lambda^2 + 
\lambda (4 a^2 + {\cal J}^2 + \bar{\Delta}^2) + 2 a \bar{\Delta}^2
\end{equation}
\begin{equation}
\lambda^3 + 4 a \lambda^2 + 
\lambda (4 a^2 + {\cal J}^2 + \bar{\Delta}^2) + \bar{\Delta}^2 (2a \pm
{\cal J})
\end{equation}
The structure of these equations is completely similar to the equation 
obtained 
by Weiss and Wollensack for the single biased spin-boson system
\cite{weissA}. The
equations have two imaginary roots if $T > \Delta/\alpha$, which
corresponds to oscillating terms in $P_{\tau_1 \tau_2} (t)$. The 
form of the decay rates and oscillation frequency can be expressed 
easily in terms of the relaxation rates and
oscillation frequency of the single spin-boson problem, 
Eq. (\ref{mutualsingle}) and yield the values cited in the text. 

We can now discuss the values of the coefficients $C_{ij}$. 
The coefficients $C_{2}$, $C_{3}$ and $C_{4}$ are related to the dynamics of
the two systems together. $C_{2}$ is the coefficient of the pure relaxation
term, with value
\begin{equation}
C_{2} = \frac{\bar{\Delta}^{2}/2}{(\Gamma_{R}-\Gamma)^{2}+\mu^{2}}
\left( 1 - \frac{8aA_{+}}{\Gamma_{R}} \right)
\end{equation}
In the limit ${\cal J} \gg \bar{\Delta}$ this coefficient is of order
$O(1)$, with limiting value
\begin{equation}
C_{2} \sim A_{-}  \; \; \; \; \; ({\cal J} \gg \bar{\Delta})
\end{equation}
which makes the connection with the overdamped relaxation of the correlated
phase. The coefficients $C_{3}$ and $C_{4}$ corresponds to oscillating terms
with frequency $\mu$. It simpler to write them directly as a sum of cosine and
sine since it avoids the complication of introducing a phase in the 
oscillations. We obtain 
\begin{equation}
C_{3}= \frac{\bar{\Delta}^{2}}{2} 
\frac{\Gamma^{2}+\mu^{2} + 8aA_{+}(\Gamma_{R}-2\Gamma)}{(\Gamma^{2}
+\mu^{2}) ((\Gamma-\Gamma_{R})^{2}+\mu^{2})}
\end{equation}
\begin{equation}
C_{4} = \frac{\bar{\Delta}^{2}}{2} 
\frac{8aA_{+}(\Gamma(\Gamma_{R}-\Gamma)+\mu^{2})-
(\Gamma_{R}-\Gamma)(\Gamma^{2}+\mu^{2})}{(\Gamma^{2}+\mu^{2})
((\Gamma-\Gamma_{R})^{2}+\mu^{2})}
\end{equation}
The main point about these coefficients is their smallness. In the limit
${\cal J} \gg \bar{\Delta}$, they are
\begin{equation}
C_{3} \sim \frac{\bar{\Delta}^{2}}{2{\cal J}^{2}}  
\end{equation}
\begin{equation}
C_{4} \sim \frac{\bar{\Delta}^{2}(\pi \alpha T)}{{\cal J}^{3}} 
\end{equation}

We now get to the coefficients pertaining to the individual dynamics of the
two spins. In this case, the Laplace transform of $P_{\tau_{1} \tau_{2}}(t)$
includes terms with denominator of the form
$(\lambda + g(\lambda))^{2} - h^{2}(\lambda)$, giving rise to very complex
coefficients. There is the apparition of two individual relaxation rate
$\gamma_{R \pm}$, with a coefficient $C_{5 \sig}^{\tau}$, where the index
$\tau= \pm 1$ refers to $P_{\up \up}(t)$ ($\tau = +1$) and 
$P_{\down \down}(t)$ ($\tau = -1$). Its expression is
\begin{equation}
C_{5\sig}^{\tau}= \frac{\bar{\Delta}^{2}}{2} \frac{\gamma_{R\sig}-8aA_{+}}{
(\gamma_{R-\sig}-\gamma_{R\sig})( \nu_{\sig}^{2} + (\gamma_{\sig}-
\gamma_{R\sig})^{2})
(\nu_{-\sig}^{2} + (\gamma_{-\sig}-\gamma_{R\sig})^{2})}
\left( {\cal J}^{2}+(2a-\gamma_{R\sig})^{2} + \tau \frac{\bar{\Delta^{2}}}
{2 \gamma_{R\sig}}(8aA_{-} - \gamma_{R\sig} \right)
\end{equation}
In the limit of interest to us, ${\cal J} \gg \Delta$, 
$C_{5\sig}^{\tau}$ is given by
\begin{equation}
C_{5 \sig}^{\tau} = \frac{1}{2} A_{+} \frac{1}{A_{-\sig}-A_{\sig}} 
[1 + \frac{\tau}{2} \frac{A_{-}}{A_{+}} ]
\end{equation}
Which is of order $O(1)$, as is expected since we are close to the 
totally overdamped limit.

The remaining terms are the oscillations of frequency $\nu_{\pm}$ and
damped at rates $\gamma_{\pm}$, with 
coefficients $C_{6\sig}^{\tau}$ and $C_{7\sig}^{\tau}$, where $\tau$ refers
once again to $P_{\up \up}$ or $P_{\down \down}$. Of course the exact 
form of these coefficients is fairly complex: 
\begin{equation}
C_{6\sig}^{\tau} = \frac{\bar{\Delta}^{2}}{2\nu_{\sig}|D^{\sig}|^{2}} 
\left( N_{2}^{\sig} D_{1}^{\sig} - D_{2}^{\sig} N_{1}^{\sig} + \tau
\frac{\bar{\Delta}^{2}}{\gamma_{\sig}^{2} + \nu_{\sig}^{2}}
(N_{4}^{\sig}D_{1}^{\sig}-N_{3}^{\sig}D_{2}^{\sig}) \right)
\end{equation}
 
\begin{equation}
C_{7\sig}^{\tau} = \frac{\bar{\Delta}^{2}}{2\nu_{\sig}|D^{\sig}|^{2}} 
\left( N_{1}^{\sig} D_{1}^{\sig} + D_{2}^{\sig} N_{2}^{\sig} + \tau
\frac{\bar{\Delta}^{2}} {\gamma_{\sig}^{2} +
\nu_{\sig}^{2}} (N_{3}^{\sig}D_{1}^{\sig}+N_{4}^{\sig}D_{2}^{\sig}) \right)
\end{equation}
The different terms appearing in the numerators and denominators 
are given by, 
\begin{eqnarray}
N_{1}^{\sig} &=& [ -\gamma_{\sig} + 8aA_{+} ][(2a-\gamma_{\sig})^{2}-
\nu_{\sig}^{2} + {\cal J}^{2} ] -2 \nu_{\sig}^{2} (2a-\gamma_{\sig})
\nonumber \\
& \sim & -8 a \bar{\Delta}_{\sig}^{2} A_{+} 
\end{eqnarray}
 
\begin{eqnarray}
N_{2}^{\sig} &=& 2 \nu_{\sig} (2a-\gamma_{\sig}) [ -\gamma_{\sig} + 8aA_{+} ]
+ \nu_{\sig} [ (2a-\gamma_{\sig})^{2} -\nu_{\sig}^{2} + {\cal J}^{2} ]
\nonumber \\
& \sim & - {\cal J} \bar{\Delta}_{\sig}^{2}
\end{eqnarray}
 
\begin{eqnarray}
N_{3}^{\sig} &=& -\gamma_{\sig} [ (2a-\gamma_{\sig})^{2} -\nu_{\sig}^{2}
- 4a^{2}\tanh^{2}({\cal J}/2T)] + 2 \nu_{\sig}^{2}(2a-\gamma_{\sig})
\nonumber \\
& \sim & - \gamma_{\sig} {\cal J}^{2} \sim 8 {\cal J}^{2} a A_{+}
\end{eqnarray}
 
\begin{eqnarray}
N_{4}^{\sig} &=& -\nu_{\sig} [  (2a-\gamma_{\sig})^{2} -\nu_{\sig}^{2}
- 4a^{2}\tanh^{2}({\cal J}/2T)] - 2\gamma_{\sig}\nu_{\sig}
(2a-\gamma_{\sig})
\nonumber \\
& \sim & {\cal J}^{3}
\end{eqnarray}
The dominant term in then $N_{4\sig}$, and the difference between $N_{4+}$ and
$N_{4-}$ can be neglected at this level.
 
The remaining terms are given by
\begin{eqnarray}
D_{1}^{\sig} &=& [(\gamma_{-}-\gamma_{+})^{2}+\nu_{-\sig}^{2} -\nu_{\sig}^{2}]
[ (\gamma_{R+}-\gamma_{\sig})(\gamma_{R-}-\gamma_{\sig}) -\nu_{\sig}^{2} ]
-2 \nu_{\sig}^{2} (\gamma_{-\sig}-\gamma_{\sig})[ \gamma_{R+}+\gamma_{R-} -2
\gamma_{\sig} ] \nonumber \\
& \sim & -4 \bar{\Delta}^{2} {\cal J}^{2} (A_{-\sig}-A_{\sig})
\end{eqnarray}
 
\begin{eqnarray}
D_{2}^{\sig} &=& 2 \nu_{\sig} (\gamma_{-\sig}-\gamma_{\sig})
[ (\gamma_{R+}-\gamma_{\sig})(\gamma_{R-}-\gamma_{\sig}) -\nu_{\sig}^{2} ]
+ \nu_{\sig}  [ \gamma_{R+}+\gamma_{R-} -2 \gamma_{\sig} ]
[ (\gamma_{-}-\gamma_{+})^{2} + \nu_{-\sig}^{2} -\nu_{\sig}^{2} ]
\nonumber \\
& \sim & -8 a {\cal J} \bar{\Delta}^{2}  (A_{-\sig}-A_{\sig})
\end{eqnarray}
 
\begin{eqnarray}
|D^{\sig}|^{2} &=& [(\gamma_{R+}-\gamma_{\sig})^{2}+\nu_{\sig}^{2}]
[(\gamma_{R-}-\gamma_{\sig})^{2}+\nu_{\sig}^{2}]
[ (\gamma_{-}-\gamma_{+})^{4} + 2(\nu_{-}^{2}+\nu_{+}^{2})
(\gamma_{-}-\gamma_{+})^{2} +(\nu_{+}^{2}-\nu_{-}^{2})^{2}]
\nonumber \\
& \sim & 16  {\cal J}^{4}  \bar{\Delta}^{4}  (A_{-\sig}-A_{\sig})^{2}
\end{eqnarray}

Thus we see that both $C_{6\sig}^{\tau}$ are quite small, so that in this
case again, not only is the damping of the oscillations quite rapid compared
to the general decay of the probabilities, but they have a very small amplitude
so that the oscillations takes place against are very large 
``background''.

\section{Mutual Coherence Regime: Overdamped plus Underdamped}
\resetcounters

The Laplace inversion in the Overdamped plus Underdamped case is very 
similar in spirit to the fully Underdamped case. 
The global denominator Eq. (\ref{denom1}) is given by
\begin{equation}
(\lambda + \Gamma_{2})((2a_{1} + \lambda)^2 + {\cal J}^2) + 
\bar{\Delta}_{1}^{2} ( 2 a_{1} + \lambda)
\end{equation}
and by defining $\bar{a} \equiv a_{1} - \Gamma_{2}/2$, the
roots for the variable $\lambda + \Gamma_{2}$ are exactly given by
Eq. (\ref{mutualsingle}) with $a$ replaced by $\bar{a}$. 

In the text, we have argued that we could
separate the individual denominator, Eq. (\ref{denom2}) into the
individual products $(\lambda + g_{2})(\lambda + g_{1})$, ie., we
neglect the correlation between the two spins
when calculating decay rates and oscillation
frequency. It is then clear that we can directly use Eq. (\ref{mutualsingle})
for the underdamped spin.

The decay rates
$\Gamma_{2}+\bar{\gamma}_{R}$ and 
$\Gamma_{2}+\bar{\gamma}$ are fully expressed as
\begin{equation}
\Gamma_{2}+\bar{\gamma}_{R} = \gamma_{R,s} + \Gamma_{2} 
\frac{{\cal J}^{2}}{{\cal J}^{2} + \bar{\Delta}^{2}} \sim 
\gamma_{R,s}+\Gamma_{2}
\end{equation}
and
\begin{equation}
\Gamma_{2}+\bar{\gamma} = 2 a - \frac{1}{2} \gamma_{R,s} (\bar{a},
{\cal J}^2,\bar{\Delta}_{1}^{2}) \sim 2a - (a -\Gamma_{2}/2)
\frac{\bar{\Delta}_{1}^{2}}{{\cal J}^2}
\end{equation} 
where the last forms are in the limit ${\cal J} \gg \Delta_{1}$. The 
coefficient $C_{2}$ is given by
\begin{equation}
C_{2} = \frac{-1}{(\bar{\gamma}_{R}+\Gamma_{2})}
\frac{1}{((\bar{\gamma}_{1}-\bar{\gamma}_{R})^{2}+\bar{\nu}_{s}^{2})}
\left[ \Gamma_{2}A_{+}(
(2\bar{a}_{1}-\bar{\gamma}_{R})^{2}+{\cal J}^{2}) +
\frac{\bar{\Delta}_{1}^{2}}{4}
(8a_{1}A_{+}-\bar{\gamma}_{R} +\Gamma_{2} {\cal J}/2T ) \right]
\end{equation}
which, in the limit ${\cal J} \gg \Delta_{1}$ reduces to
\begin{equation}
C_{2} \sim - \frac{\Gamma_{2}A_{+}}{\gamma_{R,s}+\Gamma_{2}}
\end{equation}
which is of $O(1)$ is $\Gamma_{2} \gg \gamma_{R,s}$ and negligible
otherwise. The coefficients relating to the terms oscillating at a 
frequency $\bar{\mu}$ are 
\begin{equation}
C_{3} = \frac{1}{\bar{\mu}} \frac{1}{|D^{2}|} [N_{1}D_{2}+N_{2}D_{1}]
\end{equation}
and 
\begin{equation}
C_{4} = -\frac{1}{\bar{\mu}} \frac{1}{|D^{2}|} [N_{1}D_{1}-N_{2}D_{2}]
\end{equation}
given in a form similar to the coefficients in the underdamped case. The
various factors are: 
\begin{equation}
N_{1} =  \Gamma_{2} A_{+} [(2a-\gamma)^{2}-\bar{\mu}^{2}
+ {\cal J}^{2}] + \frac{\bar{\Delta}_{1}^{2}}{4} (8a_{1}A_{+} -
\gamma_{1} + \Gamma_{2} {\cal J}/2T ) \sim 
\frac{1}{4} \bar{\Delta}^{2} [(a{\cal J}/T - \Gamma_{2}]
\end{equation}
 
\begin{equation}
N_{2} = -2 \Gamma_{2} A_{+} \bar{\mu} (2a_{1}-\gamma_{1})
+ \frac{\bar{\Delta}_{1}^{2}}{4} \nu_{1} \sim 
\frac{1}{4} \bar{\Delta}^{2} {\cal J}
\end{equation}

The remaining terms are given by
\begin{equation}
D_{1}= (\bar{\gamma}-\Gamma_{2})(\bar{\gamma}_{R}-\bar{\gamma})]
+\bar{\mu}^{2} \sim {\cal J}^{2}
\end{equation}
 
\begin{equation}
D_{2}=  \bar{\mu} (2\bar{\gamma}-\bar{\gamma}_{R}
+ \Gamma_{2}) \sim 2 a {\cal J} 
\end{equation}
 
\begin{equation}
|D^{2}|= [(\bar{\gamma}+\Gamma_{2})^{2}+\bar{\mu}^{2}]
[(\bar{\gamma}_{R}-\gamma_{1})^{2}+\bar{\mu}^{2}] \sim {\cal J}^{4}
\end{equation}
 
The coefficients $C_{5}=C(\Gamma_{2},\gamma_{1,R})$ and
$C_{6}=C(\gamma_{1,R},\Gamma_{2})$ where the function
$C(x,y)$ is defined as
\begin{equation}
C(x,y)=\frac{1}{(y-x)} \frac{1}{((\gamma_{1}-x)^{2}+\nu_{1}^{2})}
\left[ \Gamma_{2}A_{+}( (2a_{1}-x)^{2}+{\cal J}^{2}) + \frac{1}{4}
\bar{\Delta}_{1}^{2} (x+8a_{1}A_{-}) \right]
\end{equation}

The coefficients of the oscillating terms are defined similarly to the
Equivalent Spin case
\begin{equation}
C_{7} = \frac{-1}{\nu_{1}}
\frac{1}{\gamma_{1}^{2}+\nu_{1}^{2}} \frac{1}{D^{2}}
[N_{3}D_{2}+N_{4}D_{1}] -\frac{1}{\nu_{1}} \frac{1}{D^{2}}
[N_{1}D_{2}+N_{2}D_{1}]
\end{equation}

\begin{equation}
C_{8} = \frac{1}{\nu_{1}}
\frac{1}{\gamma_{1}^{2}+\nu_{1}^{2}} \frac{1}{D^{2}}
[N_{3}D_{1}-N_{4}D_{2}] +\frac{1}{\nu_{1}} \frac{1}{D^{2}}
[N_{1}D_{1}-N_{2}D_{2}]
\end{equation}

With numerators and denominators given by

\begin{equation}
|D^{2}|= [(\gamma_{1,R}-\gamma_{1})^{2}+\nu_{1}^{2}]
[(\Gamma_{2}-\gamma_{1})^{2}+\nu_{1}^{2}]
\end{equation}

\begin{equation}
N_{1} = \Gamma_{2} A_{+} [(2a_{1}-
\gamma_{1})^{2}-\nu_{1}^{2}
+ {\cal J}^{2} + \frac{\bar{\Delta}_{1}^{2}}{4} (8a_{1}A_{+} - 
\gamma_{1})
\end{equation}

\begin{equation}
N_{2} = -2 \Gamma_{2} A_{+} \nu_{1} (2a_{1}-\gamma_{1})
+ \frac{\bar{\Delta}_{1}^{2}}{4} \nu_{1}
\end{equation} 

\begin{equation}
N_{3} = \frac{\bar{\Delta}_{1}^{2}}{2} \Gamma_{2} 
[ \nu_{1}^{2} - \gamma_{1}
(2a_{1}-\gamma_{1}) ]
\end{equation}

\begin{equation}
N_{4} = \Gamma_{2} \bar{\Delta}_{1}^{2} \nu_{1} a_{1}
\end{equation}

\newpage

{\bf Figure Captions} 

\vspace{1cm}

Fig1: The probabilities $P_{\tau_{1} \tau_{2}}(t)$ for a system of spins,
coupled by $J_{0}$, to start in the state $| \up \up \rangle $ at time
$t=0$, and finish at time $t$ in state $ | \tau_{1} \tau_{2} \rangle $. We
assume $\Delta_{2}=2.5\Delta_{1}$, and a fairly strong coupling $|J_{0}| =
5 \Delta_{1}$. $P_{\tau_{1} \tau_{2}}(t)$ is plotted as a function of
$t/\Delta_{1}$. Oscillations between the 2 states $| \up \up \rangle$ and
$| \down \down \rangle$ occur at the slow ``beat'' frequency $\Delta_{1}
\Delta_{2} / 2 |J_{0}|$, with very weak high frequency oscillations superposed,
coming from the weak mixing with the states $| \up \down \rangle$ and
$| \down \up \rangle$. These high-frequency oscillations dominate
$P_{\up \down}(t)$ and $P_{\down \up}(t)$, albeit with amplitude reduced by a
factor $ \sim (\Delta_{1}+\Delta_{2})/2 J_{0}$.

\vspace{1cm}

Fig2: The same probabilities as in Fig. 2, but now for weak coupling; we
have $\Delta_{2} =2.5 \Delta_{1}$ again, but now $|J_{0}| =\Delta_{1}/10$.

\vspace{1cm}

Fig3: Diagrammatic interpretation of the equation of motion 
for the density matrix. 
We show equation 4.2 in Feynman diagram form; $J$ is the propagator (or Green
function) for $\rho$. 

\vspace{1cm}

Fig4: Contributions to the single spin influence functional. The label
$\alpha = 1,2$ labels one of the 2 spins, and $\gamma_{\alpha} \equiv
\Gamma_{\alpha} -i \Phi_{\alpha}$, (see Eq. (4.15) and (4.16))
These contributions are exponentiated to give the influence functional
Eq. (4.22).

\vspace{1cm}

Fig5: Contributions to the interaction functional $F_{12}$ (Eq. 4.23)). 
There are only interactions between the path associated with the spin 1 
and that of the spin 2. 
Here, $\gamma_{12} \equiv \Gamma_{12} - i \Phi_{12}$, as
given by Eq. (4.17) and (4.18). 

\vspace{1cm}

Fig6: Example of a typical path for the 2-spin system. The top path is for the 
1st spin, and shows transitions between eigenstates of $\tau_1^z$, at 
times $t_j$; the 
bottom path is for the 2nd spin, with transitions at times $u_k$.

\vspace{1cm}

Fig7: The various regimes in which the PISCES model dynamics are solved for
analytically in the text. We assume in the figure that the 2 spins are
equivalent, ie., that $\Delta^{*}=\Delta_{1}^{*}=\Delta_{2}^{*}$, and
$\alpha_{1}=\alpha_{2}=\alpha$. We also assume weak damping, so that
$\alpha \ll 1$. If $\alpha$ is much larger, ie., $\alpha \sim O(1)$, there is
no mutual coherence phase, and motion in the locked phase (where the 2 spins
rotate rigidly together) is overdamped at any $T$. We do not discuss the
perturbative regime in this paper - in this regime correlations are very
weak between the 2 spins.
If the 2 spins are different (ie., $\alpha_{1} \neq \alpha_{2}$, and/or
$\Delta_{1}^{*} \neq \Delta_{2}^{*}$), then one must draw 2 separates diagrams
of this type, one for each spin - then it is possible for the 2 spins to be
in different phases.
The Mutual Coherence regime is strictly defined by the conditions
$\Delta^{*}/\alpha \gg T \gg ({\cal J}^{2}+\Delta^{* 2})^{1/2}$, but since
we always assume ${\cal J} \gg \Delta^{*}$ in this paper (ie., well away from
the perturbative regime), the inequality
$T \gg ({\cal J}^{2}+\Delta^{* 2})^{1/2}$ is equivalent to the condition
$T \gg {\cal J}$ used in the text.

\vspace{1cm}

Fig8: The various dynamical regimes which describe the motion of the PISCES
system in the locked phase, as functions of $T$ and of the coupling
$\alpha_{c}$ between the locked 2-spin complex and the sea of oscillators.
The overdamped relaxation phase is divided into 2 parts by the ``Toulouse line''
$\alpha_{c}=1/2$. On the ``weak relaxation'' side, the damping rate
$\Gamma_{c}$ decreases with increasing $T$, and is small; on the
``strong relaxation'' side, $\Gamma_{c}$ is large and increases with increasing
$T$. When $T=0$, the motion is still damped for any finite $\alpha_{c}$ (with
oscillations for $ \alpha_{c} < 1/2$), but for $\alpha_{c} > 1$, the system
is completely frozen by the coupling to the oscillators.

\vspace{1cm}

Fig9: Probabilities of occupation for a system in the "correlated relaxation"
regime, with parameters
$\alpha_{1} = 1.5$, $\alpha_{2} =2$, at a temperature such that
$T/\Delta_{1}=100$, $T/\Delta_2=300$ and a ferromagnetic coupling
${\cal J}/2T = -0.02$; the time is expressed in units of inverse
temperature. The system starts at $t=0$ in the state
$ | \uparrow \uparrow \rangle$; $P_{\alpha \beta}$ is then the probability
that at time $t$ the system has $\tau_{1}^{z} = \alpha$ and $\tau_{2}^{z} =
\beta$. We plot the probabilities as a function of $ln(t)$ in order to
clearly show the different relaxation times involved.

\vspace{1cm}

Fig10: Imaginary part of the Fourier transform of $P_{\up \up}(t)$ in the
mutual coherence case. The peaks represent the oscillation frequencies,
and their width is proportional to their damping rates. We use the values
$\bar{\Delta}/{\cal J} = 0.6$, ${\cal J}/T = 0.1$, $T=100$ (in arbitrary
units) and $a=\pi \alpha T = 0.25$. The real frequency $i \lambda$ is in
the same units as the temperature. 

\vspace{1cm}

Fig11: Imaginary part of the Fourier transform of $P_{\up \up}(t)$ in the
Overdamped plus Underdamped case. The parameters are 
$\bar{\Delta}_{1}/{\cal J} = 0.6$, $\bar{\Delta}_{2}/{\cal
J} = 0.1$, $a_1 = 0.25$ and $a_2 = 15$ (where $a_{\beta} = \pi
\alpha_{\beta} T$). Again ${\cal J}/T = 0.1$,
$T=100$.  Notice that we still have $\alpha_2 \ll 1$, although the spin
$2$ is overdamped. The strong broadening of the oscillations
caused by the environment does not allow a clear separation of the two
different oscillation frequencies. 

\vspace{1cm}
Fig12: Paths for $P_{\up \up}$ with $n_{1}=n_{2}=1$ (cf. equation (A2); 
there are 6 possible orderings of the transitions, which must be summed 
over. In Fig 12(a) we also show the states associated with the periods between 
the transitions.

\vspace{1cm}

Fig13: Schematic representation of the path integral for $P_{\up \up}$.
We sum over all possible chains of coupled clusters. The charges of the
boundary sojourns are fixed at $\eta = +1$ by boundary conditions and 
there is a summation over the charges of all the internal sojourns. 

\vspace{1cm}

Fig14: Representation of a cluster, used for the summation of 
$P_{\tau_1 \tau_2}$. In this graph, the charge of the lower path 
sojourn is fixed and constant. The charges of the boundary sojourns
of the upper paths are fixed, and a summation is done over the charges
of the internal sojourns. 
This summation then results in Eq. (\ref{cluster}).

\newpage

\newpage

\newpage

\newpage

\begin{figure}
\epsfxsize=7.0in
\epsfbox{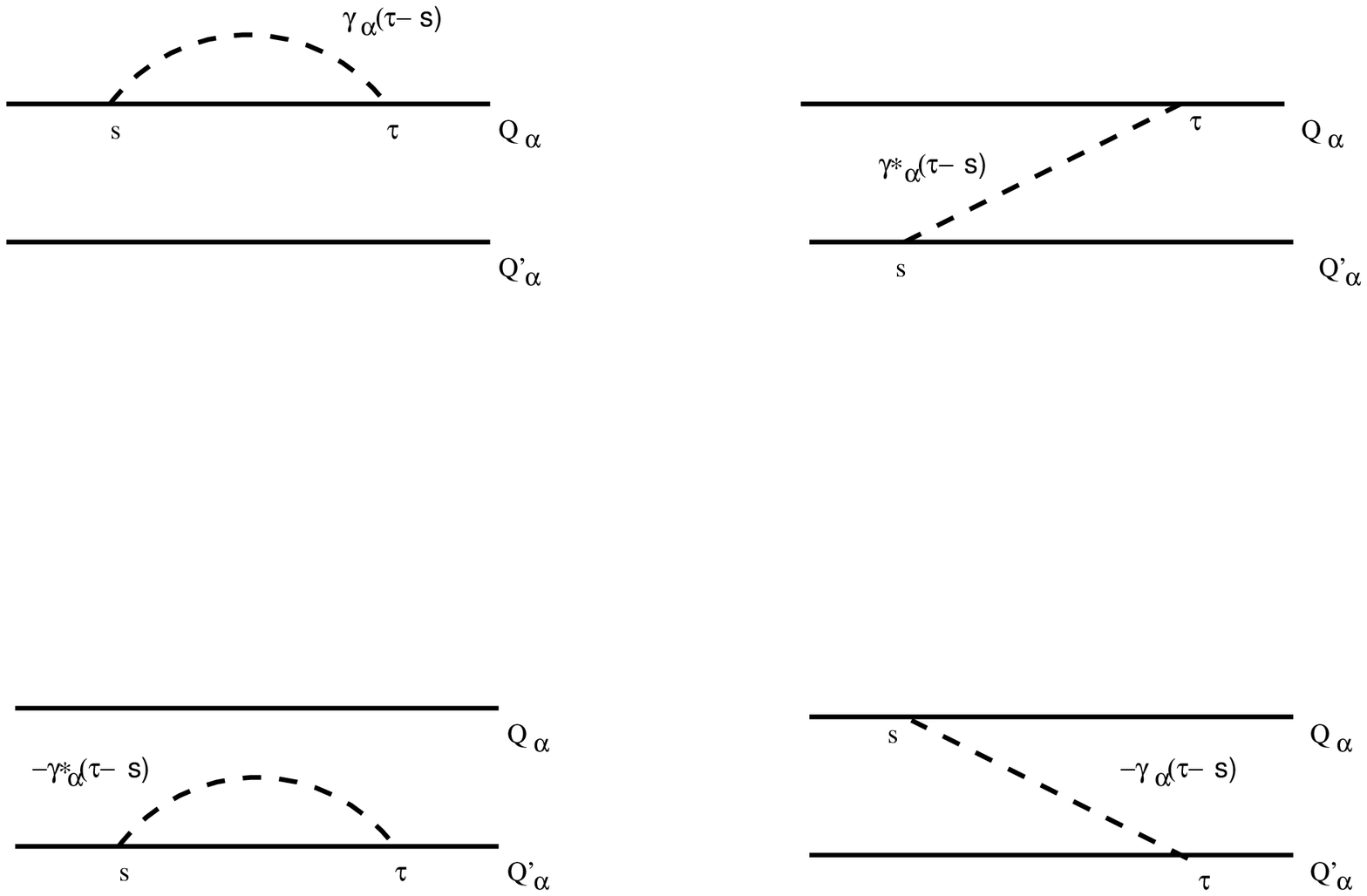}
\caption{}
\end{figure}

\newpage

\begin{figure}
\epsfxsize=7.0in
\epsfbox{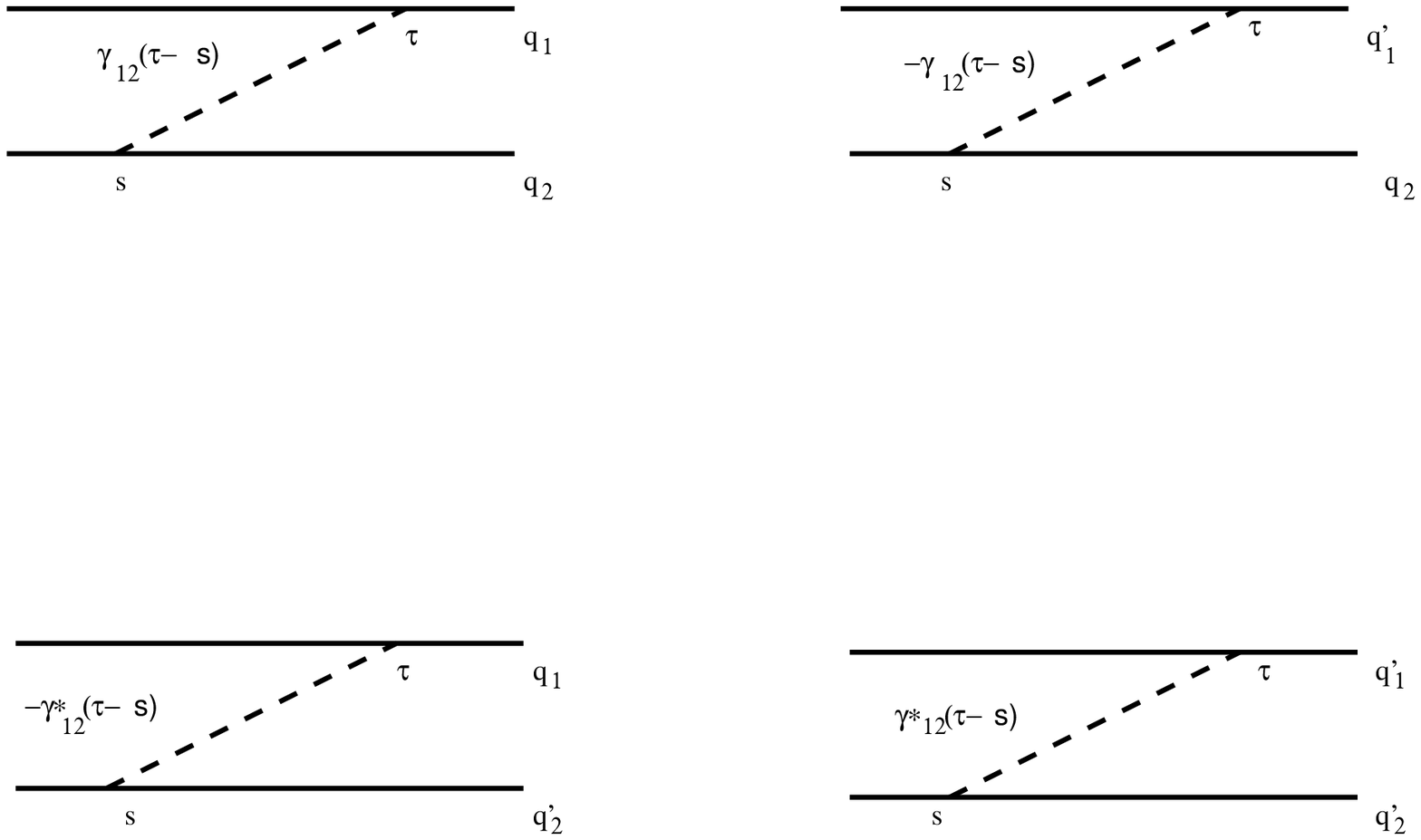}
\caption{}
\end{figure}

\newpage

\begin{figure}
\epsfxsize=6.5in
\epsfbox{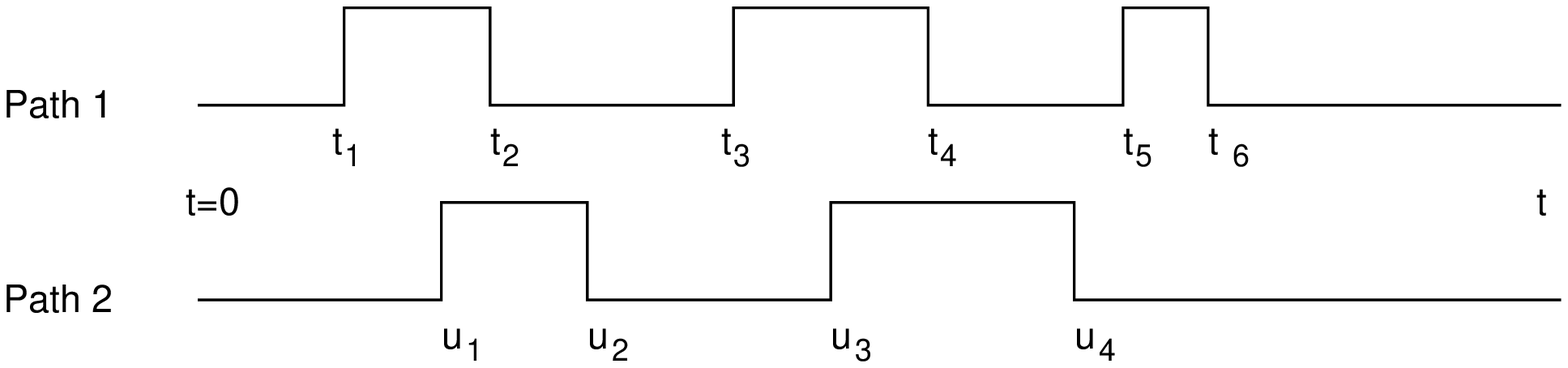}
\caption{}
\label{typpath}
\end{figure}

\newpage

\begin{figure}[t]
\epsfysize=6.0in
\epsfbox[15 200 580 700]{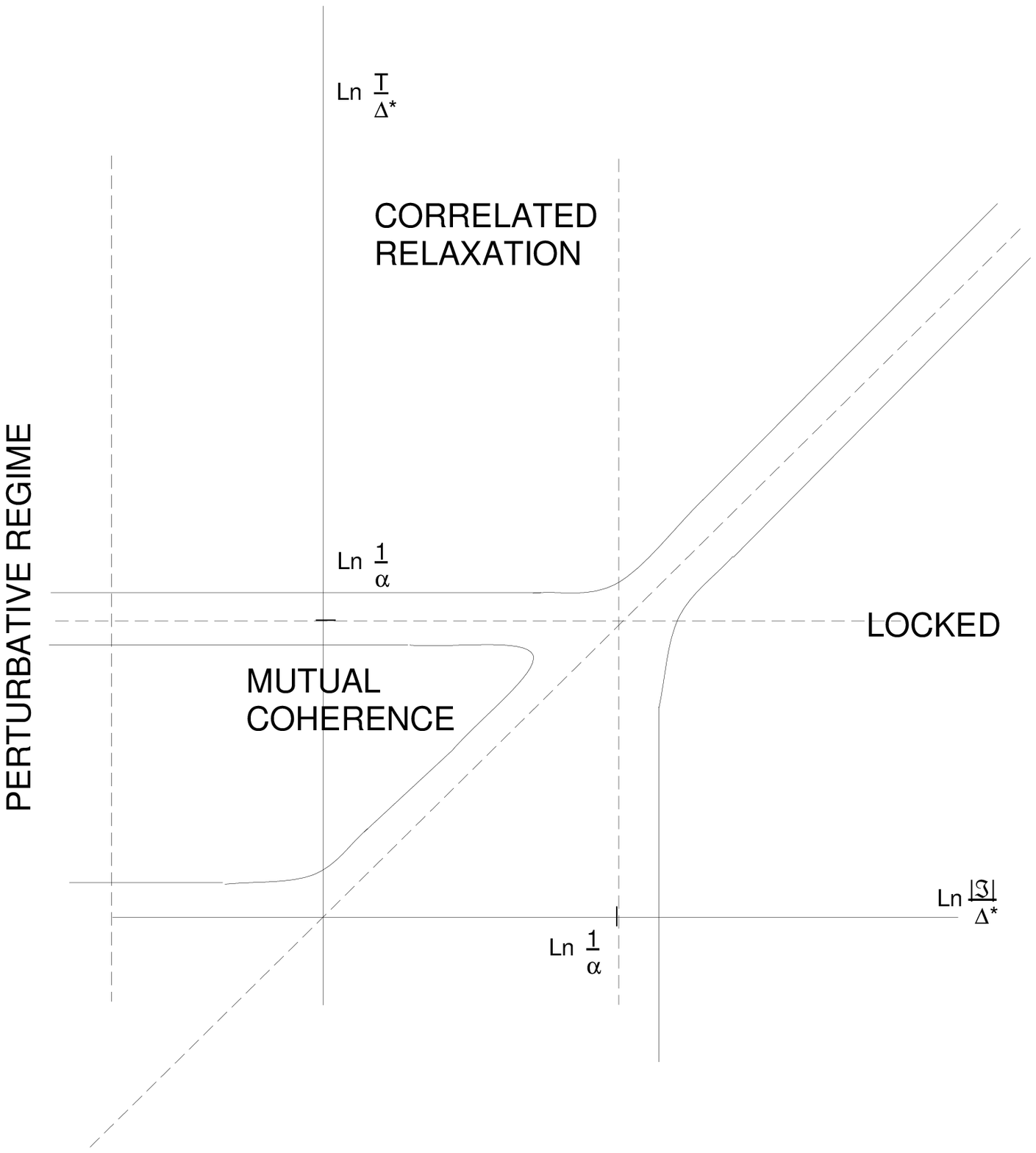}
\caption{}
\label{pphase}
\end{figure}

\newpage

\begin{figure}[t]
\epsfysize=6.0in
\epsfbox[20 200 580 700]{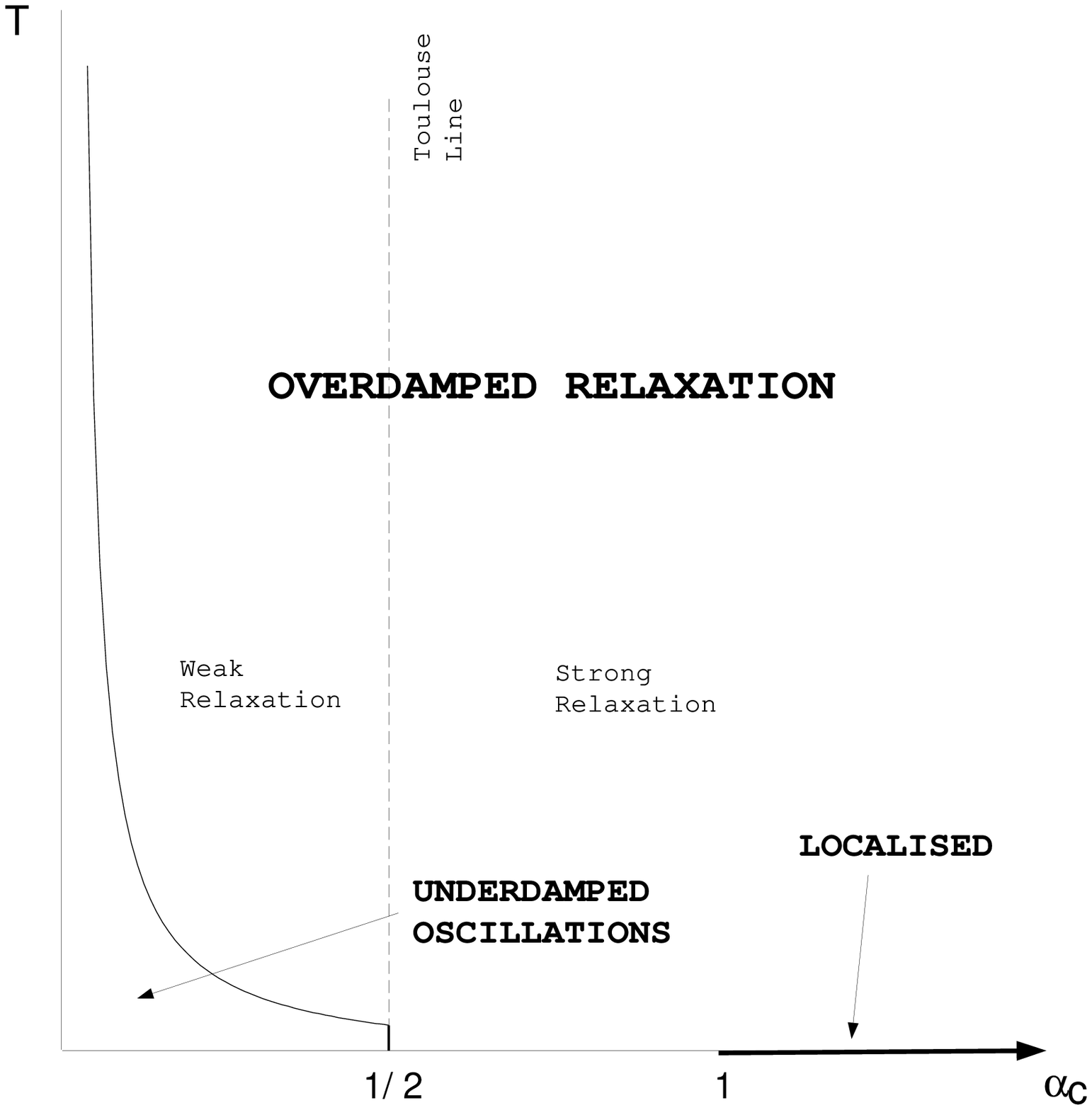}
\caption{}
\label{spbosphase}
\end{figure}

\newpage

\begin{figure}[t]
\epsfysize=6.0in
\epsfbox[0 150 550 760]{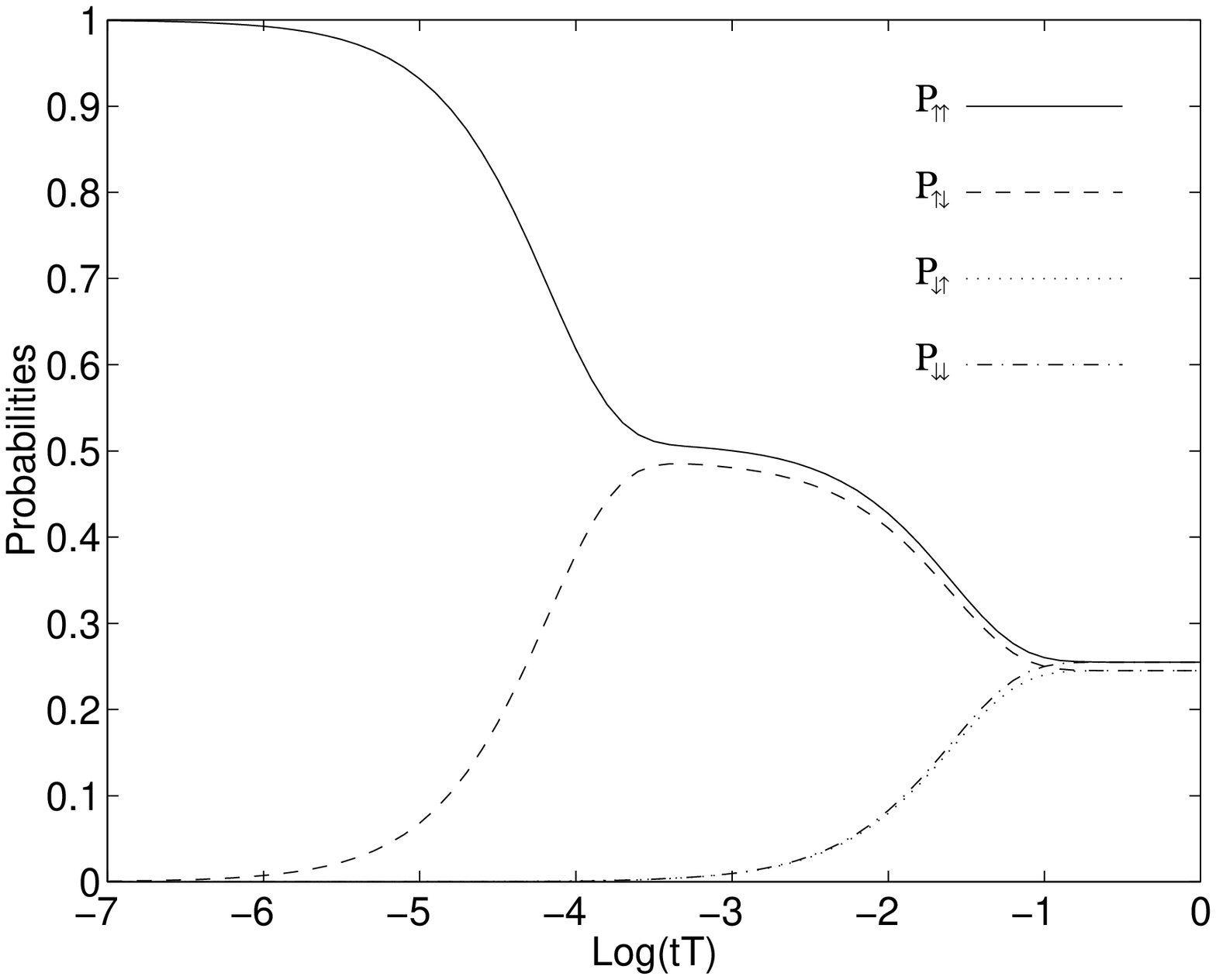}
\caption{}
\label{relaxfig}
\end{figure}

\newpage

\begin{figure}
\epsfysize=6.0in
\epsfbox[0 150 550 760]{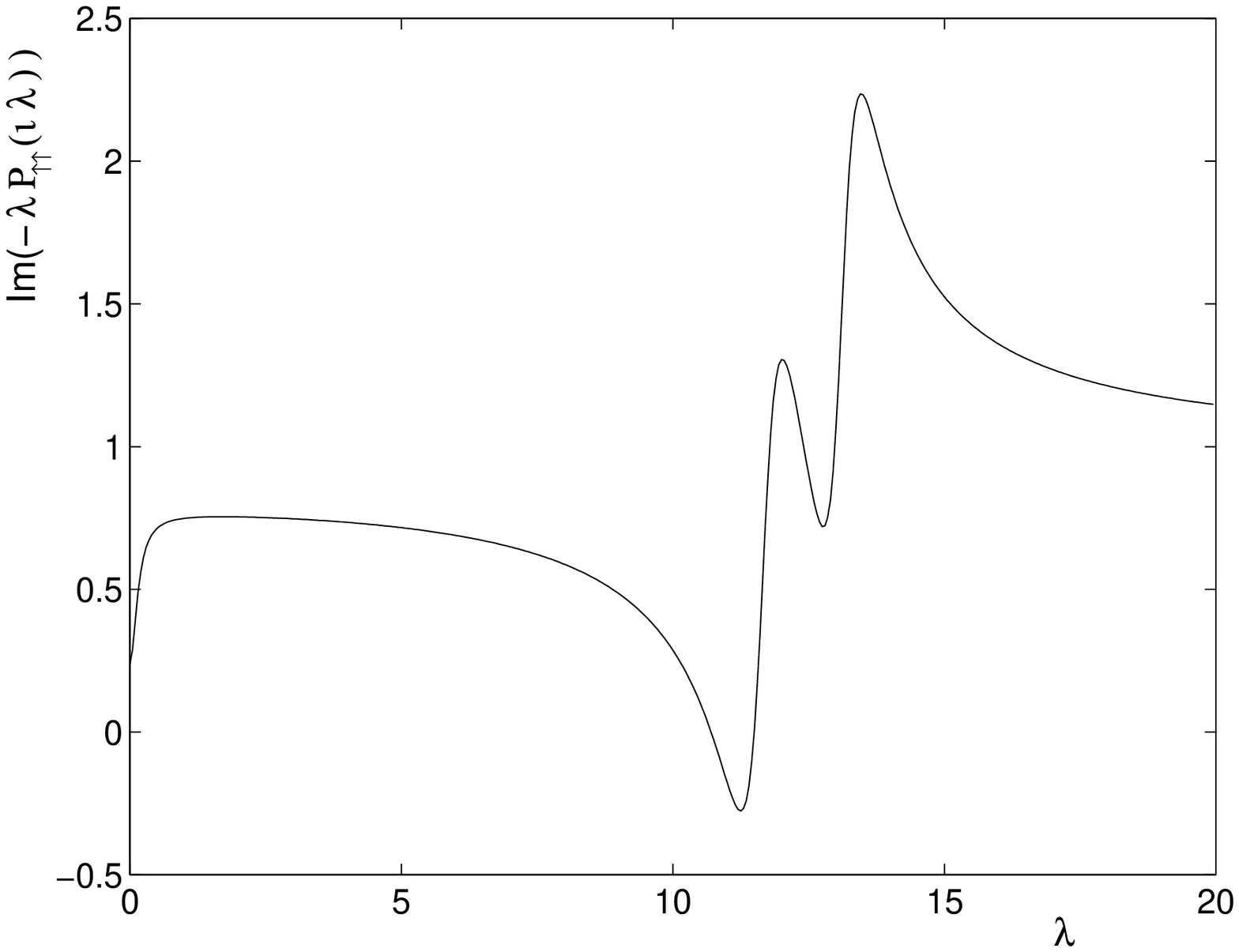}
\caption{}
\label{mutualfig}
\end{figure}

\newpage

\begin{figure}
\epsfxsize=6.0in
\epsfbox[0 150 550 760]{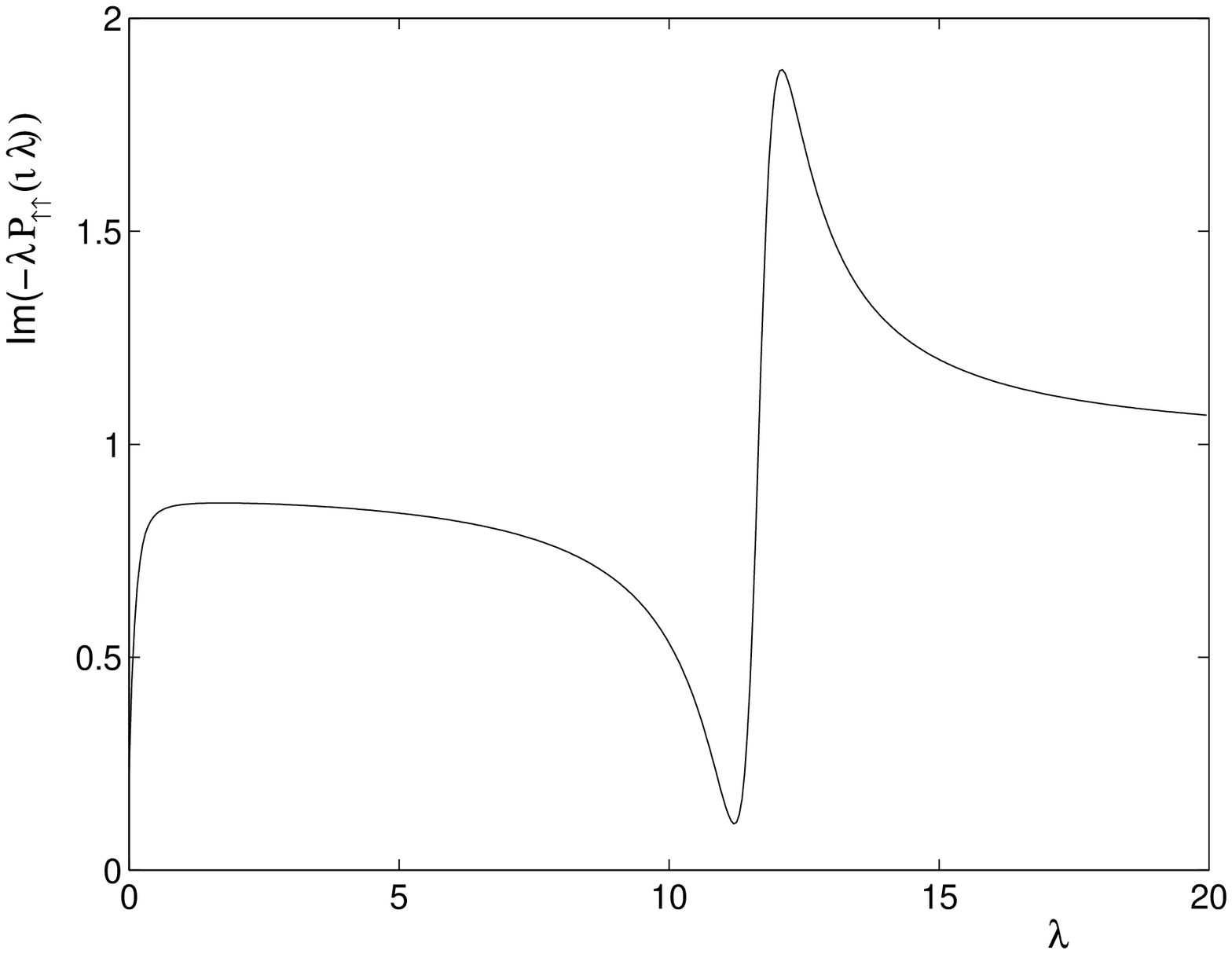}
\caption{}
\label{overpunderfig}
\end{figure}

\newpage

\begin{figure}
\epsfxsize=7.0in
\epsfbox{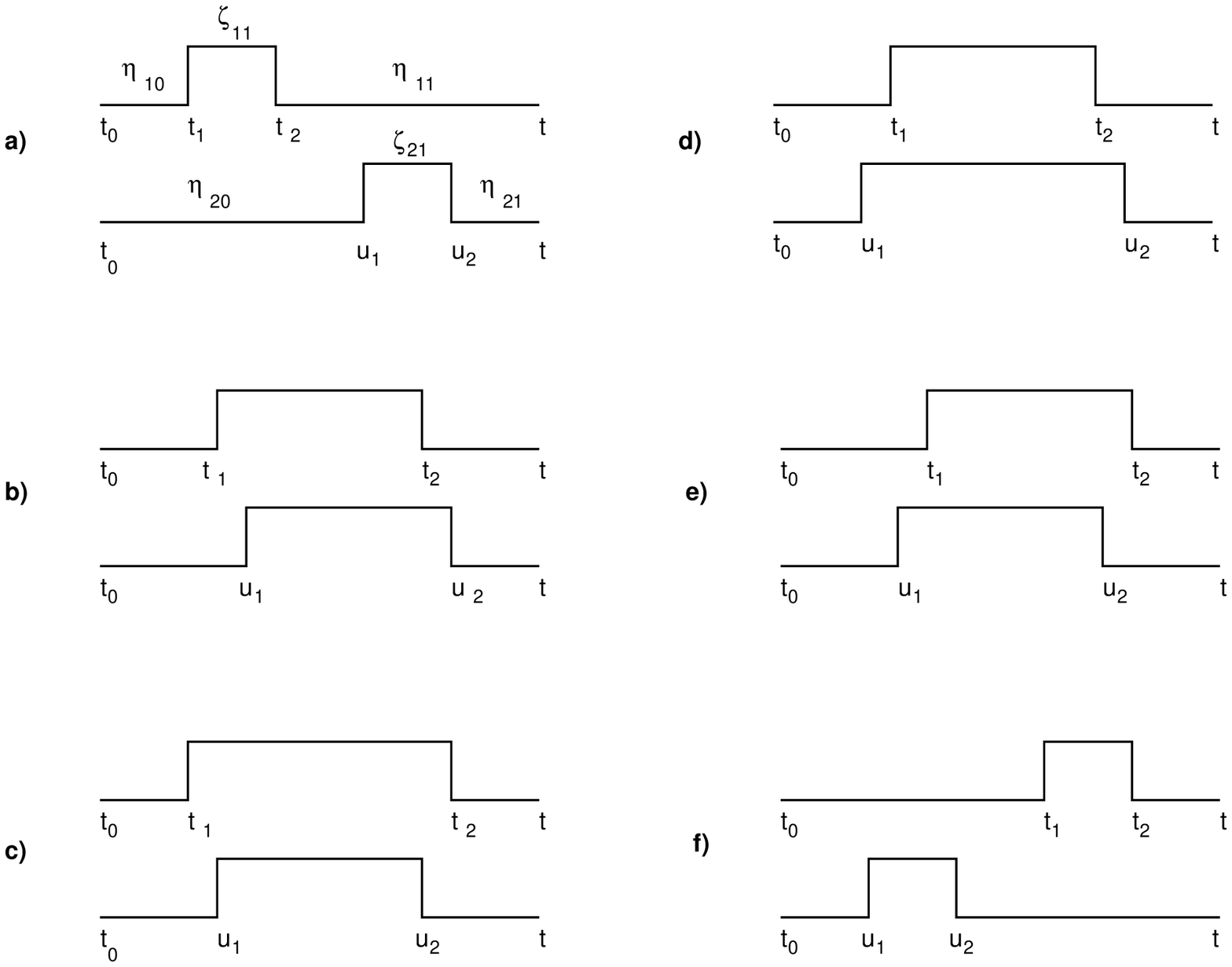}
\caption{}
\label{pathupup}
\end{figure}

\newpage

\begin{figure}
\epsfxsize=6.0in
\epsfbox{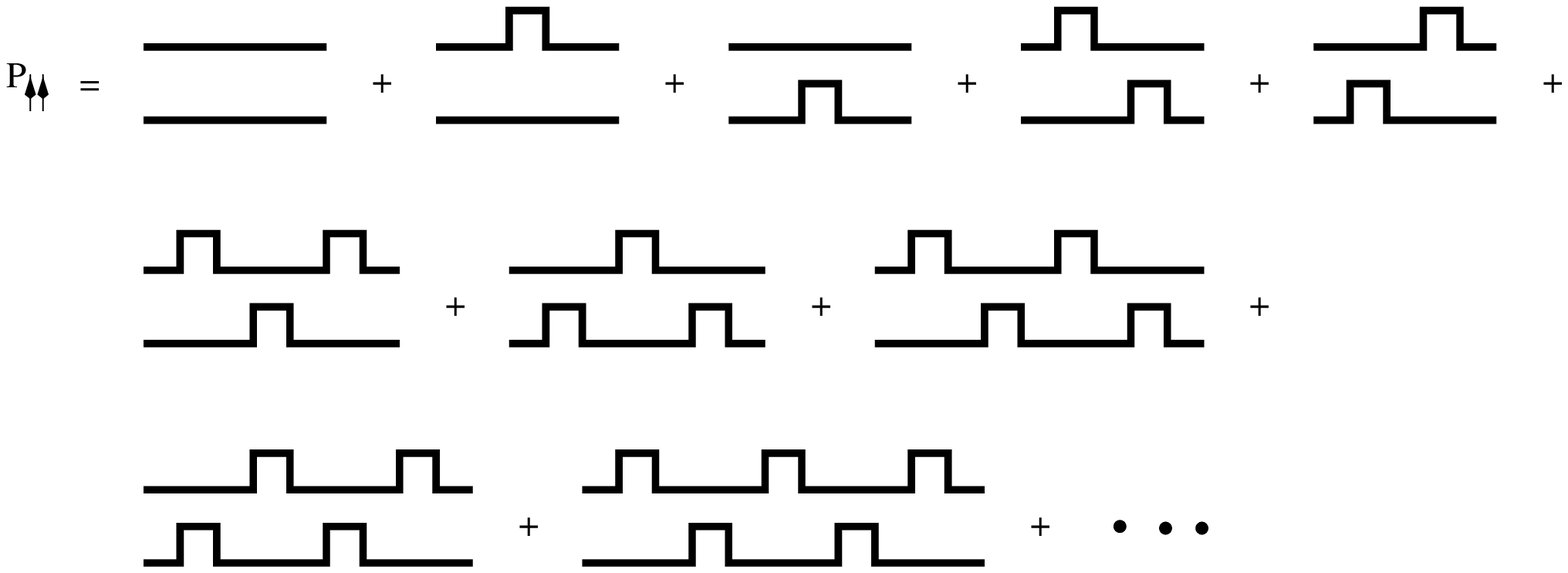}
\caption{}
\label{chainfig}
\end{figure}

\newpage

\begin{figure}
\epsfxsize=6.5in
\epsfbox{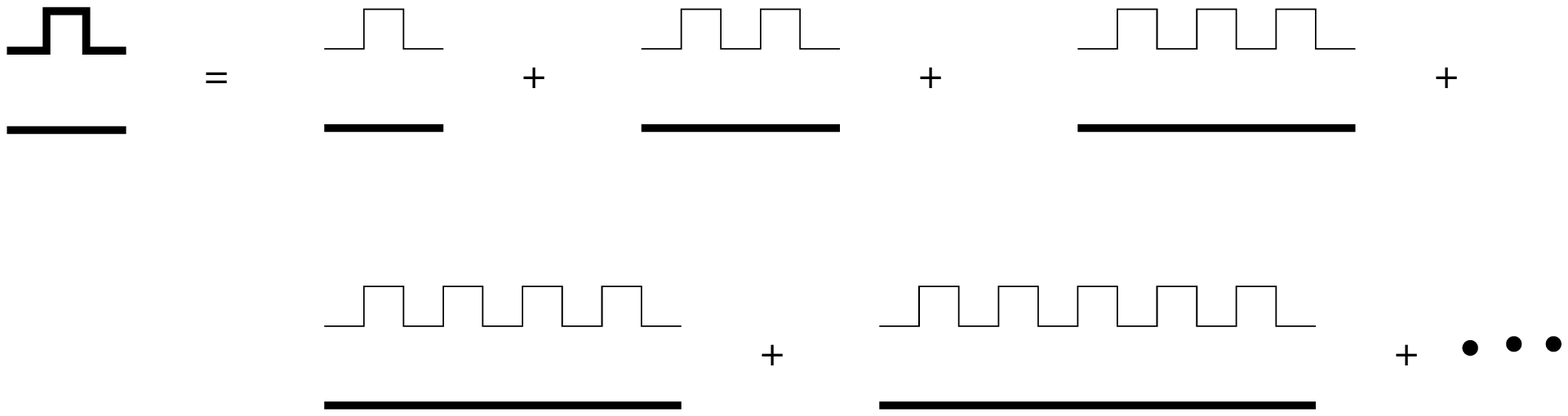}
\caption{}
\label{clusterfig}
\end{figure}

\end{document}